\documentclass[12pt]{article}

\usepackage[margin=1in]{geometry}

\usepackage{setspace}
\usepackage{amsmath, amssymb, graphicx, natbib}
\usepackage{eurosym}
\usepackage{bbm}
\usepackage[figuresright]{rotating}
\usepackage{algorithm}
\usepackage{algpseudocode}
\usepackage{xr-hyper} 
\usepackage[colorlinks=true, citecolor=blue]{hyperref}

\usepackage{multirow}
\usepackage{booktabs}
\usepackage{threeparttable}
\usepackage{siunitx}
\usepackage{booktabs}
\usepackage{array}
\usepackage{colortbl}
\usepackage{xcolor}
\usepackage{adjustbox}

\usepackage{caption}
\usepackage{booktabs}
\usepackage{threeparttable}
\usepackage[nomarkers,nolists]{endfloat}

\usepackage{etoolbox}
\AtBeginEnvironment{equation}{\setlength{\abovedisplayskip}{3pt}\setlength{\belowdisplayskip}{3pt}}

\makeatletter
\let\oldthebibliography\thebibliography
\let\endoldthebibliography\endthebibliography
\renewenvironment{thebibliography}[1]%
  {\oldthebibliography{#1}%
   \setlength{\itemsep}{0.85ex}   
   \setlength{\parskip}{0pt}
   \setlength{\parsep}{0pt}}
  {\endoldthebibliography}
\makeatother

\usepackage{lmodern}
\usepackage[T1]{fontenc}
\usepackage{amsthm, mathtools}
\usepackage{bm}
\usepackage{microtype}
\usepackage[nameinlink,noabbrev]{cleveref}
\usepackage{float}
\usepackage{placeins}
\usepackage{needspace}
\usepackage{chngcntr}

\hypersetup{
  colorlinks=true,
  linkcolor=blue,
  citecolor=blue,
  urlcolor=blue
}

\theoremstyle{definition}

\theoremstyle{remark}

\title{
  \Large \textbf{Supplementary Material for}\\[0.3em]
  \large \textbf{Doubly Robust Estimators of Quantile Treatment Effects}\\
  \large \textbf{With Semiparametric Cumulative Probability Models}
}
\begin{document}

\begin{center} {\large \bfseries Doubly Robust Estimators of Quantile Treatment Effects\\ With Semiparametric Cumulative Probability Models} 

\vspace{0.8cm}

{
Hao Wu$^{1}$, Chun Li$^{2}$, and Bryan E. Shepherd$^{1}$
}

\vspace{0.3cm}

\small
$^{1}$Department of Biostatistics, Vanderbilt University Medical Center,
Nashville, Tennessee, USA\\
$^{2}$Division of Biostatistics and Health Data Science, Department of
Population and Public Health Sciences, Keck School of Medicine of the
University of Southern California, Los Angeles, California, USA

\vspace{0.2cm}

\normalsize
\noindent Corresponding author: Bryan E. Shepherd (bryan.shepherd@vanderbilt.edu)

\end{center}
\textbf{Abstract}\\
The causal inference literature has traditionally focused on estimating the mean of the potential outcome, whereas 
evaluating how a treatment affects the entire outcome distribution can provide additional information in biomedical research. Quantile treatment effect (QTE) captures such distributional 
differences, particularly when outcomes are skewed. However, existing approaches for estimating QTE make distributional assumptions about the outcome and are thus sensitive to model misspecification. Motivated by an HIV study with skewed outcomes, one of which is subject to detection limits, we propose a doubly robust framework for estimating QTE based on the cumulative 
probability model (CPM), which is a rank-based, semiparametric linear transformation model. We develop two CPM-based estimation strategies: (1) an inverse-cumulative distribution function (CDF) approach 
that first estimates the marginal CDF of potential outcomes using the efficient influence 
function (EIF) and then obtains marginal quantiles via weighted quantile interpolation by inverting the distribution, and (2)
a direct approach that solves the EIF of potential marginal quantiles. The proposed estimators are doubly robust and asymptotically normal. We further extend the framework to probability treatment effects (PTEs) and their conditional
counterparts. For statistical inference, we investigate several variance estimation procedures, including EIF-based estimators, sandwich estimators, and the nonparametric bootstrap. Simulation studies illustrate that the empirical sandwich estimator and the nonparametric bootstrap provide doubly robust variance estimation with stable finite-sample performance under nuisance-model misspecification. The proposed methods are evaluated through extensive Monte Carlo simulations and illustrated using an HIV data application.

\newpage
\section{Introduction\label{sec1}}

Evaluating how a treatment affects the entire outcome distribution, rather than just its mean, is becoming increasingly appealing to biomedical researchers. 
For example, drug developers 
may be interested in how a treatment affects the lower or upper tail of the outcome 
distribution (\citealt{meyerson2023use}). 
In HIV research, outcomes such as CD4 count and viral load are highly skewed, so mean-based analyses may be unstable. In addition, viral load is subject to detection limits, making mean-based analyses untenable unless one makes strong additional assumptions about values below detection limits. 
Consequently, analyses focusing solely on the mean may fail to accurately characterize treatment effects (\citealt{yirga2022application}, \citealt{heller2022distributional}).

One can compare the quantiles of outcome 
distributions under treatment and control conditions. A parameter of interest is the quantile treatment effect (QTE), which was originally defined by \citealt{doksum1974empirical} and 
\citealt{lehmann1975statistical} as the difference between the unconditional quantiles of 
potential outcomes under treatment and under control in a setting with binary treatment over a range of quantiles. 
Closely related to the QTE is the probability treatment effect (PTE), which measures treatment-induced differences in the cumulative distribution functions of potential outcomes at fixed outcome values. While PTEs characterize distributional effects on the probability scale, QTEs summarize distributional differences on the quantile scale.

Growing interest in QTE has motivated several different estimation procedures.
\citealt{firpo2007efficient} first 
developed efficient estimation of QTE and proposed an inverse probability weighting (IPW) estimator based on propensity scores estimated using a logistic power series approximation.
However, its performance heavily depends on the correct specification of propensity scores.
\citealt{zhang2012causal} proposed an outcome regression estimator for the QTE, which avoids 
modeling propensity scores and
is more efficient than the IPW estimator in finite samples under correct model specification. 
However, their method requires proper specification for consistency and assumes a normal linear model, potentially after applying a Box–Cox transformation, for
the outcome variable.

Doubly robust estimators have become increasingly
popular in causal inference because they provide consistent point estimation even 
under misspecification of one of the propensity score or outcome regression 
models (\citealt{robins1994estimation}, \citealt{bang2005doubly}, \citealt{kang2007demystifying}, 
\citealt{funk2011doubly}, \citealt{kennedy2024semiparametric}). Doubly robust estimators for QTE have been proposed. \citealt{zhang2012causal} proposed an augmented inverse 
probability weighted (AIPW) estimator that is doubly robust and enhances the 
efficiency of the IPW estimator by augmenting with a term that 
involves the residuals from the outcome regression model. However, this approach heavily relies on the assumption of normality for the outcome distribution. \citealt{diaz2017efficient} proposed a semiparametric
targeted maximum likelihood estimation (TMLE) approach for estimating the QTE (see also \citealt{koenker2016tmle} for an accessible exposition), which shares the same asymptotic properties as the AIPW 
estimator, but has slightly better finite-sample performance. The TMLE framework can use conditional quantile regression (\citealt{koenker2005quantile}) as
the outcome regression model, which allows for more flexible 
handling of outcomes that deviate from normality. However, 
conditional quantile regression estimates specific 
quantile levels rather than the entire conditional outcome 
distribution. 

In this manuscript, we propose a  cumulative 
probability model (CPM)-based doubly robust framework for estimating QTE that is valid for both normally and non-normally distributed outcome data.
Early work by \citealt{liu2017modeling} showed that continuous outcomes can be effectively modeled using the CPM, which was originally developed for 
ordinal outcomes and is also known as the cumulative link model (\citealt{walker1967estimation}, \citealt{mccullagh1980regression}).
CPMs are a type of semiparametric linear transformation model 
(\citealt{zeng2007maximum}), in which an unspecified monotone
transformation of the outcome is assumed to follow a linear 
model. The CPM is highly flexible because the transformation 
function is estimated nonparametrically, eliminating the need to 
specify a transformation in advance. Moreover, its estimation 
relies only on the ranking of the outcomes rather than their 
exact numeric values.

The remainder of this paper is organized as follows. Section~$\ref{sec2}$ provides a detailed 
review of the CPM. Section~$\ref{sec3}$ introduces the proposed CPM-based 
doubly robust framework for estimating QTEs. In Section~$\ref{sec4}$, we 
evaluate the performance of the proposed method and compare it 
with existing approaches through simulation studies. Section~$\ref{sec5}$ 
illustrates the method using a real HIV study. The final section
discusses the strengths and limitations of our approach and 
outlines directions for future research.

\section{Cumulative Probability Model\label{sec2}}

With continuous outcomes, the cumulative probability model (CPM) is a class of semiparametric linear 
transformation models (\citealt{zeng2006efficient}, \citealt{de2019semiparametric}). 
The observed continuous response variable $Y$ is modeled through an unknown 
monotonically increasing transformation $H(\cdot)$ applied  to a latent continuous 
variable $Y^*$, where $Y = H(Y^*)$, $ Y^*
= \boldsymbol{\beta}^T \mathbf{X} + \epsilon$, and $\epsilon$ represents an error 
term assumed to follow a prespecified distribution $F_\epsilon$. The model can be 
expressed as:
\begin{equation}\label{eq:CPM1}
Y = H(\boldsymbol{\beta}^T\mathbf{X} + \epsilon), \quad \epsilon \sim F_\epsilon.
\end{equation}
Under this framework, the cumulative distribution function (CDF) of $Y$ conditional on covariates $\mathbf{X}$ is given by:
\begin{equation}\label{eq:cpm2}
\begin{aligned}
F(y \mid \mathbf{X}) 
  &= P\!\left(Y \leqslant y \mid \mathbf{X}\right) = P\!\left[H(\boldsymbol{\beta}^\top \mathbf{X} + \epsilon) \leqslant y \mid \mathbf{X}\right] \\
  &= P\!\left[\epsilon \leqslant H^{-1}(y) - \boldsymbol{\beta}^\top \mathbf{X} \mid \mathbf{X}\right] = F_\epsilon\!\left(H^{-1}(y) - \boldsymbol{\beta}^\top \mathbf{X}\right).
\end{aligned}
\end{equation}
For simplicity and clarity of notation, let $G = F_\epsilon^{-1}$ denote a link function, and define $\alpha(y) = H^{-1}(y)$ as an intercept function. Then, equation \eqref{eq:cpm2} can be written as: 
\begin{equation}\label{eq:cpm3}
   G\{F(y \mid \mathbf{X})\} = \alpha(y)-\boldsymbol{\beta}^T \mathbf{X}. 
\end{equation}
The intercept function $\alpha(y) = G\{F(y \mid \mathbf{X} = \mathbf{0})\}$
represents the link-transformed CDF evaluated at $\mathbf{X} = 
\mathbf{0}$ (i.e., the baseline CDF), and $\boldsymbol{\beta}^T \mathbf{X}$ 
quantifies shifts in this baseline CDF due to covariates. 
Holding the remaining covariates fixed, a positive 
(negative) $\beta_j$ indicates that an increase in $X_j$ is associated with a 
stochastic increase (decrease) in the distribution of the outcome. The
specific interpretation of 
$\boldsymbol{\beta}$ depends on the chosen link function. For example, selecting a
logistic distribution for $F_\epsilon$ (logit link) yields a log-odds ratio 
interpretation, whereas choosing an extreme value distribution (complementary log-log
link) results in a log hazard ratio interpretation (\citealt{liu2017modeling}). 

In the CPM $\eqref{eq:cpm3}$, the intercept $\alpha(y)$ and coefficients 
$\boldsymbol{\beta}$ can be jointly estimated via nonparametric 
maximum likelihood estimation (\citealt{zeng2007maximum}, \citealt{liu2017modeling}). 
 For an independent and identically distributed (i.i.d.) sample $\left\{\left(y_i, \mathbf{X}_i\right): i=1, \ldots, n\right\}$, the 
 likelihood contribution of subject $i$ is given by $F\left(y_i \mid \mathbf{X}_i\right)-
 F\left(y_i^{-} \mid \mathbf{X}_i\right),$ where $F\left(y_i^{-} \mid \mathbf{X}_i\right)=\lim _{t 
 \uparrow y_i} F\left(t \mid \mathbf{X}_i\right)$. Therefore, the likelihood is 
$L(\boldsymbol{\beta}, \boldsymbol{\alpha}) 
= \prod_{i=1}^n \left[ F\!\left(y_i \mid \mathbf{X}_i\right) - F\!\left(y_i^{-} \mid \mathbf{X}_i\right) \right]$.
Let $t_{1} \cdots t_J$ be the unique ordered values of $\{y_i\}$, where $J$ denotes the number of distinct response values. Let $S  = \{ t_1, \cdots, t_J\}$. These serve as the anchor points for the nonparametric likelihood. Let  $\alpha_j = \alpha_{t_{j}}$; then $\alpha_1 < \cdots < \alpha_J $. 
The nonparametric likelihood is
\begin{align}
L(\boldsymbol{\beta}, \boldsymbol{\alpha}) =\; & \prod_{i: y_i=t_{1}} G^{-1}\!\left(\alpha_1- \boldsymbol{\beta}^T \mathbf{X}_i\right) \notag \\[6pt]
&\times \prod_{j=2}^{J-1} \prod_{i: y_i=t_{j}} 
   \Bigl[ G^{-1}\!\left(\alpha_j-\boldsymbol{\beta}^T \mathbf{X}_i\right) 
   - G^{-1}\!\left(\alpha_{j-1}-\boldsymbol{\beta}^T \mathbf{X}_i\right) \Bigr] \notag \\[6pt]
& \times \prod_{i: y_i=t_{J}} 
   \Bigl[ 1 - G^{-1}\!\left(\alpha_{J-1}-\boldsymbol{\beta}^T \mathbf{X}_i\right) \Bigr] 
\label{eq:NPMLE2}
\end{align}
The likelihood in $\eqref{eq:NPMLE2}$ is equivalent to the likelihood of an ordinal regression model, where each distinct observed outcome value is regarded as a separate ordinal category. Maximizing it, we obtain nonparametric maximum likelihood estimates (NPMLEs)
$(\hat{\boldsymbol{\beta}},\hat{\boldsymbol{\alpha}})$, where $\hat{\boldsymbol{\alpha}} = 
(\hat{\alpha}_1, \cdots, \hat{\alpha}_{J-1})$. $\alpha_J$ is not estimated because the probabilities in a 
multinomial form of the likelihood $\eqref{eq:NPMLE2}$ sum to one across categories. Under mild regularity 
conditions including boundedness of the response variable, the NPMLEs $(\hat{\boldsymbol{\beta}}, 
\hat{\boldsymbol{\alpha}})$ obtained from CPMs are consistent and asymptotically normal, with variance estimates readily obtainable from the inverse of the observed 
information matrix (\citealt{li2023asymptotic}). In practice, a concern with estimating the NPMLEs 
$(\hat{\boldsymbol{\beta}}, \hat{\boldsymbol{\alpha}})$ with continuous outcomes is that the number of parameters increases with
the number of unique outcome values, which can make estimation computationally intensive. 
However, the \texttt{orm()} function in the \texttt{rms} package in R addresses this issue by exploiting the 
sparse structure of the score function and Hessian matrix, allowing efficient inversion through Cholesky decomposition in the 
Newton–Raphson algorithm (\citealt{rms}).
For example, fitting a CPM with $N=300{,}000$ levels (no ties) and 20 covariates, which requires
299,999 intercepts, takes only 2.5 seconds on a standard desktop computer.

CPMs have several attractive properties.
First, regression analyses for continuous outcomes often require strong distributional assumptions or a 
prespecified transformation, which may be difficult to justify (\citealt{box1964analysis}). The CPM does not have
this limitation because it directly estimates a transformation function rather than assuming a fixed form 
(\citealt{tian2020empirical}). Second, estimation of $\boldsymbol{\beta}$ in 
the CPM depends only on the relative ordering of the outcomes, not on their exact numeric values. This makes
the CPM invariant to any monotonic transformation of the response and thus suitable for ordinal, continuous, 
or mixed outcomes. Such robustness is particularly valuable when outcomes are 
subject to detection limits, where values above the limit are measured continuously but values below the limit
are only categorized (\citealt{tian2024addressing}). Third, unlike traditional regression models that 
typically focus on a single aspect of the conditional distribution, such as the conditional mean, the CPM models the entire conditional distribution. This allows direct derivation of 
multiple summaries, such as conditional exceedance probabilities, conditional expectations and conditional quantiles. The CPM framework further allows flexible modeling through the inclusion of spline terms and interactions, enabling rich covariate effects. Together, these properties make the CPM an excellent choice for constructing estimators of the QTE.  In this paper, the CPM is used as a convenient and flexible working model for estimating conditional distribution functions.

\section{CPM-based Doubly Robust Estimators of QTE\label{sec3}}
\subsection{Notation and Identification of Causal Estimand\label{sec3.1}}

Let $A$ denote the treatment indicator taking values in $\{0,1\}$, where $A=1$ indicates treatment and $A=0$ indicates control. Let $Y$ be an orderable outcome (i.e., continuous, ordinal, or a mixture of the two)
and $\mathbf{X}$ a vector of covariates. Suppose the true joint distribution of the observed data $\mathbf{Z} = (Y, A, \mathbf{X})$ is denoted by $P_0$. We observe an i.i.d. sample $Z_i= (Y_i, A_i, \mathbf{X}_i)$, $i=1, \dots, n$, with empirical distribution denoted by $P_n$. Let $Y_a$ denote the potential outcome under treatment $a$ for a generic subject, and define $F_{a}(y) = \Pr(Y_a \leq y)  = \mathbb{E}\left[\mathbb{I}(Y_a \leq y)\right]$ as the marginal CDF of $Y_a$, where $a \in \{0,1\}$ and $\mathbb{I}(\cdot)$ denotes the indicator function. (With a slight abuse of notation, we use $Y_i$ for observed values, but $Y_a$ for potential outcomes. It should be clear from the context.) 

The $p^{th}$ quantile of the distribution of $Y_a$ is $q_{a}(p) = F_{a}^{-1}(p) = \inf \{y : F_{a}(y) \geq p\}$ for $p \in (0,1)$. The Quantile Treatment Effect (QTE) at quantile level $p$ is defined as the difference in the $p^{th}$ quantiles of the distributions of potential outcomes under treatment and control:
\begin{equation}\label{eq:QTE}
    \Delta(p)  = \operatorname{QTE}(p) = q_{1}(p) - q_{0}(p) = F_{1}^{-1}(p) - F_{0}^{-1}(p), \quad p \in (0,1).
\end{equation}
An important complementary estimand to the QTE is the probability treatment effect (PTE), which measures treatment effects on the cumulative probability scale. For a given threshold $y$, the PTE is defined as
\begin{equation}\label{eq:PTE}
   \mathrm{PTE}(y) = F_{1}(y) - F_{0}(y).
\end{equation}
In fact, the PTE is effectively a special case of average treatment effect (ATE) because of $F_{a}(y) =  \mathbb{E}\left[\mathbb{I}(Y_a \leq y)\right]$. Because QTE is obtained by inverting the marginal CDFs, estimation of $F_{a}(y)$ provides a unified framework for both QTE and PTE inference. In this work, we focus primarily on QTE estimation, with PTE serving as a secondary causal estimand.

To identify $F_{a}(y)$, and thus the QTE and PTE, we make four commonly imposed identification assumptions (\citealt{firpo2007efficient}, \citealt{hernan2020causal}, \citealt{naimi2023defining}): 
\begin{itemize}
    \item \textit{Consistency}: The observed outcome corresponds exactly to the potential outcome under the actual treatment received: $Y = A Y_1 + (1 - A) Y_0.$

    \item \textit{Unconfoundedness}: Conditional on observed covariates, the treatment assignment is independent of potential outcomes: $(Y_0, Y_1 ) \perp A \mid \mathbf{X}.$

    \item \textit{Positivity}: $0 < \Pr(A = 1 \mid \mathbf{X}) < 1.$

    \item \textit{No interference}: The treatment assignment for any subject does not affect the potential outcomes of other subjects.
\end{itemize}

Under these assumptions, $F_{a}(y)$ is identifiable and can be expressed as 
\begin{equation}
    F_{a}(y) = \mathbb{E}\!\Bigl[\Pr\bigl(Y \le y\mid A=a,\;\mathbf X\bigr)\Bigr] = \mathbb{E}\Bigl[ \frac{\mathbb{I}(A=a) \mathbb{I}(Y \leq y)}{\Pr(A = a \mid  \mathbf{X})} \Bigr].
    \label{Fy}
\end{equation}
Additional technical details regarding this identification argument are provided in Supplementary Material S1. Because $F_{a}(y)$ is identifiable, PTE is identifiable. Substituting the identified CDF into the quantile functional yields $q_{a}(p)=\inf \left\{y: F_{a}(y) \geq p\right\} $, which is therefore identifiable for any $p \in(0,1)$. The QTE at level $p$, $\Delta(p)=q_{1}(p)-q_{0}(p)$, is thus also identifiable.

\subsection{Estimation\label{sec3.2}}
\subsubsection{Estimation through the Distribution \label{sec3.2.1}}
A straightforward approach to estimate $F_{a}(y) $ is to model the conditional distribution of $Y$ given $(A, \mathbf{X})$,  then average over the empirical distribution of covariates, following the representation in the middle expression of ($\ref{Fy}$). Specifically, we specify a working model for the conditional CDF $F_{Y \mid A=a, \mathbf{X}}(y)
= \Pr(Y \le y \mid A=a, \mathbf{X}),$
and obtain an estimator of the marginal distribution by plugging in the fitted conditional CDF values. We posit a working model $F^*_{Y \mid A=a, \mathbf{X}}(y; \boldsymbol{\xi})$, indexed by a parameter vector $\boldsymbol{\xi}$. We refer to this plug-in strategy as the outcome regression (OR) approach, $\widehat{F}_{a}^{\mathrm{OR}}(y)
 =
 \frac{1}{n}\sum_{i=1}^{n}
 {F}^*_{Y \mid A=a, \mathbf{X}_i}(y; \hat{\boldsymbol{\xi}})$. In this work, ${F}^*_{Y \mid A=a, \mathbf{X}}(y; \boldsymbol{\xi})$ is modeled using the CPM described in Section~\ref{sec2}. The CPM framework allows substantial modeling flexibility. One may fit a single model including the treatment indicator as a covariate, or alternatively fit separate CPMs within the $A=0$ and $A=1$ strata, permitting group-specific covariate effects and distributional shapes across treatment arms.
 
An alternative estimation strategy is based on modeling the propensity score $\pi(a \mid \mathbf{X}) = \Pr(A=a \mid \mathbf{X})$. We posit a working model $\pi^*(a \mid \mathbf{X}; \boldsymbol{\psi})$, indexed by a parameter vector $\boldsymbol{\psi}$. In practice, logistic regression is often used, although alternative models may also be employed. The marginal CDF is obtained through inverse probability weighting of the observed outcomes, following the representation in the right hand side of \eqref{Fy}. We refer to this weighting-based estimator as the inverse probability weighting (IPW) approach, denoted as $\widehat{F}^{IPW}_{a }(y)  = \frac{1}{n}
   \sum_{i=1}^n
   [
     \frac{\mathbb{I}(A_i = a)\,\mathbb{I}\{Y_i \le y\}}
          {\pi^*(a \mid \mathbf{X}_i; \hat{\boldsymbol{\psi}})}]$.

A doubly robust estimator of $F_{a}(y)$ can be constructed by solving the estimating equation derived from the efficient influence function (EIF) of $F_{a}(y)$. For any regular and asymptotically linear estimator $\hat{\boldsymbol{\theta}}$ of the parameter of interest $\boldsymbol{\theta}$ with true value $\boldsymbol{\theta}_0$, we have $\sqrt{n} \bigl(\hat{ \boldsymbol{\theta}} - \boldsymbol{\theta}_0\bigr)  =  \frac{1}{\sqrt{n}}\sum_{i =1}^n{\varphi}(Z_i) + \mathbf{o}%
_{p}(1)$,   
where ${\varphi}(Z_i)$ is a mean-zero measurable function with finite and
nonsingular variance, referred to as the influence function (IF) (\citealt{newey1994asymptotic}, \citealt{robins1994estimation}). The EIF is the unique IF that yields the lowest asymptotic variance. The EIF of $F_{a}(y)$ under $P_0$ is 
\begin{equation}
\begin{aligned}
\varphi_{F}(Z; F_a (y))
&= 
     \frac{\mathbb{I}(A=a)}{\pi(a\mid\mathbf{X})}
       \mathbb{I}(Y\leq y)
     + \frac{\pi(a\mid\mathbf{X}) - \mathbb{I}(A=a)}
            {\pi(a\mid\mathbf{X})}
       F_{Y \mid A=a, \mathbf{X} }\!\bigl(y \bigr) - F_{a}(y).
\end{aligned}
\label{eq:phi_F}
\end{equation}
This expression is the same form as the EIF of the population mean (\citealt{hahn1998role}, \citealt{van2011targeted}, \citealt{kennedy2016semiparametric}). In practice, both nuisance components, $ \pi(a\mid\mathbf{X})$ and $ F_{Y 
\mid A=a, \mathbf{X} }\!\bigl(y\bigr)$ are unknown and are replaced by 
their estimates $ \pi^*(a \mid \mathbf{X}; \hat{\boldsymbol{\psi}})$ and $ F^*_{Y \mid A=a, \mathbf{X}}(y;
\hat{\boldsymbol{\xi}})$, respectively. Then, a doubly robust estimator of the marginal CDF $F_{a}(y)$ can be obtained by
solving the empirical estimating equation
$    \frac{1}{n} \sum_{i=1}^n 
\varphi_{F}\!\left(Z_i ; F_{a}(y), \hat{\boldsymbol{\psi}}, \hat{\boldsymbol{\xi}} \right)=0$ for $F_{a}(y)$, yielding:
\begin{equation}
    \widehat F_{a}^{DR}(y)
= \frac{1}{n}
   \sum_{i=1}^n
   \Biggl[
     \frac{\mathbb{I}(A_i = a)\,\mathbb{I}\{Y_i \le y\}}
          {\pi^*(a \mid \mathbf{X}_i; \hat{\boldsymbol{\psi}})}
     \;+\;
     \frac{\pi^*(a \mid \mathbf{X}_i;\hat{\boldsymbol{\psi}}) - \mathbb{I}(A_i = a) }
          {\pi^*(a \mid \mathbf{X}_i; \hat{\boldsymbol{\psi}})}
     \; F^*_{Y\mid A=a,\mathbf{X}}\bigl(y; \hat{\boldsymbol\xi}\bigr)
   \Biggr].
\label{eq:cdf_dr}
\end{equation}
This estimator is consistent if either the propensity score model $\pi^*(a \mid \mathbf{X}; \hat{\boldsymbol{\psi}})$ or the outcome regression model $F^*_{Y \mid A=a, \mathbf{X}}(y; \hat{\boldsymbol{\xi}})$ is correctly specified. As with the outcome regression estimator described earlier, we use the CPM to robustly model $F^*_{Y \mid A=a, \mathbf{X}}(y; \hat{\boldsymbol{\xi}})$. Throughout, we use logistic regression for the propensity score model.

In finite samples, $\widehat F_{a}^{DR}(y)$ is not guaranteed to satisfy the monotonicity and range constraints of a CDF (\citealt{van2006targeted}; \citealt{gruber2012tmle}; \citealt{zhang2012causal}). To address this, we apply a post-processing step to obtain a valid CDF estimator. Specifically, we define the truncated estimator $\widetilde F_{a}(y)
= \min\bigl\{1, \max\{0, \widehat F^{DR}_{a}(y)\}\bigr\}$, and enforce monotonicity by $\widehat F_{a}^{DR^*}(y) = \sup_{t \le y} \widetilde F_{a}(t)$, which corresponds to the cumulative maximum over ordered outcome values. These modifications are generally very minor and do not affect the asymptotic 
distribution of the estimator.

Given an estimated marginal CDF $\widehat F_{a}(y)$, such as 
$\widehat F_{a}^{OR}(y)$, $\widehat F_{a}^{IPW}(y)$, or $\widehat F_{a}^{DR^*}(y) $, 
the PTE at threshold $y$ can be estimated as 
$\widehat F_{1}(y) - \widehat F_{0}(y)$. The $p$th quantile can be estimated with the plug-in estimator $\widehat Q(p)=\inf\{y:\widehat F_{a}(y)\ge p\}$. Because $\hat F_{a}(y)$ is a step function with jumps only at observed outcome levels, $\widehat Q(p)$ is similarly a step function, which may be undesirable for continuous outcomes, particularly when large gaps exist between adjacent values. To address this issue with continuous $Y$, we adopt the linearly interpolated quantile estimators for CPMs proposed by \citealt{liu2017modeling} and \citealt{tian2024addressing} to obtain a weighted unconditional quantile estimator $\widehat Q_a(p)$ (see Supplementary Material S2.1). The QTE at quantile level $p$ is then estimated as $\widehat{\Delta}^F(p)=\widehat Q_{1}(p)-\widehat Q_{0}(p)$. The superscript $F$ denotes that $\widehat{Q}_a(p)$, and hence $\widehat{\Delta}^F(p)$, is derived by inverting $\widehat{F}_{a}(y)$. In the following presentation, we focus on the estimator based on $\widehat{F}_a^{DR^{*}}(y)$, but we have evaluated the performance of all three versions and will present the results in Section \ref{sec4}.

 \subsubsection{Directly Estimating Quantiles\label{sec3.2.2}}
 
Alternatively, one may solve a different estimating equation that leads to a similar doubly robust estimator as that obtained by inverting \eqref{eq:cdf_dr}. Let $f_{a}(y)$ denote the density of the potential outcome $Y_a$ at $y$. The EIF of $q_{a}(p)$ under $P_0$ is equal to 
\begin{equation}
\begin{aligned}
\varphi_{q}(Z; q_{a}(p))
&= -\frac{1}{f_{a}\!\bigl(q_{a}(p)\bigr)}
   \Biggl[
     \frac{\mathbb{I}(A=a)}{\pi(a\mid\mathbf{X})}
        \mathbb{I}(Y\le q_{a}(p))
     + \frac{\pi(a\mid\mathbf{X}) - \mathbb{I}(A=a)}
            {\pi(a\mid\mathbf{X})}
        F_{Y \mid A=a, \mathbf{X} }\!\bigl(q_{a}(p) \bigr) - p 
   \Biggr].
\end{aligned}
\label{eq:phi_q}
\end{equation}
The same EIF expressions were also derived by \citealt{zhang2012causal} and \citealt{cheng2025inverting}. In finite samples, the estimator $\hat{q}_{a}(p)$ is defined as the solution to 
\begin{equation}
  \frac{1}{n} \sum_{i=1}^n 
\varphi_{q}\!\left(Z_i; q_{a}(p), \hat{\boldsymbol{\psi}}, \hat{\boldsymbol{\xi}} \right)=0,\label{directEE}
\end{equation}
plugging in CPM estimates $\hat{\boldsymbol{\xi}}$ for the outcome regression component and an estimated $\hat{\boldsymbol{\psi}}$ from a separate propensity score model. The same fitted CPM can be reused across different quantile levels $p$, as it models the entire conditional distribution rather than a single quantile. Notably, when solving the empirical estimating equation in \eqref{directEE}, 
the density term $f_{a}(q_{a}(p))$ appears only as a multiplicative constant 
and therefore does not affect the solution of the estimating equation, and can be ignored. The doubly robust estimator of the QTE is then defined as
$\widehat{\Delta}^E(p) = \widehat{q}_{1}(p) - \widehat{q}_{0}(p)$. The superscript $E$ denotes that this QTE estimator is estimated directly from the estimating equation \eqref{directEE}, without first estimating the marginal CDF, in contrast to $\widehat{\Delta}^F(p)$. 

Note that the EIFs for the marginal CDF $F_{a}(y)$ and the marginal quantile $q_{a}(p)$ are directly related. Specifically, comparing \eqref{eq:phi_F} with \eqref{eq:phi_q} and setting $F_{a}(y) = p$ and $y = q_{a}(p)$, we obtain
\[
\varphi_{q}(Z; q_{a}(p))
=
-\frac{1}{f_{a}(q_{a}(p))}
\varphi_{F}(Z; F_a(y)).
\] 
Hence, the two EIF-based estimators share the same doubly robust structure and are therefore conceptually equivalent. However, from an implementation standpoint, solving \eqref{directEE} differs from obtaining the quantile via inversion of the doubly robust estimator of the marginal CDF, $\widehat F^{DR}_{a}(y)$, in \eqref{eq:cdf_dr}. 

In addition, the TMLE approach (\citealt{diaz2017efficient}) is also formulated as a plug-in estimator and may offer improved finite-sample performance compared with AIPW (\citealt{zhang2012causal}). In practice, the conditional CDF component in existing TMLE procedures can be modeled using the CPM, leading to a CPM-based TMLE. Additional implementation details for this CPM-based TMLE are provided in the Supplementary Material S2.2, and its finite-sample performance is examined in the simulation study in Section~\ref{sec4}.

\subsection{Inference\label{3.3}}
In this section, we summarize the asymptotic properties of the proposed CPM-based doubly robust estimators and develop variance estimation procedures for the corresponding estimators. In particular, we derive variance estimators for two proposed QTE estimators: one, $\widehat{\Delta}^F(p)$, obtained through the inverse-CDF approach in Section~\ref{sec3.2.1}, and the other, $\widehat{\Delta}^E(p)$,  through the direct estimating-equation approach in Section~\ref{sec3.2.2}. We additionally develop variance estimation for the PTE estimator.

Under standard regularity conditions, the proposed doubly robust estimators of the marginal quantiles $q_{a}(p)$ are $\sqrt{n}$ consistent and asymptotically normal:
\begin{equation}
\sqrt{n}\,(\hat{q}_{a}(p) - q_{a}(p))
\xrightarrow{d}
\mathcal{N}\!\left(
0,\;
\mathbb{E}\!\big[
\varphi_{q}(Z; q_{a}(p))\varphi^\top_{q}(Z; q_{a}(p))
\big]
\right).
\label{eq:asymptotic_normality_q} 
\end{equation}
Consequently, the QTE estimator $\widehat{\Delta}^E(p)=\widehat{q}_{1}(p)-\widehat{q}_{0}(p)$ is also asymptotically normal. A commonly used approach for variance estimation of $\hat{q}_{a}(p)$ is to empirically estimate the variance in $\eqref{eq:asymptotic_normality_q}$, $\widehat{\mathrm{Var}}(\hat{q}_{a}(p))
=
\frac{1}{n}
\sum_{i=1}^n
\left\{
\varphi_{q}(Z_i; \hat q_{a}(p), \hat{\boldsymbol{\psi}}, \hat{\boldsymbol{\xi}})
-
\bar{\varphi}_{q}
\right\}^2$, where $\bar{\varphi}_{q}
=
\frac{1}{n}
\sum_{i=1}^n
\varphi_{q}(Z_i; \hat q_{a}(p), \hat{\boldsymbol{\psi}}, \hat{\boldsymbol{\xi}})$ (\citealt{tsiatis2006semiparametric,gruber2012tmle,kennedy2024semiparametric,cheng2025inverting}). From now on, we call this approach as EIF-based variance estimator. However, recent work by \citep{shook2025double} showed that, for doubly robust estimators of the ATE, the variance estimators obtained from this approach are generally consistent only when both the propensity score model and the outcome regression model are correctly specified. Our simulation studies in Section~\ref{sec4} similarly illustrate that this phenomenon extends to doubly robust estimators of QTE. Therefore, although the corresponding point estimator retains the double robustness property, the associated EIF-based variance estimator fails to do so under nuisance model misspecification.

In contrast, sandwich variance estimators derived from stacked estimating equations within the general M-estimation framework can preserve double robustness for variance estimation (\citealt{shook2025double,wu2025double}). Let the full parameter vector be $\boldsymbol{\theta}
=
\left(
\boldsymbol{\psi}^\top,
\boldsymbol{\xi}^\top,
q_{1}(p),
q_{0}(p)
\right)^\top$, where $\boldsymbol{\psi}$ parameterizes the propensity score model and 
$\boldsymbol{\xi}$ parameterizes the CPM outcome model. Let $\boldsymbol{\theta}_0$ denote the true value of the parameter vector. Define the stacked estimating equations
\[
\boldsymbol{\Phi}(Z_i;\boldsymbol{\theta})
=
\begin{pmatrix}
\boldsymbol{\phi}_\psi(Z_i;\boldsymbol{\psi}) \\
\boldsymbol{\phi}_\xi(Z_i;\boldsymbol{\xi}) \\
\phi_{1,p}(Z_i;q_{1}(p),\boldsymbol{\psi},\boldsymbol{\xi}) \\
\phi_{0,p}(Z_i;q_{0}(p),\boldsymbol{\psi},\boldsymbol{\xi})
\end{pmatrix},
\]
where $\boldsymbol{\phi}_\psi$ denotes the score function associated with the propensity score model, $\boldsymbol{\phi}_\xi$ denotes the score function corresponding to the CPM likelihood, and $\phi_{1,p}$ and $\phi_{0,p}$ denote the estimating equations for the marginal quantiles $q_{1}(p)$ and $q_{0}(p)$, respectively. Under standard regularity conditions for M-estimation (\citealt{stefanski2002calculus}), the estimator $\widehat{\boldsymbol{\theta}}$ is asymptotically normal:
\[
\sqrt{n}
\left(
\widehat{\boldsymbol{\theta}}
-
\boldsymbol{\theta}_0
\right)
\xrightarrow{d}
\mathcal{N}
\left(
0,
\boldsymbol{\Sigma}
\right),
\]
where $\boldsymbol{\Sigma}
=
\boldsymbol{A}^{-1}
\boldsymbol{B}
(\boldsymbol{A}^{-1} )^{ \top}$,
with $\boldsymbol{A}
=
\mathbb{E}
\left[
-
\frac{\partial}{\partial \boldsymbol{\theta}^\top}
\boldsymbol{\Phi}(Z_i;\boldsymbol{\theta}_0)
\right],
\boldsymbol{B}
=
\mathbb{E}
\left[
\boldsymbol{\Phi}(Z_i;\boldsymbol{\theta}_0)
\boldsymbol{\Phi}(Z_i;\boldsymbol{\theta}_0)^\top
\right]$. In practice, these matrices are estimated empirically by $
\widehat{\boldsymbol{\Sigma}}
=
\boldsymbol{A}_n(\widehat{\boldsymbol{\theta}})^{-1}
\boldsymbol{B}_n(\widehat{\boldsymbol{\theta}})
\left(\boldsymbol{A}_n(\widehat{\boldsymbol{\theta}})^{-1}\right)^\top,$
where
\[
\boldsymbol{A}_n (\widehat{\boldsymbol{\theta}})
=
\frac{1}{n}
\sum_{i=1}^n
-
\frac{\partial}{\partial \boldsymbol{\theta}^\top}
\boldsymbol{\Phi}(Z_i;\widehat{\boldsymbol{\theta}}),
\qquad
\boldsymbol{B}_n (\widehat{\boldsymbol{\theta}})
=
\frac{1}{n}
\sum_{i=1}^n
\boldsymbol{\Phi}(Z_i;\widehat{\boldsymbol{\theta}})
\boldsymbol{\Phi}(Z_i;\widehat{\boldsymbol{\theta}})^\top.
\]
The asymptotic variance of the QTE estimator can subsequently be obtained via the Delta method. Nevertheless, the presence of indicator functions $\mathbb{I}(Y_i \le q_{a}(p))$ within the estimating equations introduces non-smoothness, which complicates both analytical derivations and numerical implementation. To address these challenges, we investigated several approaches, including interchanging differentiation and expectation as well as kernel-based density smoothing techniques. Additional details are provided in Supplementary Material Section~S2.3, and the finite-sample performance of these approaches is evaluated through simulation studies.

In addition, for the estimator of QTE, $\widehat{\Delta}^F(p)=\widehat Q_{1}(p)-\widehat Q_{0}(p)$, variance estimation is less straightforward because  $\widehat Q_{1}(p)$ and  $\widehat Q_{0}(p)$ are not obtained as solutions to estimating equations. We therefore employ a nonparametric bootstrap procedure to estimate their sampling variability. Specifically, the bootstrap procedure involves drawing $B$ independent samples of size $n$ with replacement from the observed data. 
For each bootstrap sample $b = 1, \ldots, B$, the estimated QTE is recomputed using the inverse-CDF approach, yielding $\widehat{\Delta}^F_b(p)$. 
The bootstrap variance estimator is then defined as $    \widehat{\mathrm{Var}}\{\widehat{\Delta}^F(p)\}
= \frac{1}{B-1} \sum_{b=1}^B 
\left\{\widehat{\Delta}^F_b(p) - \widehat{\Delta}^{F^*}(p)\right\}^2,$ where $\widehat{\Delta}^{F^*}(p) = \frac{1}{B} \sum_{b=1}^B \widehat{\Delta}^F_b(p)$.
Under standard regularity conditions, the nonparametric bootstrap provides a consistent estimator of the asymptotic variance of $\widehat{\Delta}^F(p)$ in large samples (\citealt{davison1997bootstrap}). This bootstrap-based strategy will also be applied for inference on conditional treatment effects estimators (described in Section~\ref{sec3.4}). Confidence intervals can be constructed using the empirical percentiles of the bootstrapped estimates in a standard manner. Similar bootstrap procedures can be applied for $\widehat{\Delta}^E(p)$.

The PTE estimator is also asymptotically normal under standard regularity conditions,
\begin{equation}
\sqrt{n}
\left\{
\widehat{\mathrm{PTE}}(y) - \mathrm{PTE}(y)
\right\}
\xrightarrow{d}
\mathcal{N}\!\left(0, \mathbb{E}
\big[
\varphi_{\mathrm{PTE}}(Z)
\varphi_{\mathrm{PTE}}^\top(Z)
\big]\right),
\nonumber
\end{equation}
where $\varphi_{\mathrm{PTE}}(Z)$
is the EIF of the PTE at threshold $y$, defined as $\varphi_{F}(Z; F_1(y)) - \varphi_{F}(Z; F_0(y))$, where the two components are defined in \eqref{eq:phi_F}. However, similar to what we discussed for the QTE estimator $ \widehat \Delta^E (p)$, EIF-based variance estimator is not generally doubly robust.
A sandwich variance estimator can be constructed within the
standard M-estimation framework (\citealt{stefanski2002calculus}). More derivation details are provided in Supplementary
Material Section S2.4. A nonparametric bootstrap procedure can also be used.

\subsection{Conditional Quantile Treatment Effects\label{sec3.4}}

In many applications, interest lies not only in marginal quantile treatment effects but also
in how treatment effects vary across covariate-defined subpopulations.
This motivates the study of conditional quantile treatment effects (CQTE). Let $\boldsymbol{V} \subseteq \mathbf{X}$ denote a vector of possibly effect-modifying covariates. In this work, we focus on discrete $\boldsymbol{V}$ satisfying $\Pr(\boldsymbol{V}=\boldsymbol{v})>0$. In practice, this requires a non-negligible fraction of the sample to satisfy $\boldsymbol{V}=\boldsymbol{v}$ exactly. Define the CDF of the potential outcome $Y_a$ given
$\boldsymbol{V}=\boldsymbol{v}$ as $F_{a \mid \boldsymbol{V}}(y \mid \boldsymbol{v})
= \Pr(Y_a \le y \mid \boldsymbol{V}=\boldsymbol{v}),$
and the corresponding conditional $p$-th quantile as $q_{a}(p,\boldsymbol{v}) = F_{a \mid \boldsymbol{V}}^{-1}(p, \boldsymbol{v})$. Then the CQTE at the quantile level $p$ and covariate value
$\boldsymbol{v}$ is defined as
$\Delta(p , \boldsymbol{v})
= q_{1}(p, \boldsymbol{v}) - q_{0}(p,\boldsymbol{v}). \nonumber$

Under the causal assumptions stated in Section~\ref{sec3.1}, the conditional
distribution function $F_{a \mid \boldsymbol{V}}(y \mid \boldsymbol{v})$ is identifiable
from the observed data. In particular, it can be represented using the OR approach as
$\mathbb{E}\!\left[
  F_{Y \mid A=a, \mathbf{X}}(y)
  \,\big|\, \boldsymbol{V}=\boldsymbol{v}
\right],$
or using the IPW approach as
$\mathbb{E}\!\left[
  \frac{\mathbb{I}(A=a)\,\mathbb{I}(Y \le y)}
       {\pi(a \mid \mathbf{X})}
  \,\big|\, \boldsymbol{V}=\boldsymbol{v}
\right]$. In addition, a doubly robust representation is given by
$$
F^{DR}_{a \mid \boldsymbol{V}}(y \mid \boldsymbol{v})
=
\mathbb{E}\!\left[
  \frac{\mathbb{I}(A=a)\,\mathbb{I}(Y \le y)}
       {\pi(a \mid \mathbf{X})}
  +
  \frac{\pi(a \mid \mathbf{X}) - \mathbb{I}(A=a)}
       {\pi(a \mid \mathbf{X})}
  F_{Y \mid A=a, \mathbf{X}}(y)
  \,\big|\, \boldsymbol{V}=\boldsymbol{v}
\right].
$$
In practice, for a fixed $\boldsymbol{v}$, $F_{a \mid \boldsymbol{V}}(y \mid \boldsymbol{v})$ can be estimated by taking the empirical average of the corresponding quantities among observations satisfying $\boldsymbol{V}=\boldsymbol{v}$ using the OR, IPW or doubly robust representations above. Although all three representations can be used for estimation, we focus on the doubly robust estimator of CQTE throughout this section, $\widehat{F}^{DR}_{a \mid \boldsymbol{V}}(y \mid \boldsymbol{v})
=
\frac{1}{n}\!\left[
  \frac{\mathbb{I}(A=a)\,\mathbb{I}(Y \le y)}
       {\pi^*(a \mid \mathbf{X}; \hat{\boldsymbol{\psi}})}
  +
  \frac{\pi^*(a \mid \mathbf{X}; \hat{\boldsymbol{\psi}}) -\mathbb{I}(A=a)}
       {\pi^*(a \mid \mathbf{X}; \hat{\boldsymbol{\psi}})}
  F^*_{Y \mid A=a, \mathbf{X}}(y; \hat{\boldsymbol{\xi}})
  \,\big|\, \boldsymbol{V}=\boldsymbol{v}
\right]$. Again, we recommend modeling $\pi^*(a \mid \mathbf{X}; \hat{\boldsymbol{\psi}})$ and $F^*_{Y \mid A=a, \mathbf{X}}(y; \hat{\boldsymbol{\xi}})$ using a logistic regression and a CPM, respectively. 
The conditional PTE follows directly from the estimated conditional CDFs: $\widehat{\operatorname{PTE}}(y \mid \boldsymbol{v})=\widehat{F}^{DR}_{1 \mid \boldsymbol{V}}(y \mid \boldsymbol{v})-\widehat{F}^{DR}_{0 \mid \boldsymbol{V}}(y \mid \boldsymbol{v})$. The conditional quantile, $\widehat{q}_{a}^{F}(p, \boldsymbol{v})$, is obtained by inverting $\widehat{F}^{DR}_{a \mid \boldsymbol{V}}(y \mid \boldsymbol{v})$ using the weighted interpolation procedure described in Section~\ref{sec3.2.1}. The corresponding CQTE estimator is then given by $\widehat{\Delta}^F(p, \boldsymbol{v})=\hat{q}_{1}^F(p, \boldsymbol{v})-\hat{q}_{0}^F(p, \boldsymbol{v})$.

Alternatively, the direct estimating equation approach in Section~\ref{sec3.2.2} can be extended to the conditional setting as well, yielding a doubly robust conditional quantile estimator $\hat{q}_{a}^{E}(p, \boldsymbol{v})$. 
For a fixed $\boldsymbol{v}$, this is defined as the solution to 
$\mathbb{E}\!\left[\varphi_{q}(Z; q_{a}(p)) \mid \boldsymbol{V}=\boldsymbol{v}\right] = 0$, 
where $\varphi_{q}(Z; q_a{(p)})$ denotes the doubly robust estimating function for the marginal quantile $q_{a}(p)$ introduced in Section~\ref{sec3.2.2}. 
In practice, we replace the nuisance functions with their estimates including estimates from the fitted CPM and logitic regression, and solve the empirical analogue of this equation by taking the empirical average of the estimating function among observations satisfying $\boldsymbol{V}_i=\boldsymbol{v}$. The corresponding CQTE estimator is then given by $\widehat{\Delta}^E(p, \boldsymbol{v})=\hat{q}_{1}^E(p, \boldsymbol{v})-\hat{q}_{0}^E(p, \boldsymbol{v})$.

\section{Simulation Study\label{sec4}}
We conducted a series of Monte Carlo simulation studies to illustrate the finite 
sample performance of the proposed CPM-based doubly robust estimators for QTE($p$), in comparison
with several existing approaches. 

\subsection{Data Generating Mechanism\label{sec4.1}}
In each simulation replicate, we generated independent covariates $X_1 \sim \mathrm{Bernoulli}(0.5)$ and $ X_2 \sim \mathcal{N}(0, 1).$
Treatment assignment was determined using a logistic model:
\begin{equation}
    \pi(1 \mid \mathbf{X}; \boldsymbol{\psi}) = P(A = 1 \mid \mathbf{X}) = \mathrm{expit}(\psi_{1} X_1 + \psi_{2} X_2), \quad \text{where } \psi_{1} = 0.5,  \psi_{2} = 0.35. \nonumber
\end{equation}
 Across 1,000 replicates, the resulting propensity scores ranged from 
0.09 to 0.94, indicating no extreme values. The potential outcomes were 
generated via a linear model:
\begin{equation}
    Y^*_a = \beta_1 X_1 + \beta_2 X_2 + A\delta + \varepsilon,
\quad \text{where } \beta_1 = -2, \ \beta_2 = 3, \ \delta = 2, \ \varepsilon \sim \mathcal{N}(0,1) \nonumber
\end{equation}
The latent outcome was defined as:
$Y^* = A Y^*_1 + (1 - A) Y^*_0.$ The observed outcome $Y$ was generated by applying a transformation to $Y^*$, $Y=H\left(Y^*\right)$, where $H(y) = \text{Inv-}\chi^2(\Phi(y), 5)$.
Here, $\Phi(\cdot)$ denotes the CDF of the standard normal distribution, and $\mathrm{Inv}\text{-}\chi^2(\cdot, 5)$ denotes the inverse CDF of a chi-square distribution with 5 degrees of freedom. This transformation is intentionally designed to mimic realistic situations in which the outcome is skewed and the true transformation is unlikely to be correctly specified by an analyst. The true values of marginal quantiles $q_{a}(p)$ and $\text{QTE}(p)$ were computed using a large Monte Carlo sample of size $10^8$. The empirical distribution of the observed outcome $Y$ is displayed in Supplementary Figure S3.1.

\vspace{-2ex}
\subsection{Estimators Considered and Model Specifications \label{sec4.2}}
\vspace{-2ex}

Our CPM-based doubly robust estimators are built upon two strategies described in Section~\ref{sec3}: $\widehat{\Delta}^F(p)$ and $\widehat{\Delta}^E(p)$, which we refer to as \textbf{AIPW-CPM-icdf} and \textbf{AIPW-CPM}, respectively. We additionally evaluated the outcome regression and inverse probability weighting estimators under the inverse-CDF approach, denoted as \textbf{OR-CPM-icdf} and \textbf{IPW-icdf}.

To benchmark the proposed methods against existing approaches, we also considered several competing estimators introduced in Section~\ref{sec1}. 
First, we included the inverse probability weighted estimator proposed by \citealt{firpo2007efficient} and the AIPW estimator proposed by \citealt{zhang2012causal}, which are referred to \textbf{IPW-Firpo} and \textbf{AIPW}, respectively. In addition, we implemented two TMLE variants following \citealt{diaz2017efficient}: one using linear regression as the outcome regression component, corresponding to their simulation study (denoted \textbf{TMLE}), and one using conditional quantile regression as the outcome regression component, corresponding to their real-data application (denoted \textbf{TMLE-cqr}). We also developed a CPM-based TMLE variant, denoted \textbf{TMLE-CPM}, as described in Section~\ref{sec3.2.2}.

We also examined the performance of the AIPW, TMLE, and TMLE-cqr estimators after transforming the observed outcome variable.
First, we applied a logarithmic transformation to $Y$, a commonly adopted strategy for handling right-skewed outcomes. The resulting estimators are denoted as \textbf{log-AIPW}, \textbf{log-TMLE}, and \textbf{log-TMLE-cqr}. Second, because the data generating mechanism is fully known in the simulation study, the latent outcome $Y^*$ can be exactly recovered from the observed outcome via the inverse transformation $H^{-1}(Y) = \Phi^{-1}\!\left\{ F_{\chi^2(5)}(Y) \right\}$.
Applying the estimators to this correctly transformed outcome yields an oracle benchmark. These estimators are referred to as \textbf{ct-AIPW}, \textbf{ct-TMLE}, and \textbf{ct-TMLE-cqr}, respectively. The empirical distribution of both the log-transformed and correctly transformed outcomes is shown in Supplementary Figure S3.1. 

All estimators were evaluated under four scenarios:
(a) both the outcome regression and propensity score models are correctly specified;
(b) only the outcome regression model is correctly specified;
(c) only the propensity score model is correctly specified; and
(d) both models are misspecified.
Correct specification refers to cases where the linear predictors in the fitted models are the same as
that in the true data-generating process, which does not extend to the correctness of the other 
aspects, such as the link function or error distribution. Both the misspecified outcome regression and propensity score models omit the continuous covariate $X_2$. 

The number of simulation replicates was 1,000. All estimators were evaluated at the 10th, 25th, 50th, 75th, and 90th percentiles, with sample sizes of 500 and 1,000. However, we present here only the results for $\text{QTE}(0.5)$ with sample size 1,000. Results at other quantile levels and those based on a sample size of 500 exhibit qualitatively similar patterns (see Supplementary Material S3.2).

\subsection{Simulation Results\label{sec4.4}}
\subsubsection{Primary Results\label{sec4.4.1}}
Table~\ref{table1} reports the finite-sample performance of all estimators across the four model specification scenarios. 
When both the outcome regression and propensity score models were correctly specified, all estimators produced unbiased estimates. 
Among them, the OR-CPM-icdf estimator achieved the smallest variance, consistent with the findings of \citealt{kang2007demystifying} and others regarding the efficiency of outcome regression models for estimating ATE. All doubly robust estimators were more efficient than the IPW-based estimators, consistent with semiparametric efficiency theory (\citealt{robins1994estimation, van2006targeted, gabriel2024inverse}). The CPM-based doubly robust estimators showed lower variance than AIPW, TMLE, and TMLE-cqr applied to the original untransformed outcome Y, as well as 
their log-transformed counterparts (log-AIPW, log-TMLE, and log-TMLE-cqr).

When the outcome regression model was correctly specified but the propensity score model was misspecified, both IPW-icdf and IPW-Firpo became biased, as expected given their lack of double robustness. Notably, the AIPW and TMLE estimators also showed pronounced bias, reflecting their sensitivity to distributional assumptions imposed on the outcome model. Applying a log transformation did not adequately resolve this issue: the log-AIPW, log-TMLE, and log-TMLE-cqr remained biased, with estimated biases of 0.283, 0.274, and 0.163, respectively. In contrast, when the correct transformation was applied, the latent outcome $Y^*$ satisfied the normality assumptions, and the corresponding ct-AIPW, ct-TMLE, and ct-TMLE-cqr estimators showed negligible bias. Such idealized conditions, however, are rarely attainable in practice because the true transformation linking the observed and latent outcomes is typically unknown. By comparison, the CPM-based doubly robust estimators (AIPW-CPM, AIPW-CPM-icdf, and TMLE-CPM) consistently produced unbiased estimates without requiring a prespecified outcome transformation. Moreover, their performance was comparable to that of AIPW and TMLE applied to the correctly transformed outcome, exhibiting similar variance, root mean squared error (RMSE), and median absolute error (MedAE) across all simulation scenarios.

When only the propensity score model was correctly specified, all doubly robust estimators remained unbiased with comparable variances. When both the propensity score and outcome regression models were misspecified, all estimators showed appreciable bias, consistent with theoretical expectations. In addition, across all four scenarios, the inverse-CDF and direct implementations (AIPW-CPM-icdf vs. AIPW-CPM) produced nearly identical point estimates and empirical variances, RMSE and MedAE. The TMLE-CPM also had similar performance to AIPW-CPM-icdf and AIPW-CPM.
\subsubsection{Misspecified Link Function\label{sec4.3.2}}
In practice, the latent error distribution is unlikely to coincide exactly with the CDF corresponding to the inverse of the chosen link function for CPMs. Therefore, it is important to assess the robustness of CPMs under link function misspecification. For example, the logit link function is frequently adopted because of its convenient interpretation in terms of odds ratios, even when the underlying error distribution may deviate from the logistic form. 

We fitted the CPM using a logit link instead of the correctly specified probit link. The resulting estimators are denoted as AIPW-CPM-mislink, AIPW-CPM-icdf-mislink, and TMLE-CPM-mislink in Table~\ref{table1}. When either the outcome regression model or the propensity score model was misspecified, all CPM-based doubly robust estimators still remained unbiased. Moreover, compared with their correctly specified link function counterparts (AIPW-CPM, AIPW-CPM-icdf and TMLE-CPM), the mislink estimators showed nearly identical performance in terms of both point estimation and variance estimation. These results indicate that CPM-based doubly robust approaches are highly robust to moderate misspecification of the link function.
\subsubsection{Inference\label{sec4.4.4}}
We compared three variance estimation strategies for the AIPW-CPM estimator: the EIF--based variance estimators, two sandwich estimators, and the nonparametric bootstrap (with 500 bootstrap replications). Sandwich-I refers to the estimator obtained by interchanging differentiation and expectation, whereas Sandwich-S is based on direct differentiation with a smoothing approximation; technical details are provided in Supplementary Material Section~S2.3. Their performance was evaluated against the empirical Monte Carlo variance (Table~\ref{table2}). When both the outcome regression (CPM) and propensity score models were correctly specified, the bootstrap, EIF--based, and sandwich variance estimators all performed well, producing variance estimates close to the empirical Monte Carlo variance and confidence interval coverage probabilities near the nominal level. Similar performance was observed when only the outcome regression model was correctly specified or when both nuisance models were misspecified, with all approaches continuing to provide reasonably accurate variance estimates. However, when the propensity score model was correctly specified but the outcome regression model was misspecified, the EIF--based variance estimator substantially overestimated the variance. This finding suggests that, in finite samples, EIF--based variance estimation for doubly robust estimators may be more sensitive to misspecification of the outcome regression model than to misspecification of the propensity score model, although the underlying theoretical explanation remains unclear. Similar phenomena have been noted in other doubly robust settings (\citealt{shook2025double}). 

In contrast, the nonparametric bootstrap and both sandwich estimators demonstrated stable performance across all simulation scenarios, producing variance estimates close to the empirical Monte Carlo variance and confidence interval coverage probabilities near the nominal 95\% level. Although Sandwich-S occasionally yielded coverage probabilities slightly closer to the nominal level than Sandwich-I, its implementation depends on a smoothing approximation to the derivative of the indicator function $\mathbb{I}(Y_i \leq q_{a}(p))$, which may introduce numerical instability. This behavior is further illustrated in the additional simulation results presented in Supplementary Material Section~S3.3.1. Considering both empirical performance and computational stability, we recommend Sandwich-I for practical inference. For the AIPW-CPM-icdf estimator, the nonparametric bootstrap likewise provided stable and reliable inference under all nuisance-model specification scenarios.

Simulation results for the point and variance estimation of the doubly robust estimator of PTE are provided in Supplementary Material Section~S3.3.2. Similar to the findings for QTE, the EIF--based variance estimator was sensitive to nuisance model misspecification, particularly when the outcome regression was misspecified while the propensity score model was correctly specified. In comparison, the nonparametric bootstrap and sandwich variance estimators remained comparatively robust across all nuisance-model specification settings.

\subsubsection{Computational Efficiency\label{sec4.4.5}}
We evaluated the average computational time under the scenario in which both the outcome regression and propensity score models were correctly specified. In general, the IPW estimators showed the shortest runtimes, followed by the AIPW (mean 0.5 seconds), then the CPM-based estimators (mean approximately 0.7 seconds), and finally the TMLE estimators (mean 5–6 seconds). Additional details are provided in Section S3.4 of the Supplementary Material. This simulation was conducted on a MacBook Air (M2, 2022) equipped with an Apple M2 CPU and 8 GB of RAM. The software environment was R version 4.4.2.

\section{Real Data Application\label{sec5}}

In this section,  we applied the proposed CPM-based doubly robust estimators to an observational cohort of adults living with HIV who initiated ART between January 1, 2014, and December 31, 2019, in Latin America. 
The goal was to estimate the effect of initiating an integrase strand transfer inhibitor (INSTI)–based regimen compared with a non–INSTI-based regimen on 6-month CD4 count and HIV RNA viral load after antiretroviral therapy (ART) initiation. The analytic cohort included 2,669 adults, of whom 954 initiated an INSTI-based regimen and 1,715 initiated a non-INSTI-based regimen. Potential confounders include age, sex, study site, probable route of HIV infection, an indicator of AIDS diagnosis prior to ART initiation, baseline CD4 count and viral load, calendar year of ART initiation,  and the time interval between the baseline and 6-month CD4  measurements.  Baseline CD4 count was defined 
as the measurement closest to ART initiation within a window of $-180$ to $+30$ days, and the 6-month CD4 count as the measurement closest to 180 days within $\pm 90$ days. Baseline viral load was defined 
as the measurement closest to ART initiation within a window of $-180$ to $0$ days, and the 6-month viral 
load as the measurement closest to 180 days within $\pm 90$ days.
Baseline CD4 count was square-root
transformed, and baseline viral load was log$_{10}$ transformed. Detection limits for viral load measurements varied across study sites due to differences in laboratory assays and equipment. We assigned all undetectable measurements and all detectable measurements $<40$ copies/\mbox{mL} to the category $<40$ copies/\mbox{mL}, corresponding to the highest detection limit of 40 copies/\mbox{mL}. 

We applied the proposed CPM-based doubly robust estimators to evaluate the distributional effect of treatment on CD4 cell count. All reported estimates of QTE are $\widehat{\Delta}^E(p)$; results were nearly identical for $\widehat{\Delta}^F(p)$. The 95\% confidence intervals were obtained using the percentile method based on 1{,}000 nonparametric bootstrap replicates. For the outcome regression component, we fit separate CPMs within the INSTI and non-INSTI groups. In each model, we incorporated restricted cubic splines with four knots for baseline CD4 count, baseline viral load, and the time interval between the baseline and 6-month CD4 measurements to allow for flexible nonlinear effects. For the propensity score model, we used logistic regression; restricted cubic splines with four knots were included for age and for the time interval between the baseline and 6-month CD4 measurements to account for potential nonlinearity.

\begin{figure}[b]
\centerline{\includegraphics[width=6.6in]{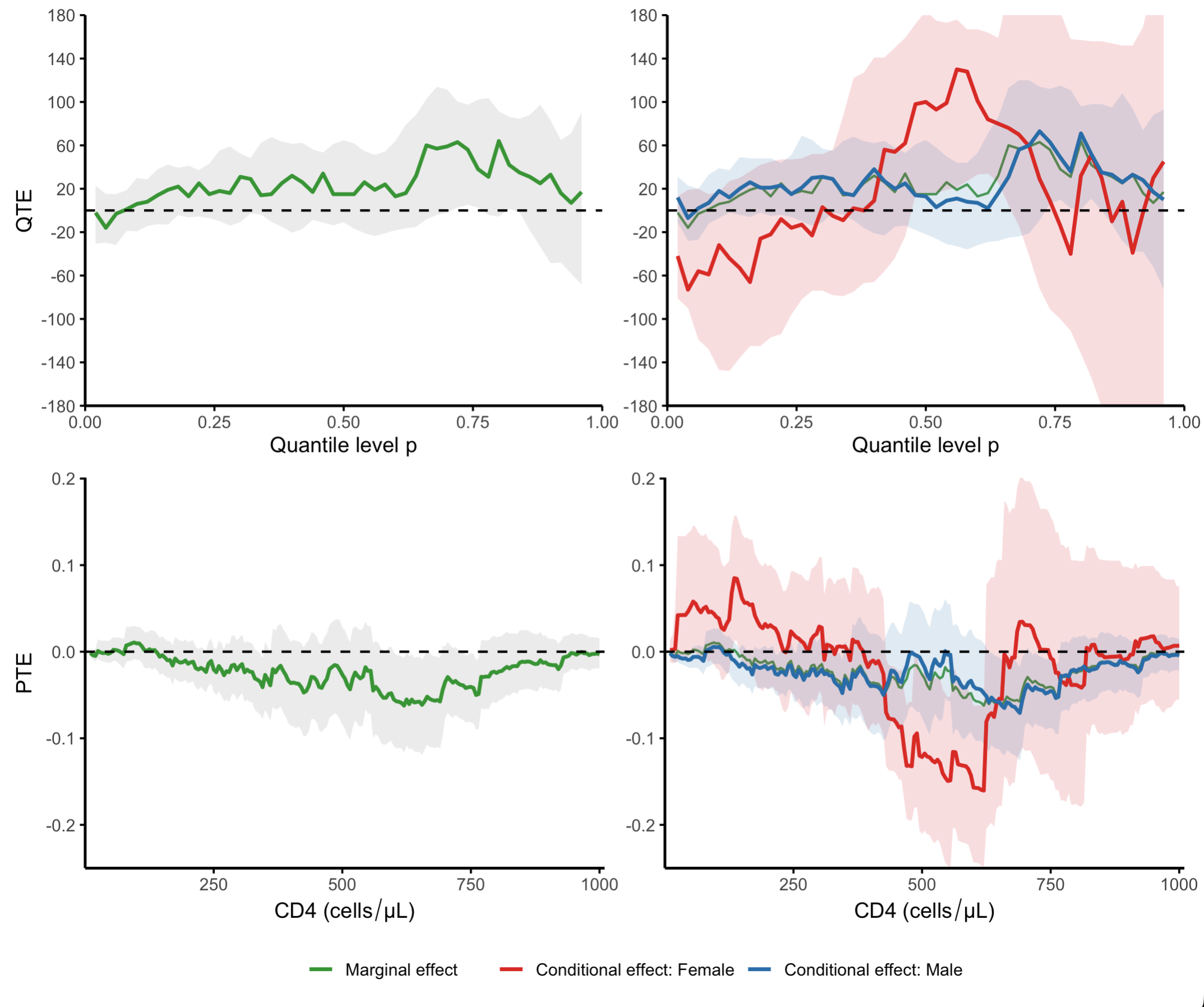}} \vspace*{-3pt}
\caption{
Estimated marginal and conditional distributional treatment effects of INSTI-based versus non--INSTI-based regimens on 6-month CD4 cell count. 
The top panels display the quantile treatment effect (QTE) as a function of the quantile level, while the bottom panels show the probability treatment effect (PTE) as a function of the CD4 threshold. 
The left panels present marginal effects (green), and the right panels present sex-specific conditional effects, with red lines corresponding to females and blue lines corresponding to males. 
Shaded bands represent 95\% bootstrap percentile confidence intervals.
}
\label{fig:figure2}
\end{figure}

The estimated probability treatment effect curve (Figure~\ref{fig:figure2}, lower left) is negative across much of the central range of 6-month CD4 values, indicating that individuals would have a lower probability of falling below a given CD4 threshold had they initiated an INSTI-based regimen than had they initiated a non-INSTI-based regimen. For example, at a CD4 threshold of 350 cells/\mbox{$\mu$L}, the estimated PTE is $-2.7\%$ (95\% CI: $-6.7\%$, $1.1\%$), indicating that if all individuals were assigned to initiate INSTI-based therapy, the probability of having a CD4 count below 350 cells/\mbox{$\mu$L} at 6 months after ART initiation would be 2.7 percentage points lower than if all individuals were instead assigned to non-INSTI regimens. At 500 cells/\mbox{$\mu$L}, the estimated PTE is $-2.7\%$ (95\% CI: $-8.4\%$, $2.8\%$), and at 600 cells/\mbox{$\mu$L}, the estimated PTE is $-5.5\%$ (95\% CI: $-10.9\%$, $-0.1\%$). Consistently, across most quantile levels, the CD4 counts that would have been observed under INSTI-based regimens were higher than those that would have been observed under non--INSTI regimens, as reflected by QTE estimates greater than zero (Figure~\ref{fig:figure2}, upper left). For example, at the 25th percentile, the estimated QTE is 15 cells/\mbox{$\mu$L} (95\% CI: $-7.0$, $43.9$), indicating that had everyone in the study population 
initiated an INSTI-based regimen, the 25th percentile of the 6-month CD4 count would be 15 cells/\mbox{$\mu$L} higher than if they all had instead initiated a non--INSTI regimen. At the 50th percentile, the estimated QTE is 15 cells/\mbox{$\mu$L} (95\% CI: -10.8, 64.7), and at the 75th percentile, the estimated QTE is 56 cells/\mbox{$\mu$L} (95\% CI: 5.1, 97.6), suggesting that the treatment effect is more pronounced in the upper tail of the potential CD4 outcome distribution. For comparison, the average treatment effect estimated using the \textit{tmle} package was 24.2 cells/\mbox{$\mu$L} (95\% CI: 8.5, 39.9), less than half of the estimated effect at the 75th percentile, suggesting that mean-based analyses may underestimate improvements among patients with higher 6-month CD4 counts.

To further assess treatment effect heterogeneity, we estimated sex-specific conditional PTE and QTE (Figure~\ref{fig:figure2}, right bottom). The analytic cohort included 1{,}481 males (86\%) in the non-INSTI group and 848 males (89\%) in the INSTI group. We used the same modeling strategy as in the marginal effect analysis. Specifically, CPMs with restricted cubic splines were fitted separately within each treatment group. However, when estimating the marginal CDFs and corresponding causal contrasts, averaging was performed within sex subgroups (i.e., over males and females separately) rather than over the entire study population. Among males, initiation of an INSTI-based regimen was estimated to lower the probability of having 6-month CD4 counts below various thresholds and to increase CD4 counts across a broad range of the distribution. For example, at a CD4 threshold of 350 cells/\mbox{$\mu$L}, the estimated PTE among males was $-3\%$ (95\% CI: $-7.5\%$, $1.2\%$), suggesting that, among males, the probability of remaining below this clinically relevant threshold would be lower if they were to initiate an INSTI-based regimen rather than a non--INSTI regimen. In contrast, the corresponding estimate among females was $0.6\%$ (95\% CI: $-6.3\%$, $8.6\%$), indicating little difference between treatment strategies. At the 25th percentile, the QTE among males was 15 cells/\mbox{$\mu$L} (95\% CI: $-6$, $51$), suggesting that, among individuals with poorer immune response, CD4 recovery would be higher under an INSTI-based regimen than under a non--INSTI regimen. In contrast, among females, the estimated QTE was $-16$ cells/\mbox{$\mu$L} (95\% CI: $-96$, $47$), with wide uncertainty and no clear pattern of benefit. 

We also applied our methods to investigate the effect of INSTI-based regimens on 6-month viral load. A substantial proportion (77\%) of patients had viral load below the detection limit (i.e., $\leq 40$ copies/mL). Viral load follows a mixture distribution, with a discrete component representing undetectable values ($\leq 40$ copies/mL) and a highly skewed continuous distribution above the detection limit. With such an outcome, the ATE, as well as estimates of the QTE based on normality assumptions, are dubious. In contrast, the CPM can naturally handle mixed-type data of this nature and can be used to estimate both the QTE and PTE while preserving the outcome on its original scale. At the clinically relevant threshold of 40 copies/mL, corresponding to the limit of detection, the estimated probability of achieving viral suppression at 6 months was 0.87 under an INSTI-based regimen, compared with 0.76 under a non--INSTI regimen. This corresponds to an estimated $\text{PTE}(40)=0.11$ (95\% CI: 0.04, 0.19), indicating that, at the population level, initiation of an INSTI-based regimen would increase the proportion achieving viral suppression by approximately 11 percentage points relative to a non--INSTI regimen. At the upper end of the viral load distribution, the 95th percentile 
was lower under an INSTI-based regimen (QTE = $-3723$ copies/mL; 
95\% CI: $-14123$, $2519$), suggesting a potential reduction in the 95th percentile of the viral load distribution with the use of INSTI-based 
regimens, although the estimate is imprecise with substantial uncertainty.

\vspace{-2ex}
\section{Discussion\label{sec6}}
\vspace{-2ex}
In this paper, we developed a CPM-based doubly robust framework for estimating marginal quantiles $q_{a}(p)$ and the corresponding QTE. We illustrated the proposed framework
using an application that examines the effect of starting an INSTI–based regimen versus a non-INSTI–
based regimen on 6-month CD4 count and viral load after ART initiation.
CPM estimators have several advantages over commonly used approaches for addressing such
skewed outcomes.
They are invariant to monotone transformations of the outcome, eliminating the need to 
identify a suitable transformation prior to analysis, they model the entire conditional
distribution, rather than targeting a single moment or quantile, and they can handle mixed-type response variables such as viral load. From simulation studies we 
saw that our proposed CPM-based doubly robust estimators retained unbiasedness even when the
link function was moderately misspecified. Moreover, they were both more robust and similarly efficient to existing AIPW- and TMLE- based estimators that rely on strong distributional
assumptions. We investigated several variance estimation procedures,
including EIF-based estimators, sandwich estimators, and the nonparametric bootstrap. The empirical
sandwich estimator and the nonparametric bootstrap provide doubly robust variance estimation with stable finite-sample performance under nuisance-model misspecification.

Several additional extensions of this work merit further investigation. First, although our 
study focused on binary treatments, the proposed framework can be naturally generalized to 
multivalued or continuous exposures through modeling of the generalized propensity score (\citealt{imai2004causal}, \citealt{kluve2012evaluating}). Second, when machine learning or ensemble methods such as the super learner (\citealt{van2011targeted}) are employed to estimate nuisance models, they provide greater modeling flexibility. Incorporating double machine learning procedures may further improve finite-sample performance and reduce bias due to overfitting. We are currently developing CPM-based machine learning methods (e.g. random forest, boosting and neural networks) to more flexibly estimate outcome regression models in a rank-based manner. Third, extending the CPM-based doubly robust estimation framework to cluster data settings would broaden its applicability to 
complex biomedical and epidemiologic studies.

\section*{Code Availability}
All code used for the simulation studies and real data analysis in this paper is publicly available at \url{https://github.com/haowu1701/doubly-robust-qte.git}.

\section*{Acknowledgment}
This study was supported in part by grants from the United States National Institutes of Health: R01AI093234 (funding for HW, CL, and BES) and U01AI069923 (CCASAnet data collection and funding for BES). We thank CCASAnet and its participants for the use of their data for this study.

\bibliographystyle{biom}
\bibliography{biomsample}

\begin{table}[p]
\centering
\scriptsize
\setlength{\tabcolsep}{3.6pt}
\renewcommand{\arraystretch}{1.10}

\caption{Simulation performance metrics across nuisance-model specification scenarios. Target parameter is $QTE(0.5)$, true value is 5.9. Each scenario was evaluated using 1,000 simulation replicates. Sample size is 1,000.
\textit{OR}: outcome regression. 
\textit{PS}: propensity score. 
\textit{Var}: empirical variance over 1,000 simulation replicates. 
\textit{RMSE}: root mean squared error. 
\textit{MedAE}: median absolute error.}
\label{table1}

\begin{threeparttable}

\begin{adjustbox}{max width=\linewidth}
\begin{tabular}{@{}l
  *{4}{S[table-format=2.3,table-text-alignment=center]}
  @{\hspace{3.6em}}
  *{4}{S[table-format=2.3,table-text-alignment=center]}@{}}
\toprule
& \multicolumn{4}{c}{\textbf{Correct OR, Correct PS}}
& \multicolumn{4}{c}{\textbf{Correct OR, Misspecified PS}}\\
\cmidrule(r){2-5}\cmidrule(l){6-9}
\textbf{Estimators}
& {\textbf{Bias}}
& {\textbf{Var}}
& {\textbf{RMSE}}
& {\textbf{MedAE}}
& {\textbf{Bias}}
& {\textbf{Var}}
& {\textbf{RMSE}}
& {\textbf{MedAE}}\\
\midrule

AIPW                 & 0.014 & 0.331 & 0.576 & 0.384 & 0.525 & 0.342 & 0.786 & 0.523\\
TMLE                 & 0.011 & 0.317 & 0.563 & 0.368 & 0.472 & 0.331 & 0.744 & 0.490\\
TMLE-cqr             & 0.017 & 0.296 & 0.545 & 0.356 & 0.379 & 0.312 & 0.675 & 0.440\\
\midrule
log-AIPW             & 0.012 & 0.306 & 0.554 & 0.361 & 0.283 & 0.331 & 0.642 & 0.413\\
log-TMLE             & 0.015 & 0.294 & 0.542 & 0.355 & 0.274 & 0.310 & 0.620 & 0.390\\
log-TMLE-cqr         & 0.020 & 0.280 & 0.530 & 0.340 & 0.163 & 0.285 & 0.559 & 0.364\\
\midrule
ct-AIPW              & 0.024 & 0.271 & 0.521 & 0.328 & 0.026 & 0.267 & 0.518 & 0.318\\
ct-TMLE              & 0.022 & 0.270 & 0.520 & 0.319 & 0.026 & 0.271 & 0.521 & 0.324\\
ct-TMLE-cqr          & 0.022 & 0.271 & 0.521 & 0.330 & 0.026 & 0.271 & 0.521 & 0.322\\
\midrule
AIPW-CPM             & 0.023 & 0.269 & 0.519 & 0.324 & 0.030 & 0.269 & 0.520 & 0.319\\
AIPW-CPM-icdf        & 0.025 & 0.271 & 0.521 & 0.330 & 0.031 & 0.270 & 0.520 & 0.318\\
TMLE-CPM             & 0.024 & 0.272 & 0.522 & 0.329 & 0.025 & 0.273 & 0.523 & 0.320\\
\midrule
AIPW-CPM-mislink      & 0.023 & 0.268 & 0.518 & 0.324 & 0.037 & 0.271 & 0.522 & 0.317\\
AIPW-CPM-icdf-mislink & 0.026 & 0.271 & 0.521 & 0.328 & 0.039 & 0.271 & 0.522 & 0.317\\
TMLE-CPM-mislink      & 0.025 & 0.272 & 0.522 & 0.330 & 0.029 & 0.272 & 0.523 & 0.326\\
\midrule
OR-CPM-icdf          & 0.031 & 0.213 & 0.463 & 0.305 & 0.031 & 0.213 & 0.463 & 0.305\\
IPW-icdf             & 0.008 & 0.357 & 0.597 & 0.401 & 2.690 & 0.781 & 2.831 & 2.660\\
IPW-Firpo          & 0.004 & 0.354 & 0.595 & 0.398 & 2.681 & 0.780 & 2.823 & 2.642\\

\bottomrule
\end{tabular}
\end{adjustbox}

\vspace{7pt}

\begin{adjustbox}{max width=\linewidth}
\begin{tabular}{@{}l
  *{4}{S[table-format=2.3,table-text-alignment=center]}
  @{\hspace{3.6em}}
  *{4}{S[table-format=2.3,table-text-alignment=center]}@{}}
\toprule
& \multicolumn{4}{c}{\textbf{Misspecified OR, Correct PS}}
& \multicolumn{4}{c}{\textbf{Misspecified OR, Misspecified PS}}\\
\cmidrule(r){2-5}\cmidrule(l){6-9}
\textbf{Estimators}
& {\textbf{Bias}}
& {\textbf{Var}}
& {\textbf{RMSE}}
& {\textbf{MedAE}}
& {\textbf{Bias}}
& {\textbf{Var}}
& {\textbf{RMSE}}
& {\textbf{MedAE}}\\
\midrule

AIPW                 & 0.005 & 0.357 & 0.598 & 0.402 & 2.681 & 0.780 & 2.823 & 2.642\\
TMLE                 & 0.011 & 0.356 & 0.597 & 0.404 & 2.691 & 0.790 & 2.834 & 2.647\\
TMLE-cqr             & 0.005 & 0.356 & 0.597 & 0.398 & 2.681 & 0.781 & 2.823 & 2.649\\
\midrule
log-AIPW             & 0.004 & 0.354 & 0.595 & 0.397 & 2.681 & 0.780 & 2.823 & 2.642\\
log-TMLE             & 0.009 & 0.357 & 0.598 & 0.400 & 2.685 & 0.785 & 2.827 & 2.648\\
log-TMLE-cqr         & 0.005 & 0.356 & 0.597 & 0.398 & 2.681 & 0.781 & 2.823 & 2.649\\
\midrule
ct-AIPW              & 0.005 & 0.356 & 0.597 & 0.398 & 2.681 & 0.780 & 2.823 & 2.642\\
ct-TMLE              & 0.014 & 0.356 & 0.597 & 0.409 & 2.677 & 0.780 & 2.819 & 2.634\\
ct-TMLE-cqr          & 0.005 & 0.356 & 0.597 & 0.398 & 2.681 & 0.781 & 2.823 & 2.649\\
\midrule
AIPW-CPM             & 0.002 & 0.356 & 0.597 & 0.398 & 2.681 & 0.780 & 2.823 & 2.642\\
AIPW-CPM-icdf        & 0.006 & 0.355 & 0.596 & 0.397 & 2.689 & 0.781 & 2.831 & 2.660\\
TMLE-CPM             & 0.009 & 0.354 & 0.595 & 0.396 & 2.681 & 0.775 & 2.821 & 2.631\\
\midrule
AIPW-CPM-mislink      & 0.002 & 0.356 & 0.597 & 0.398 & 2.681 & 0.780 & 2.823 & 2.642\\
AIPW-CPM-icdf-mislink & 0.006 & 0.356 & 0.596 & 0.397 & 2.690 & 0.781 & 2.831 & 2.661\\
TMLE-CPM-mislink      & 0.010 & 0.358 & 0.599 & 0.406 & 2.685 & 0.774 & 2.826 & 2.653\\
\midrule
OR-CPM-icdf          & 2.843 & 0.679 & 2.960 & 2.836 & 2.843 & 0.679 & 2.960 & 2.836\\
IPW-icdf             & 0.008 & 0.357 & 0.597 & 0.401 & 2.690 & 0.781 & 2.831 & 2.660\\
IPW-Firpo          & 0.004 & 0.354 & 0.595 & 0.398 & 2.681 & 0.780 & 2.823 & 2.642\\

\bottomrule
\end{tabular}
\end{adjustbox}

\end{threeparttable}
\end{table}

\begin{table*}[!t]
\centering
\caption{Empirical standard deviations, estimated standard errors, and 95\% confidence interval coverage for $QTE(0.5)$ under different variance estimators and nuisance-model specifications.}
\label{table2}

\begin{threeparttable}
\small
\setlength{\tabcolsep}{4.2pt}
\renewcommand{\arraystretch}{1.08}

\begin{tabular}{l
                S[table-format=1.3] S[table-format=3.1]
                S[table-format=1.3] S[table-format=3.1]
                S[table-format=1.3] S[table-format=3.1]
                S[table-format=1.3] S[table-format=3.1]
                S[table-format=1.3] S[table-format=3.1]}
\toprule

& \multicolumn{2}{c}{Empirical}
& \multicolumn{2}{c}{Bootstrap}
& \multicolumn{2}{c}{EIF-based}
& \multicolumn{2}{c}{Sandwich-I}
& \multicolumn{2}{c}{Sandwich-S} \\
\cmidrule(lr){2-3}\cmidrule(lr){4-5}\cmidrule(lr){6-7}\cmidrule(lr){8-9}\cmidrule(lr){10-11}

Estimators
& {SD} & {Cov,\%}
& {SE} & {Cov,\%}
& {SE} & {Cov,\%}
& {SE} & {Cov,\%}
& {SE} & {Cov,\%} \\
\midrule

\multicolumn{11}{c}{Correct OR, Correct PS}\\

AIPW-CPM
& 0.519 & 94.5
& 0.532 & 95.7
& 0.501 & 91.9
& 0.502 & 91.9
& 0.497 & 92.9 \\
AIPW-CPM-icdf
& 0.521 & 94.5
& 0.529 & 95.2
& \multicolumn{2}{c}{--}
& \multicolumn{2}{c}{--}
& \multicolumn{2}{c}{--} \\

\midrule
\multicolumn{11}{c}{Correct OR, Misspecified PS}\\

AIPW-CPM
& 0.519 & 94.4
& 0.529 & 95.4
& 0.497 & 91.1
& 0.495 & 90.9
& 0.504 & 93.6 \\
AIPW-CPM-icdf
& 0.520 & 94.7
& 0.526 & 94.9
& \multicolumn{2}{c}{--}
& \multicolumn{2}{c}{--}
& \multicolumn{2}{c}{--} \\

\midrule
\multicolumn{11}{c}{Misspecified OR, Correct PS}\\

AIPW-CPM
& 0.597 & 94.9
& 0.593 & 95.5
& 0.799 & 99.0
& 0.562 & 91.3
& 0.577 & 94.6 \\
AIPW-CPM-icdf
& 0.596 & 95.1
& 0.590 & 95.4
& \multicolumn{2}{c}{--}
& \multicolumn{2}{c}{--}
& \multicolumn{2}{c}{--} \\

\midrule
\multicolumn{11}{c}{Misspecified OR, Misspecified PS}\\

AIPW-CPM
& 0.883 & 13.8
& 0.866 & 7.8
& 0.825 & 10.9
& 0.825 & 10.9
& 0.910 & 12.8 \\
AIPW-CPM-icdf
& 0.883 & 13.6
& 0.864 & 7.9
& \multicolumn{2}{c}{--}
& \multicolumn{2}{c}{--}
& \multicolumn{2}{c}{--} \\

\bottomrule
\end{tabular}

\begin{tablenotes}[flushleft]
\footnotesize
\item Target parameter is $QTE(0.5)$, with true value equal to 5.9. \textit{OR} denotes outcome regression, and \textit{PS} denotes propensity score. \textit{SD} denotes the empirical standard deviation of the point estimates across 1,000 simulation replicates. \textit{SE} denotes the average estimated standard error across simulation replicates. 
\textit{Bootstrap} denotes standard errors estimated via the nonparametric bootstrap. \textit{EIF-based} denotes standard errors estimated using the efficient influence function. 
\textit{Sandwich-I} refers to the sandwich variance estimator obtained by interchanging differentiation and expectation, whereas \textit{Sandwich-S} denotes the version based on direct differentiation with smoothing approximation. 
\textit{Cov}, \% denotes the empirical coverage of the 95\% confidence interval, that is, the proportion of simulated samples for which the interval contained the true value. Bootstrap confidence intervals are constructed using the percentile method, whereas all other intervals are based on normal approximations.
\end{tablenotes}

\end{threeparttable}
\end{table*}

\clearpage
\makeatletter
\processdelayedfloats           
\renewenvironment{figure}{\@float{figure}}{\end@float}   
\renewenvironment{table}{\@float{table}}{\end@float}     
\makeatother

\onehalfspacing

\setcounter{section}{0}
\renewcommand{\thesection}{S\arabic{section}}
\renewcommand{\thesubsection}{S\arabic{section}.\arabic{subsection}}
\renewcommand{\thefigure}{S\arabic{figure}}
\renewcommand{\thetable}{S\arabic{table}}
\counterwithin{figure}{section}
\counterwithin{table}{section}
\counterwithin{theorem}{section}

\renewcommand{\theHsection}{S\arabic{section}}
\renewcommand{\theHfigure}{S\arabic{section}.\arabic{figure}}
\renewcommand{\theHtable}{S\arabic{section}.\arabic{table}}

\date{}
\maketitle
\vspace{-1.5cm}
\begin{center}

Hao Wu$^{1}$, Chun Li$^{2}$, and Bryan E. Shepherd$^{1}$

\vspace{0.3cm}

\small
$^{1}$Department of Biostatistics, Vanderbilt University Medical Center,
Nashville, Tennessee, USA\\
$^{2}$Division of Biostatistics and Health Data Science, Department of
Population and Public Health Sciences, Keck School of Medicine,
University of Southern California, Los Angeles, California, USA

\vspace{0.2cm}

\normalsize
Corresponding author: Bryan E. Shepherd (bryan.shepherd@vanderbilt.edu)

\end{center}


\clearpage
\FloatBarrier

\section{Identification of the Marginal CDF}
\begin{align}
F_{a}(y)
&= \Pr(Y_a \le y) \nonumber\\
&= \mathbb{E}[\mathbb{I}(Y_a \le y)] \nonumber\\
&= \mathbb{E}\!\left[\mathbb{E}\{\mathbb{I}(Y_a \le y)\mid \mathbf X\}\right] \nonumber\\
&= \mathbb{E}\!\left[\mathbb{E}\{\mathbb{I}(Y_a \le y)\mid A=a,\mathbf X\}\right]
\quad \text{(by unconfoundedness)} \nonumber\\
&= \mathbb{E}\!\left[\mathbb{E}\{\mathbb{I}(Y \le y)\mid A=a,\mathbf X\}\right]
\quad \text{(by consistency)} \nonumber\\
&= \mathbb{E}\!\left[\Pr(Y \le y\mid A=a,\mathbf X)\right]. \nonumber \label{S1.1}
\end{align}
Additionally,we derive an equivalent inverse probability weighted (IPW) representation of $F_{a}(y)$.
Under the positivity condition $\Pr(A=a \mid \mathbf X)>0$, we have 
$$
\begin{aligned}
\mathbb{E}\left[\left.\frac{\mathbb{I}(A=a) \mathbb{I}(Y \le y)}{\Pr(A=a \mid \mathbf X)} \right\rvert \mathbf X\right]
&= \frac{1}{\Pr(A=a \mid \mathbf X)}
   \mathbb{E}\big[\mathbb{I}(A=a)\mathbb{I}(Y \le y) \mid \mathbf X\big] \\
&= \frac{1}{\Pr(A=a \mid \mathbf X)}
   \left(
   \Pr(A=a \mid \mathbf X)\,
   \mathbb{E}\big[\mathbb{I}(Y \le y) \mid A=a, \mathbf X\big]
   \right) \\
&= \mathbb{E}\big[\mathbb{I}(Y \le y) \mid A=a, \mathbf X\big]\\
&= \Pr(Y \le y\mid A=a,\mathbf X).
\end{aligned}
$$
Thus, 
$$
\begin{aligned}
F_{a}(y) &= \mathbb{E}\!\left[\Pr(Y \le y\mid A=a,\mathbf X)\right]\\
& =\mathbb{E}\left[\mathbb{E}\left\{\left.\frac{\mathbb{I}(A=a) \mathbb{I}(Y \leq y)}{\Pr(A=a \mid {\mathbf X})} \right\rvert\, {\mathbf X} \right\}\right] \\
& =\mathbb{E}\left[\frac{\mathbb{I}(A=a) \mathbb{I}(Y \leq y)}{\Pr(A=a \mid {\mathbf X})}\right] .
\end{aligned}
$$

\section{CPM-based Doubly Robust Estimators Details}
\subsection{Quantile Function\label{quantile function}}
\FloatBarrier

\begin{figure}[!htbp]
\centerline{\includegraphics[width=3.7in]{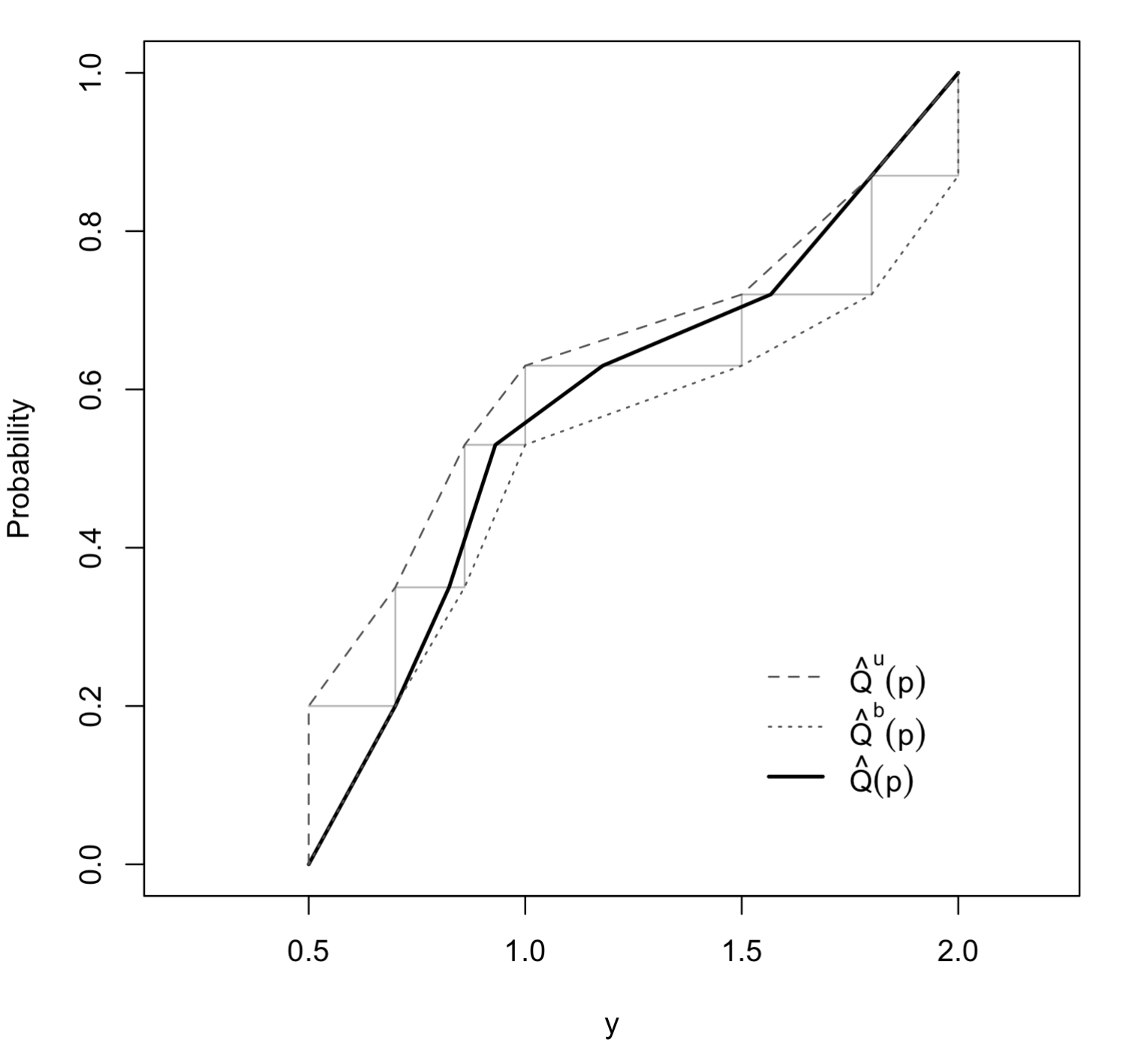}} \vspace*{-3pt}
\caption{ Illustration of three approaches for marginal quantiles. The dashed lines are for $\hat{Q}^u(p)$, the dotted lines are for $\hat{Q}^b(p)$, the solid black lines are for $\hat{Q}(p)$, and the solid gray lines are for the empirical CDF. }
\label{fig:figure1}
\end{figure}

 Let $y_{(j)}$, $j = 1, \dots, J$, denote the $j$th smallest distinct observed response value, where $J$ is the number of distinct response values. If $y < y_{(1)} $, then $F_Y(y) = 0$; if $y \geq
 y_{(J)}$, then $F_Y(y) = 1$.
 For notational simplicity, we write $P_j  = \hat{F}_Y(y_{(j)})$ for a generic marginal CDF estimator $\hat F_Y$. We first define a linear interpolation between $y_{(j-1)}$ and $y_{(j)}$:
\begin{equation}
\hat Q^u(p) =
\begin{cases}
y_{(j-1)} + \dfrac{p - P_{j-1}}{P_j - P_{j-1}}
\bigl(y_{(j)} - y_{(j-1)}\bigr), & p > P_1, \\[6pt]
y_{(1)}, & p \le P_1 . \nonumber
\end{cases}
\end{equation}
Alternatively, interpolating between $y_{(j)}$ and $y_{(j+1)}$ gives
\begin{equation}
\hat Q^b(p) =
\begin{cases}
y_{(j)} + \dfrac{p - P_{j-1}}{P_j - P_{j-1}}
\bigl(y_{(j+1)} - y_{(j)}\bigr), & p < P_J, \\[6pt]
y_{(J)}, & p \ge P_J . \nonumber
\end{cases}
\end{equation}
The two interpolation schemes are illustrated by
the dashed and dotted lines in Figure~\ref{fig:figure1}, respectively. Based on $\hat{Q}^u(p)$ and $\hat{Q}^b(p)$, we develop a weighted quantile estimator that removes this discontinuity,
\begin{equation}
    \hat{Q}(p)=w \hat{Q}^u(p)+ (1 - w) \hat{Q}^b(p), \nonumber
\end{equation}
where $w=\left(p-P_1\right) /\left(P_J-P_1\right)$ when $P_1<p<P_J$; $0$ when $p \leq P_1$; and $1$ when $p \geq P_J$. This weighted definition yields a smooth transition between $\hat{Q}^u(p)$ and $\hat{Q}^b(p)$, providing a continuous quantile curve across the full range of $p$ (solid line in Figure~\ref{fig:figure1}).

The construction above defines $\hat Q(p)$ for a generic marginal CDF estimator $\hat F_Y$. To estimate the marginal quantile of a potential outcome $Y_a$, $a \in \{0,1\}$, we apply this same construction with $\hat F_Y$ replaced by the marginal CDF estimator $\hat F_{a}$; that is, we set $P_j = \hat F_{a}(y_{(j)})$ in the definitions of $\hat Q^u(p)$, $\hat Q^b(p)$, and $\hat Q(p)$ above. We denote the resulting estimator by $\hat Q_a(p)$. The QTE estimator at quantile level $p$ is then
\begin{equation}
    \widehat\Delta(p) = \widehat{QTE}(p)=\hat{F}_{1}^{-1}(p)-\hat{F}_{0}^{-1}(p) = \hat{Q}_{1}(p) -\hat{Q}_{0}(p). \nonumber
\end{equation}

\FloatBarrier
\subsection{Direct Approach with CPM based TMLE}

\subsubsection{Review of Targeted Maximum Likelihood Estimation}

Targeted maximum likelihood estimation (TMLE), also known as targeted minimum loss estimation, targeted machine learning estimation, or targeted learning estimation (\citealt{van2006targeted}), is a general semiparametric estimation procedure that typically solves an efficient influence function (EIF) estimating equation, yielding doubly robust and locally efficient estimators. 
We briefly outline the TMLE algorithm for marginal quantiles as presented in Section~5 of \citealt{diaz2017efficient}. 

Let $\boldsymbol{\eta}$ denote the nuisance parameters, including the propensity score 
$\pi(a\mid \mathbf X)=\Pr(A=a\mid \mathbf X)$ and the conditional outcome distribution $F_{Y\mid A=a,\mathbf X}(y)=\Pr(Y\le y\mid A=a, \mathbf X)$ as shown in the main text.
For a fixed treatment level $a\in\{0,1\}$, define the initial estimates 
$\widehat{\pi}(a\mid \mathbf X)$ and $\widehat{F}_{Y\mid A=a,\mathbf X}(y)$. Let $q_{a}(p)$ denote the target parameter marginal quantile.

Given current estimates, define the estimated marginal CDF of the potential outcome $Y_a$ as
\[
\widehat F_{a}(y)=\frac{1}{n}\sum_{i=1}^n \widehat F_{Y\mid A=a,\mathbf X_i}(y),
\qquad
\widehat q_{a}(p) =\inf\{y:\widehat F_{a}(y)\ge p\}.
\]
Note that the quantile definition adopted here differs slightly from that in Section~\ref{quantile function}. Next, we update the conditional density corresponding to $\widehat F_{Y\mid A=a, \mathbf X}$ using an exponential fluctuation submodel:
\[
\widehat f_{\epsilon}(y\mid A=a,\mathbf X= \mathbf x)
=
c(\epsilon,\widehat f)\exp\!\Bigl\{\epsilon\,H_{\widehat{\boldsymbol{\eta}},\widehat q_{a}(p)}(Z)\Bigr\}\widehat f(y\mid A=a,\mathbf X= \mathbf x),
\]
where $c(\epsilon,\widehat f)$ is a normalizing constant ensuring integration to one, and
\[
H_{\widehat\eta,\widehat q_{a}(p)}(Z)
=
\frac{\mathbb{I}(A=a)}{\widehat\pi(a\mid \mathbf X)}
\Bigl\{\mathbb{I}(Y\le \widehat q_{a}(p))-\widehat F_{Y\mid A=a,\mathbf X}(\widehat q_{a}(p)\mid \mathbf X)\Bigr\}.
\]

The fluctuation parameter $\epsilon$ is estimated by maximizing the empirical log-likelihood under the fluctuation submodel,
\[
\widehat\epsilon
=
\arg\max_{\epsilon}
\sum_{i=1}^n \log \widehat f_{\epsilon}(y_i \mid A_i, \mathbf X_i).
\]

The updated conditional density is then given by $\widetilde f=\widehat f_{\widehat\epsilon}$, with the corresponding conditional CDF $\widetilde F_{Y\mid A=a,\mathbf X}$. The marginal CDF and quantile estimate $\widehat q_{a}(p)$ are recomputed using $\widetilde F_{Y\mid A=a,\mathbf X}$. These updating steps are iterated until the estimated fluctuation parameter $\widehat\epsilon$ is sufficiently close to zero.

\subsubsection{TMLE with Cumulative Probability Models}

From an implementation perspective about the initial estimates, existing TMLE procedures for quantile treatment effects often rely on parametric working models for the conditional outcome distribution $F_{Y \mid A,\mathbf X}(y)$\citep{diaz2017efficient}. 
While convenient, such specifications may be restrictive when the outcome distribution is skewed. To address this limitation, we replace the parametric outcome model with a cumulative probability model (CPM), which provides a flexible semiparametric representation of the entire conditional distribution. The CPM allows the full conditional distribution of $Y$ to be estimated without requiring ad hoc transformations.
\FloatBarrier
\subsection{Sandwich Variance Estimation of QTE \label{sec qte var}}
The efficient influence function of $q_{a}(p)$ is 
\[
\varphi_{q}(Z_i; q_{a}(p))
=
-\frac{1}{f_{a}(q_{a}(p))}
\left[
\frac{\mathbb{I}(A = a)}{\pi(A = a \mid \mathbf X)}
\left\{\mathbb{I}(Y \le q_{a}(p)) -  F_{Y \mid A=a, \mathbf{X} }\!\bigl(q_{a}(p) \bigr)\right\}
+
F_{Y \mid A=a, \mathbf{X} }\!\bigl(q_{a}(p) \bigr)- p
\right].
\]
We specify working models $m_a(q,\mathbf X;\boldsymbol{\xi}) = \Pr(Y \le q \mid A = a, \mathbf X; \boldsymbol{\xi})$ and
$\pi(A = a \mid \mathbf X; \boldsymbol{\psi}) = \Pr(A =a \mid \mathbf X; \boldsymbol{\psi})$. The outcome-regression parameter \(\boldsymbol\xi\) is estimated using the CPM,
whereas the propensity-score parameter \(\boldsymbol\psi\) is estimated from a
separate treatment model. To estimate \(q_a(p)\), we solve the empirical estimating equation $\frac{1}{n}\sum_{i=1}^n
\varphi_{a,p}\{Z_i;q_a(p),\widehat{\boldsymbol\psi},
\widehat{\boldsymbol\xi}\}=0$.
Note that the density term $f_{a}(q_{a}(p))$ appears only as a multiplicative constant and therefore does not affect the root of the estimating equation.
We thus define the estimating equation for $q_{a}(p)$ as
\[
\phi_{a,p}(Z_i; q_{a}(p); \boldsymbol{\psi}; \boldsymbol{\xi})
=
\frac{\mathbb{I}(A = a)}{\pi(A = a \mid \mathbf X; \boldsymbol{\psi})}
\left\{\mathbb{I}(Y \le q_{a}(p)) - m_a(q_{a}(p), \mathbf X; \boldsymbol{\xi})\right\}
+
m_a(q_{a}(p), \mathbf X; \boldsymbol{\xi}) - p.
\]
Let the full parameter vector be $\boldsymbol{\theta}
=
\left(
\boldsymbol{\psi}^\top,
\boldsymbol{\xi}^\top,
q_{1}(p),
q_{0}(p)
\right)^\top$, where $\boldsymbol{\psi}$ parameterizes the propensity score model and 
$\boldsymbol{\xi}$ parameterizes the CPM. Let $\boldsymbol{\theta}_0$ denote the true value of the parameter vector. Define the stacked estimating equation
\[
\boldsymbol{\Phi}(Z_i;\boldsymbol{\theta})
=
\begin{pmatrix}
\boldsymbol{\phi}_\psi(Z_i;\boldsymbol{\psi}) \\
\boldsymbol{\phi}_\xi(Z_i;\boldsymbol{\xi}) \\
\phi_{1,p}(Z_i;q_{1}(p);\boldsymbol{\psi};\boldsymbol{\xi}) \\
\phi_{0,p}(Z_i;q_{0}(p);\boldsymbol{\psi};\boldsymbol{\xi})
\end{pmatrix},
\]
where $\boldsymbol{\phi}_\psi$ is the score function of the propensity score model; $\boldsymbol{\phi}_\xi$ is the score function of the CPM likelihood; $\phi_{1,p}$ and $\phi_{0,p}$ are the estimating equations for $q_{1}(p)$ and $q_{0}(p)$, respectively. \\
Under standard regularity conditions for M-estimation (\citealt{stefanski2002calculus}), $\sqrt{n}(\widehat{\boldsymbol{\theta}}-\boldsymbol{\theta}_0)
\xrightarrow{d}
\mathcal{N}\left(
0,
\boldsymbol{\Sigma}
\right),$
where $\boldsymbol{\Sigma}
=
\boldsymbol{A}^{-1}
\boldsymbol{B}
(\boldsymbol{A}^{-1})^\top,$
with
\[
\boldsymbol{A}
=
\mathbb{E}
\left[
-
\frac{\partial}{\partial \boldsymbol{\theta}^\top}
\boldsymbol{\Phi}(Z_i;\boldsymbol{\theta}_0)
\right],
\qquad
\boldsymbol{B}
=
\mathbb{E}
\left[
\boldsymbol{\Phi}(Z_i;\boldsymbol{\theta}_0)
\boldsymbol{\Phi}(Z_i;\boldsymbol{\theta}_0)^\top
\right].
\]

Empirically, $\widehat{\boldsymbol{\Sigma}}
=
\boldsymbol{A}_n(\widehat{\boldsymbol{\theta}})^{-1}
\boldsymbol{B}_n(\widehat{\boldsymbol{\theta}})
(\boldsymbol{A}_n(\widehat{\boldsymbol{\theta}})^{-1})^\top$, where
\[
\boldsymbol{A}_n(\boldsymbol{\theta})
=
\frac{1}{n}
\sum_{i=1}^n
-
\frac{\partial}{\partial \boldsymbol{\theta}^\top}
\boldsymbol{\Phi}(Z_i;\widehat{\boldsymbol{\theta}}),
\qquad
\boldsymbol{B}_n(\boldsymbol{\theta})
=
\frac{1}{n}
\sum_{i=1}^n
\boldsymbol{\Phi}(Z_i;\widehat{\boldsymbol{\theta}})
\boldsymbol{\Phi}^\top(Z_i;\widehat{\boldsymbol{\theta}}).
\]
The asymptotic variance of the QTE can then be obtained via the Delta method.
The detailed derivation is provided below.

\subsubsection*{Block Structure of the Bread Matrix}

The Jacobian of the stacked estimating equations is block lower-triangular:

\[
\boldsymbol{A}
=
\begin{pmatrix}
\boldsymbol{A}_{\psi\psi} & 0 & 0 & 0 \\
0 & \boldsymbol{A}_{\xi\xi} & 0 & 0 \\
\boldsymbol{A}_{1\psi} & \boldsymbol{A}_{1\xi} & A_{11} & 0 \\
\boldsymbol{A}_{0\psi} & \boldsymbol{A}_{0\xi} & 0 & A_{00}
\end{pmatrix},
\]
where
\[
\boldsymbol{A}_{\psi\psi}
=
-
\mathbb{E}
\!\left[
\frac{\partial}{\partial \boldsymbol{\psi}^\top}
\boldsymbol{\phi}_\psi(Z;\boldsymbol{\psi})
\right],
\quad
\boldsymbol{A}_{\xi\xi}
=
-
\mathbb{E}
\!\left[
\frac{\partial}{\partial \boldsymbol{\xi}^\top}
\boldsymbol{\phi}_\xi(Z;\boldsymbol{\xi})
\right],
\]
\[
\boldsymbol{A}_{1\psi}
=
-
\mathbb{E}
\!\left[
\frac{\partial}{\partial \boldsymbol{\psi}^\top}
\phi_{1,p}(Z;q_{1}(p);\boldsymbol{\psi}; \boldsymbol{\xi})
\right],
\quad
\boldsymbol{A}_{0\psi}
=
-
\mathbb{E}
\!\left[
\frac{\partial}{\partial \boldsymbol{\psi}^\top}
\phi_{0,p}(Z;q_{0}(p);\boldsymbol{\psi}; \boldsymbol{\xi})
\right]
\]
\[
\boldsymbol{A}_{1\xi}
=
-
\mathbb{E}
\!\left[
\frac{\partial}{\partial \boldsymbol{\xi}^\top}
\phi_{1,p}(Z;q_{1}(p);\boldsymbol{\psi}; \boldsymbol{\xi})
\right],
\quad
\boldsymbol{A}_{0\xi}
=
-
\mathbb{E}
\!\left[
\frac{\partial}{\partial \boldsymbol{\xi}^\top}
\phi_{0,p}(Z;q_{0}(p);\boldsymbol{\psi}; \boldsymbol{\xi})
\right]
\]
\[
A_{11}
=
-
\mathbb{E}
\!\left[
\frac{\partial}{\partial q_{1}(p)}
\phi_{1,p}(Z;q_{1}(p);\boldsymbol{\psi}; \boldsymbol{\xi})
\right],
\quad
A_{00}
=
-
\mathbb{E}
\!\left[
\frac{\partial}{\partial q_{0}(p)}
\phi_{0,p}(Z;q_{0}(p);\boldsymbol{\psi}; \boldsymbol{\xi})
\right]
\]

\subsubsection*{$\boldsymbol{\phi}_\psi(Z;\boldsymbol{\psi})$ and $\boldsymbol{A}_{\psi\psi}$}

Assume a logistic propensity score model $\pi(1\mid \mathbf{X}; {\boldsymbol{\psi}})
=
\mathrm{expit}\!\left(\boldsymbol{\psi}^\top \mathbf{X}\right),$
and let $\ell_\psi(Z;\boldsymbol{\psi})$ denote the log-likelihood,
\[
\ell_\psi(Z;\boldsymbol{\psi})
=
A\log\{\pi(1\mid \mathbf{X}; {\boldsymbol{\psi}})\}
+
(1-A)\log\{\pi(0\mid \mathbf{X}; {\boldsymbol{\psi}})\}.
\]

The corresponding score function is
\[
\boldsymbol{\phi}_\psi(Z;\boldsymbol{\psi})
=
\frac{\partial}{\partial \boldsymbol{\psi}}
\ell_\psi(Z;\boldsymbol{\psi})
=
\mathbf{X}\{A-\pi(1\mid \mathbf{X}; {\boldsymbol{\psi}})\}.
\]
\[
\frac{\partial}{\partial \boldsymbol{\psi}^\top}
\boldsymbol{\phi}_\psi(Z;\boldsymbol{\psi})
=
-
\mathbf{X}\mathbf{X}^\top
\,\pi(1\mid \mathbf{X}; {\boldsymbol{\psi}})
\{1-\pi(1\mid \mathbf{X}; {\boldsymbol{\psi}})\},
\]
and therefore
\[
\boldsymbol{A}_{\psi\psi}
=
\mathbb{E}\!\left[
\mathbf{X}\mathbf{X}^\top
\,\pi(1\mid \mathbf{X}; {\boldsymbol{\psi}})
\{1-\pi(1\mid \mathbf{X}; {\boldsymbol{\psi}})\}
\right].
\]

\subsubsection*{$\boldsymbol{\phi}_\xi(Z;\boldsymbol{\xi})$ and $\boldsymbol{A}_{\xi\xi}$}
In practice, both the score function $\boldsymbol{\phi}_\xi(Z;\boldsymbol{\xi})$ of the CPM and  $\boldsymbol{A}_{\xi\xi} = -\mathbb{E}\left[\frac{\partial}{\partial \boldsymbol{\xi}^\top}\boldsymbol{\phi}_\xi(Z;\boldsymbol{\xi})\right]$ can be readily obtained using the \texttt{rms} package (\citealt{rms}).

\subsubsection*{ $\mathbf{A_{1 \psi}}$ and $\mathbf{A_{0 \psi}}$}
Since \(m_1(q_{1}(p),\mathbf X;\boldsymbol{\xi})\) does not depend on \(\boldsymbol{\psi}\),
\[
\frac{\partial}{\partial \boldsymbol\psi^\top}
\phi_{1,p}(Z;q_{1}(p);\boldsymbol{\psi};\boldsymbol{\xi})
=
-
\frac{\mathbb{I}(A=1)}{\pi(1\mid \mathbf X;\boldsymbol{\psi})^2}
\left\{\mathbb{I}(Y\le q_{1}(p))-m_1(q_{1}(p),\mathbf X;\boldsymbol{\xi})\right\}
\frac{\partial}{\partial \boldsymbol\psi^\top}
\pi(1\mid \mathbf X;\boldsymbol{\psi}),
\]
and hence \[\boldsymbol A_{1\psi}
=
-\mathbb E\!\left[
\frac{\partial}{\partial \boldsymbol\psi^\top}
\phi_{1,p}(Z;q_{1}(p);\boldsymbol{\psi};\boldsymbol{\xi})
\right]
=
\mathbb E\!\left[
\frac{\mathbb{I}(A=1)}{\pi(1\mid \mathbf X;\boldsymbol{\psi})^2}
\left\{\mathbb{I}(Y\le q_{1}(p))-m_1(q_{1}(p),\mathbf X;\boldsymbol{\xi})\right\}
\frac{\partial}{\partial \boldsymbol\psi^\top}
\pi(1\mid \mathbf X;\boldsymbol{\psi})
\right].\]
Assume the propensity score is modeled via logistic regression, $\pi(1\mid \mathbf X;\boldsymbol{\psi})
=
\operatorname{expit}(\boldsymbol{\psi}^\top \widetilde{\mathbf X})$, where $\widetilde{\mathbf X}=(1, \mathbf X^\top)^\top$. Then, by the chain rule, $\frac{\partial}{\partial \boldsymbol{\psi}^\top}\pi(1\mid \mathbf X;\boldsymbol{\psi})
=
\pi(1\mid \mathbf X;\boldsymbol{\psi})\{1-\pi(1\mid \mathbf X;\boldsymbol{\psi})\}\widetilde{\mathbf X}^\top$.
Substituting this into \(\boldsymbol A_{1\psi}\) yields
\[
\boldsymbol A_{1\psi}
=
\mathbb E\!\left[
\mathbb I(A=1)\frac{1-\pi(1\mid X;\boldsymbol{\psi})}{\pi(1\mid X;\boldsymbol{\psi})}
\left\{\mathbb I(Y\le q_{1}(p))-m_1(q_{1}(p),\mathbf X;\boldsymbol{\xi})\right\}
\widetilde{\mathbf X}^\top
\right].
\]
Similarly, 
\[
\boldsymbol A_{0\psi}
=
-\mathbb E\!\left[
\mathbb I(A=0)\frac{ 1- \pi(0\mid X;\boldsymbol{\psi})}{\pi(0\mid X;\boldsymbol{\psi})}
\left\{\mathbb I(Y\le q_{0}(p))-m_0(q_{0}(p),X;\boldsymbol{\xi})\right\}
\widetilde X^\top
\right].
\]
\subsubsection*{ $\mathbf{A_{1 \xi}}$ and $\mathbf{A_{0 \xi}}$}
Since \(\pi(1\mid \mathbf X;\boldsymbol{\psi})\) does not depend on \(\boldsymbol{\xi}\),
\[
\frac{\partial}{\partial \boldsymbol\xi^\top}
\phi_{1,p}(Z;q_{1}(p);\boldsymbol{\psi};\boldsymbol{\xi})
=
\left\{
1-\frac{\mathbb{I}(A=1)}{\pi(1\mid \mathbf X;\boldsymbol{\psi})}
\right\}
\frac{\partial}{\partial \boldsymbol\xi^\top}
m_1(q_{1}(p),\mathbf X;\boldsymbol{\xi}),
\]
so that
\[
\boldsymbol A_{1\xi}
=
-\mathbb E\!\left[
\frac{\partial}{\partial \boldsymbol\xi^\top}
\phi_{1,p}(Z;q_{1}(p);\boldsymbol{\psi};\boldsymbol{\xi})
\right]
=
\mathbb E\!\left[
\left\{
\frac{\mathbb{I}(A=1)}{\pi(1\mid \mathbf X;\boldsymbol{\psi})}-1
\right\}
\frac{\partial}{\partial \boldsymbol\xi^\top}
m_1(q_{1}(p),\mathbf X;\boldsymbol{\xi})
\right].
\]
The explicit form of $\frac{\partial}{\partial \boldsymbol\xi^\top}
m_1(q_{1}(p),\mathbf X;\boldsymbol{\xi})$ depends on how the outcome model is
specified. In causal inference, two strategies are common: fitting a single joint CPM on
all observations, or fitting separate CPMs within each treatment arm. We derive
$\boldsymbol{A}_{1\xi}$ and $\boldsymbol{A}_{0\xi}$ under each.
\paragraph{Joint model.}
A single CPM is fit on all observations: $\Pr(Y\le t_j\mid\mathbf{X},A;\boldsymbol{\xi})
=
F_\varepsilon\!\left(\alpha_j - \eta(\mathbf{X},A;\boldsymbol{\xi})\right)$, where $j=1,\ldots,J-1$. $\eta(\mathbf{X},A;\boldsymbol{\xi})$ is a predictor that may incorporate linear
effects, spline terms, and treatment--covariate interactions. $F_\varepsilon$ is a known
link CDF with density $f_\varepsilon$, and
$\boldsymbol{\xi}=(\alpha_1,\ldots,\alpha_{J-1},\boldsymbol{\beta}^\top,\beta^a)^\top$ is shared across
both arms. For $q_{1}(p)\in[t_j,t_{j+1})$, set $\alpha(q_{1}(p))=\alpha_j$ and define
$u_j^a(\mathbf{X})=\alpha_j- \boldsymbol{\beta}^\top\mathbf{X}-\beta^a$, so that
$m_1(q_{1}(p),\mathbf{X};\boldsymbol{\xi})=F_\varepsilon\{u_j^1(\mathbf{X})\}$.
By the chain rule,
\[
\frac{\partial}{\partial\boldsymbol{\xi}^\top}m_1(q_{1}(p),\mathbf{X};\boldsymbol{\xi})
=
f_\varepsilon\{u_j^1(\mathbf{X})\}\frac{\partial}{\partial\boldsymbol{\xi}^\top}u_j^1(\mathbf{X}),
\]
with component-wise derivatives, for $r = 1, \cdots, J-1$,
\[
\frac{\partial u_j^1}{\partial\alpha_r} = \mathbb{I}(r=j),
\qquad
\frac{\partial u_j^1}{\partial\boldsymbol{\beta}^\top} = -\mathbf{X}^\top,
\qquad
\frac{\partial u_j^1}{\partial\beta^1} = -1.
\]
Hence $\frac{\partial}{\partial\boldsymbol{\xi}^\top}u_j^1(\mathbf{X})=(e_j^\top,-\mathbf{X}^\top,-1)$,
where $e_j$ is the unit vector with 1 in the $j$th position, and
\[
\boldsymbol{A}_{1\boldsymbol{\xi}}
=
\mathbb{E}\!\left[
\left\{\frac{\mathbb{I}(A=1)}{\pi(1\mid\mathbf{X};\boldsymbol{\psi})}-1\right\}
f_\varepsilon\{u_j^1(\mathbf{X})\}
\bigl(e_j^\top,-\mathbf{X}^\top,-1\bigr)
\right].
\]
Similarly, for the control arm, define $u_j^0(\mathbf{X})=\alpha_j-\beta^\top\mathbf{X}$
(i.e.\ $\beta^0\cdot 0=0$), giving
\[
\boldsymbol{A}_{0\boldsymbol{\xi}}
=
\mathbb{E}\!\left[
\left\{\frac{\mathbb{I}(A=0)}{\pi(0\mid\mathbf{X};\boldsymbol{\psi})}-1\right\}
f_\varepsilon\{u_j^0(\mathbf{X})\}
\bigl(e_j^\top,-\mathbf{X}^\top,0\bigr)
\right].
\]

\paragraph{Separate models.}
Two CPMs are fit independently, one within each treatment arm, with parameters
$\boldsymbol{\xi}_1$ and $\boldsymbol{\xi}_0$, so
$\boldsymbol{\xi}=(\boldsymbol{\xi}_1^\top,\boldsymbol{\xi}_0^\top)^\top$.
Each model takes the form
\[
\Pr(Y\le t_j\mid\mathbf{X},A=a;\boldsymbol{\xi}_a)
=
F_\varepsilon\!\left(\alpha_{a,j}-\eta_a(\mathbf{X};\boldsymbol{\xi}_a)\right),
\quad j=1,\ldots,J-1,\quad a=0,1,
\]
where $\boldsymbol{\xi}_a=(\alpha_{a,1},\ldots,\alpha_{a,J-1},\boldsymbol{\beta}_a^\top)^\top$. Then,
\[
\boldsymbol{A}_{\boldsymbol{\xi}\boldsymbol{\xi}}
=
\begin{pmatrix}
\boldsymbol{A}_{\boldsymbol{\xi}_1\boldsymbol{\xi}_1} & 0\\
0 & \boldsymbol{A}_{\boldsymbol{\xi}_0\boldsymbol{\xi}_0}
\end{pmatrix}.
\]
With $u_j^a(\mathbf{X};\boldsymbol{\xi}_a)=\alpha_{a,j}-\boldsymbol{\beta}_a^\top\mathbf{X}$, the gradient
$\frac{\partial}{\partial\boldsymbol{\xi}_a^\top}u_j^a=(e^\top_{a,j},-\mathbf{X}^\top)$, and the
blocks reduce to
\[
\boldsymbol{A}_{1\boldsymbol{\xi}} = \left(\boldsymbol{A}_{1\boldsymbol{\xi}_1},\ \mathbf{0}\right),
\qquad
\boldsymbol{A}_{0\boldsymbol{\xi}} = \left(\mathbf{0},\ \boldsymbol{A}_{0\boldsymbol{\xi}_0}\right),
\]
where
\[
\boldsymbol{A}_{1\boldsymbol{\xi}_1}
=
\mathbb{E}\!\left[
\left\{\frac{\mathbb{I}(A=1)}{\pi(1\mid\mathbf{X};\boldsymbol{\psi})}-1\right\}
f_\varepsilon\{u_j^1(\mathbf{X};\boldsymbol{\xi}_1)\}
\bigl(e_{1,j}^\top,-\mathbf{X}^\top\bigr)
\right],
\]
\[
\boldsymbol{A}_{0\boldsymbol{\xi}_0}
=
\mathbb{E}\!\left[
\left\{\frac{\mathbb{I}(A=0)}{\pi(0\mid\mathbf{X};\boldsymbol{\psi})}-1\right\}
f_\varepsilon\{u_{0,j}^0(\mathbf{X};\boldsymbol{\xi}_0)\}
\bigl(e_{0,j}^\top,-\mathbf{X}^\top\bigr)
\right].
\]

\subsubsection*{ $\phi_{1,p}(Z_i;q_{1}(p);\boldsymbol{\psi};\boldsymbol{\xi})$ and $A_{11}$}
\subsubsection{Take derivation first}
By standard M-estimation theory, $A_{11}
=
-
\mathbb{E}
\!\left[
\frac{\partial}{\partial q_{1}(p)}
\phi_{1,p}(Z; q_{1}(p); \boldsymbol{\psi}; \boldsymbol{\xi})
\right]$.
Taking the derivative with respect to \(q_{1}(p)\) yields
\[
\begin{aligned}
& \frac{\partial}{\partial q_{1}(p)}
\phi_{1,p}(Z; q_{1}(p); \boldsymbol{\psi}; \boldsymbol{\xi})\\
&=
\frac{\mathbb{I}(A = 1)}{\pi(A = 1 \mid \mathbf X; \boldsymbol{\psi})}
\left\{
\frac{\partial}{\partial q_{1}(p)} \mathbb{I}(Y \le q_{1}(p))
-
\frac{\partial}{\partial q_{1}(p)} m_1(q_{1}(p), \mathbf X; \boldsymbol{\xi})
\right\}
+
\frac{\partial}{\partial q_{1}(p)} m_1(q_{1}(p), \mathbf X; \boldsymbol{\xi}) \\
&=
\frac{\mathbb{I}(A = 1)}{\pi(A = 1 \mid \mathbf X; \boldsymbol{\psi})}
\,\delta(Y-q_{1}(p))
+
\left\{
\frac{\pi(A = 1 \mid \mathbf X; \boldsymbol{\psi}) - \mathbb{I}(A = 1)}{\pi(A = 1 \mid \mathbf X; \boldsymbol{\psi})}
\right\}
\frac{\partial}{\partial q_{1}(p)} m_1(q_{1}(p), \mathbf X; \boldsymbol{\xi}),
\end{aligned}
\]
where $\delta(Y-q_{1}(p))
=
\frac{\partial}{\partial q_{1}(p)} \mathbb{I}(Y \le q_{1}(p))$. To obtain a smooth approximation, we replace the indicator by a logistic function:
\[
S_{\kappa}(t)=\operatorname{expit}\!\left(\frac{t}{\kappa}\right),
\qquad t=q_{1}(p)-Y,
\]
where \(\kappa>0\) is a bandwidth parameter. Then $ \delta(Y-q_{1}(p))
\approx
\frac{1}{\kappa}
S_{\kappa}(q_{1}(p)-Y)\{1-S_{\kappa}(q_{1}(p)-Y)\}$.
Additionally, since
\(
m_1(q_{1}(p),\mathbf X;\boldsymbol{\xi})
=
F_{Y\mid A=1,\mathbf X}(q_{1}(p)\mid \mathbf X;\boldsymbol{\xi}),
\)
we have $\frac{\partial}{\partial q_{1}(p)}m_1(q_{1}(p),\mathbf X;\boldsymbol{\xi})
=
f_{Y\mid A=1, \mathbf X}(q_{1}(p)\mid \mathbf X;\boldsymbol{\xi})$. Therefore,
\[
A_{11}
=
-
\mathbb{E}
\!\left[
\frac{\mathbb{I}(A = 1)}{\pi(A = 1 \mid \mathbf X; \boldsymbol{\psi})}
\,\delta(Y-q_{1}(p))
\right]
-
\mathbb{E}
\!\left[
\left\{
\frac{\pi(A = 1 \mid \mathbf X; \boldsymbol{\psi}) - \mathbb{I}(A = 1)}{\pi(A = 1 \mid \mathbf X; \boldsymbol{\psi})}
\right\}
f_{Y\mid A=1,\mathbf X}(q_{1}(p)\mid \mathbf X;\boldsymbol{\xi})
\right].
\]
Bandwidth selection plays a critical role in this implementation, as it can be sensitive to sample size and local features of the outcome distribution. In our approach, two types of smoothing are involved. First, to approximate the derivative of the indicator function, $\delta(Y - q_{1}(p))$, we introduce a bandwidth parameter $k$. The bandwidth is selected using the normal reference rule,
$k = 1.06\,\hat{\sigma}_Y\, n^{-1/5}$, where $\hat{\sigma}_Y$ is the sample standard deviation (\citealt{scott2015multivariate, silverman2018density}). Second, to estimate the conditional density $f_{Y \mid A=a, \mathbf X}(q_{a}(p) \mid \mathbf X)$, we differentiate a smoothed estimate of the conditional CDF obtained via local polynomial regression. For this step, we adopt the data-driven bandwidth selection method of \citet{sheather1991reliable}, implemented via \texttt{bw.SJ} function in R.

\subsubsection{Take expectation first}
An alternative route to deriving $A_{11}$ proceeds by first taking the expectation of the estimating function $\phi_{1,p}$, and then differentiating with respect to $q_{1}(p)$. Although $\mathbb{I}(Y \le q_{1}(p))$ is not differentiable, the interchange of differentiation and expectation can nonetheless be justified following \citealt[Section 4]{stefanski2002calculus}, yielding
\[
A_{11}
\;=\;
-\,\mathbb{E}\!\left[\frac{\partial}{\partial q_{1}(p)}\phi_{1,p}(Z;\,q_{1}(p);\boldsymbol{\psi};\boldsymbol{\xi})\right]
\;=\;
-\,\frac{\partial}{\partial q_{1}(p)}\,\mathbb{E}\!\left[\phi_{1,p}(Z;\,q_{1}(p);\boldsymbol{\psi};\boldsymbol{\xi})\right].
\]
We have 
\[
\begin{aligned}
& \mathbb{E}\!\left[\phi_{1,p}(Z_i;q_{1}(p);\boldsymbol{\psi};\boldsymbol{\xi})\right]\\
&=
\mathbb{E}\!\left\{
\frac{\mathbb{I}(A=1)}{\pi(A=1\mid \mathbf X;\boldsymbol{\psi})}
\bigl(\mathbb{I}(Y\le q_{1}(p))-m_1(q_{1}(p),\mathbf X;\boldsymbol{\xi})\bigr)
\right\}
+\mathbb{E}\!\left\{m_1(q_{1}(p),\mathbf X;\boldsymbol{\xi})-p\right\}
\\[0.5em]
&=
\mathbb{E}
\Bigg\{
\mathbb{E}\!\left[
\left.
\frac{\mathbb{I}(A=1)}{\pi(A=1\mid \mathbf X;\boldsymbol{\psi})}
\bigl(\mathbb{I}(Y\le q_{1}(p))-m_1(q_{1}(p),\mathbf X;\boldsymbol{\xi})\bigr)
\,\right|\, \mathbf X
\right]
\Bigg\}
+\mathbb{E}\!\left\{m_1(q_{1}(p),\mathbf X;\boldsymbol{\xi})-p\right\}
\\[0.5em]
&=
\mathbb{E}
\Bigg\{
\frac{\Pr(A=1\mid \mathbf X)}{\pi(A=1\mid \mathbf X;\boldsymbol{\psi})}
\,
\mathbb{E}\!\left[
\left.
\mathbb{I}(Y\le q_{1}(p))-m_1(q_{1}(p), \mathbf X;\boldsymbol{\xi})
\,\right|\, \mathbf X,A=1
\right]
\Bigg\}
+\mathbb{E}\!\left\{m_1(q_{1}(p),\mathbf X;\boldsymbol{\xi})-p\right\}.
\end{aligned}
\]

If the propensity score model is correctly specified, i.e., $\pi(A=1\mid \mathbf X;\boldsymbol{\psi})=\Pr(A=1\mid \mathbf X)$, then
\[
\begin{aligned}
\mathbb{E}\!\left[\phi_{1,p}(Z_i;q_{1}(p);\boldsymbol{\psi};\boldsymbol{\xi})\right]
&=
\mathbb{E}
\Bigg\{
\mathbb{E}
\left[
\left.
\mathbb{I}(Y\le q_{1}(p))-m_1(q_{1}(p),\mathbf X;\boldsymbol{\xi})
\,\right|\,\mathbf X,A=1
\right]
\Bigg\}
+
\mathbb{E}\{m_1(q_{1}(p),\mathbf X;\boldsymbol{\xi})-p\}
\\
&=
\mathbb{E}\!\left\{\mathbb{E}\!\left[\mathbb{I}(Y\le q_{1}(p))\mid \mathbf X,A=1\right]\right\}
-p
\\
&=
F_{1}(q_{1}(p))-p.
\end{aligned}
\]

If the outcome regression model is correctly specified, then $m_1(q_{1}(p),X;\boldsymbol{\xi})
=
\Pr(Y\le q_{1}(p)\mid A=1,\mathbf X)$. Then the first term vanishes because
$\mathbb{E}[\mathbb{I}(Y \le q_{1}(p)) - m_1(q_{1}(p),\mathbf{X};\boldsymbol{\xi}) \mid \mathbf{X}, A=1] = 0$. Thus,
\[
\mathbb{E}\!\left[\phi_{1,p}(Z_i;q_{1}(p);\boldsymbol{\psi};\boldsymbol{\xi})\right]
=
\mathbb{E}\{m_1(q_{1}(p),\mathbf X;\boldsymbol{\xi})\}-p
=
F_{1}(q_{1}(p))-p.
\]

Thus, under either a correctly specified propensity score model or a correctly specified outcome regression model,
\[
\mathbb{E}\!\left[\phi_{1,p}(Z_i;q_{1}(p);\boldsymbol{\psi};\boldsymbol{\xi})\right]
=
F_{1}(q_{1}(p))-p.
\]

Then,
\[
\begin{aligned}
A_{11}
&= - \frac{\partial}{\partial q_{1}(p)}
\mathbb{E}
\!\left[
\phi_{1,p}(Z_i;q_{1}(p);\boldsymbol{\psi};\boldsymbol{\xi})
\right]
=
-
\frac{\partial}{\partial q_{1}(p)}
\left\{
F_{1}(q_{1}(p))-p
\right\}
=
-
f_{1}(q_{1}(p)).
\end{aligned}
\]
Similarly,  the estimating equation for $q_{0}(p)$ is
\[
\phi_{0,p}(Z_i; q_{0}(p); \boldsymbol{\psi}; \boldsymbol{\xi})
=
\frac{\mathbb{I}(A = 0)}{\pi(A = 0 \mid \mathbf X; \boldsymbol{\psi})}
\left\{\mathbb{I}(Y \le q_{0}(p)) - m_0(q_{0}(p), \mathbf X; \boldsymbol{\xi})\right\}
+
m_0(q_{0}(p), \mathbf X; \boldsymbol{\xi}) - p.
\]
and
\[
\begin{aligned}
A_{00}
&= - \frac{\partial}{\partial q_{0}(p)}
\mathbb{E}
\!\left[
\phi_{0,p}(Z_i;q_{0}(p);\boldsymbol{\psi};\boldsymbol{\xi})
\right]
=
-
\frac{\partial}{\partial q_{0}(p)}
\left\{
F_{0}(q_{0}(p))-p
\right\}
=
-
f_{0}(q_{0}(p)).
\end{aligned}
\]
To estimate the density $F_{a}(y)$, we first obtain a smooth approximation of 
$\hat F_{a}^{DR}(y)$ (described in the main text) using local polynomial regression, 
with bandwidth selected via the data-driven method of \citealt{sheather1991reliable} 
(as implemented by \texttt{bw.SJ} in R), and then compute its derivative with respect to $y$ and evaluate it at $q_{a}(p)$.

\subsection{Sandwich Variance Estimation of PTE}
The PTE is defined as $\mathrm{PTE}(y)
=
F_{1}(y)-F_{0}(y)$,
where \(F_{a}(y)=\Pr(Y_a \le y)\) denotes the marginal CDF of the potential outcome \(Y_a\), for \(a\in\{0,1\}\). Let $m_a(y, \mathbf X;\boldsymbol{\xi})
=
\Pr(Y \le y \mid A=a, \mathbf X;\boldsymbol{\xi})$ and $
\pi(A=1\mid \mathbf X;\boldsymbol{\psi})
=
\Pr(A=1\mid \mathbf X;\boldsymbol{\psi})$.
For a fixed threshold \(y\), the estimating equation for
\(F_{1}(y)\) is the efficient influence function itself, thus,
\[
\phi_{1,y}
\left(
Z_i;
F_{1}(y),
\boldsymbol{\psi},
\boldsymbol{\xi}
\right)
=
\frac{\mathbb{I}(A=1)}
{\pi(A=1\mid \mathbf X;\boldsymbol{\psi})}
\left\{
\mathbb{I}(Y \le y)
-
m_1(y,\mathbf X;\boldsymbol{\xi})
\right\}
+
m_1(y, \mathbf X;\boldsymbol{\xi})
-
F_{1}(y).
\]

Let the full parameter vector be $\boldsymbol{\theta}
=
\left(
\boldsymbol{\psi}^\top,
\boldsymbol{\xi}^\top,
F_{1}(y),
F_{0}(y)
\right)^\top$, where \(\boldsymbol{\psi}\) parameterizes the propensity score model and
\(\boldsymbol{\xi}\) parameterizes the CPM. Let
\(\boldsymbol{\theta}_0\) denote the true parameter value.

Define the stacked estimating equation
\[
\boldsymbol{\Phi}(Z_i;\boldsymbol{\theta})
=
\begin{pmatrix}
\boldsymbol{\phi}_{\boldsymbol{\psi}}(Z_i;\boldsymbol{\psi}) \\
\boldsymbol{\phi}_{\boldsymbol{\xi}}(Z_i;\boldsymbol{\xi}) \\
\phi_{1,y}(Z_i;F_{1}(y),\boldsymbol{\psi},\boldsymbol{\xi}) \\
\phi_{0,y}(Z_i;F_{0}(y),\boldsymbol{\psi},\boldsymbol{\xi})
\end{pmatrix},
\]
where \(\boldsymbol{\phi}_\psi\) is the score function for the propensity score model,
\(\boldsymbol{\phi}_\xi\) is the score function for the CPM likelihood,
and \(\phi_{1,y}\) and \(\phi_{0,y}\) are the estimating equations for
\(F_{1}(y)\) and \(F_{0}(y)\), respectively. Under standard regularity conditions for M-estimation
(\citealt{stefanski2002calculus}),
\[
\sqrt{n}
\left(
\widehat{\boldsymbol{\theta}}
-
\boldsymbol{\theta}_0
\right)
\xrightarrow{d}
\mathcal{N}
\left(
0,
\boldsymbol{\Sigma}
\right),
\]
where $\boldsymbol{\Sigma}
=
\boldsymbol{A}^{-1}
\boldsymbol{B}
(\boldsymbol{A}^{-1})^\top$, with
\[
\boldsymbol{A}
=
\mathbb{E}
\left[
-
\frac{\partial}
{\partial \boldsymbol{\theta}^\top}
\boldsymbol{\Phi}(Z_i;\boldsymbol{\theta}_0)
\right],
\qquad
\boldsymbol{B}
=
\mathbb{E}
\left[
\boldsymbol{\Phi}(Z_i;\boldsymbol{\theta}_0)
\boldsymbol{\Phi}(Z_i;\boldsymbol{\theta}_0)^\top
\right].
\]

Empirically, the sandwich variance estimator is $\widehat{\boldsymbol{\Sigma}}
=
\boldsymbol{A}_n(\widehat{\boldsymbol{\theta}})^{-1}
\boldsymbol{B}_n(\widehat{\boldsymbol{\theta}})
(\boldsymbol{A}_n(\widehat{\boldsymbol{\theta}})^{-1})^\top$,
where
\[
\boldsymbol{A}_n(\widehat{\boldsymbol{\theta}})
=
\frac{1}{n}
\sum_{i=1}^n
\left[
-
\frac{\partial}
{\partial \boldsymbol{\theta}^\top}
\boldsymbol{\Phi}(Z_i;\widehat{\boldsymbol{\theta}})
\right],
\qquad
\boldsymbol{B}_n(\widehat{\boldsymbol{\theta}})
=
\frac{1}{n}
\sum_{i=1}^n
\boldsymbol{\Phi}(Z_i;\widehat{\boldsymbol{\theta}})
\boldsymbol{\Phi}^\top(Z_i;\widehat{\boldsymbol{\theta}}).
\]
The asymptotic variance of the PTE estimator can be obtained via the Delta method. The detailed derivation is provided below.

\subsubsection*{Block Structure of the Bread Matrix}

The Jacobian matrix of the stacked estimating equations has the block lower-triangular form
\[
\boldsymbol{A}
=
\begin{pmatrix}
\boldsymbol{A}_{\boldsymbol{\psi}\boldsymbol{\psi}} & 0 & 0 & 0 \\
0 & \boldsymbol{A}_{\boldsymbol{\xi}\boldsymbol{\xi}} & 0 & 0 \\
\boldsymbol{A}_{1\boldsymbol{\psi}} & \boldsymbol{A}_{1\boldsymbol{\xi}} & A_{11} & 0 \\
\boldsymbol{A}_{0\boldsymbol{\psi}} & \boldsymbol{A}_{0\boldsymbol{\xi}} & 0 & A_{00}
\end{pmatrix},
\]
where
\[
\boldsymbol{A}_{\boldsymbol{\psi}\boldsymbol{\psi}}
=
-
\mathbb{E}
\left[
\frac{\partial}
{\partial \boldsymbol{\psi}^\top}
\boldsymbol{\phi}_{\boldsymbol{\psi}}(Z;\boldsymbol{\psi})
\right],
\qquad
\boldsymbol{A}_{\boldsymbol{\xi}\boldsymbol{\xi}}
=
-
\mathbb{E}
\left[
\frac{\partial}
{\partial \boldsymbol{\xi}^\top}
\boldsymbol{\phi}_{\boldsymbol{\xi}}(Z;\boldsymbol{\xi})
\right],
\]
\[
\boldsymbol{A}_{1\boldsymbol{\psi}}
=
-
\mathbb{E}
\left[
\frac{\partial}
{\partial \boldsymbol{\psi}^\top}
\phi_{1,y}
\left(
Z;
F_{1}(y),
\boldsymbol{\psi},
\boldsymbol{\xi}
\right)
\right],
\qquad
\boldsymbol{A}_{0\boldsymbol{\psi}}
=
-
\mathbb{E}
\left[
\frac{\partial}
{\partial \boldsymbol{\psi}^\top}
\phi_{0,y}
\left(
Z;
F_{0}(y),
\boldsymbol{\psi},
\boldsymbol{\xi}
\right)
\right],
\]
\[
\boldsymbol{A}_{1\boldsymbol{\xi}}
=
-
\mathbb{E}
\left[
\frac{\partial}
{\partial \boldsymbol{\xi}^\top}
\phi_{1,y}
\left(
Z;
F_{1}(y),
\boldsymbol{\psi},
\boldsymbol{\xi}
\right)
\right],
\qquad
\boldsymbol{A}_{0\boldsymbol{\xi}}
=
-
\mathbb{E}
\left[
\frac{\partial}
{\partial \boldsymbol{\xi}^\top}
\phi_{0,y}
\left(
Z;
F_{0}(y),
\boldsymbol{\psi},
\boldsymbol{\xi}
\right)
\right],
\]
\[
A_{11}
=
-
\mathbb{E}
\left[
\frac{\partial}
{\partial F_{1}(y)}
\phi_{1,y}
\left(
Z;
F_{1}(y),
\boldsymbol{\psi},
\boldsymbol{\xi}
\right)
\right],
\qquad
A_{00}
=
-
\mathbb{E}
\left[
\frac{\partial}
{\partial F_{0}(y)}
\phi_{0,y}
\left(
Z;
F_{0}(y),
\boldsymbol{\psi},
\boldsymbol{\xi}
\right)
\right].
\]

The score functions
$\boldsymbol{\phi}_{\boldsymbol{\psi}}(Z;\boldsymbol{\psi})$
and
$\boldsymbol{\phi}_{\boldsymbol{\xi}}(Z;\boldsymbol{\xi})$,
together with the corresponding Jacobian blocks
$\boldsymbol{A}_{\boldsymbol{\psi}\boldsymbol{\psi}}$
and
$\boldsymbol{A}_{\boldsymbol{\xi}\boldsymbol{\xi}}$,
were derived in Section~\ref{sec qte var}.

\subsubsection*{\texorpdfstring{\(\boldsymbol A_{1\boldsymbol \psi}\)  \(\boldsymbol A_{0\boldsymbol \psi}\) and \(\boldsymbol A_{1 \boldsymbol \xi}\)  \(\boldsymbol A_{0 \boldsymbol \xi}\)}{A1psi and A1xi}}
The derivations of the nuisance-parameter blocks for the PTE estimator closely parallel those for the QTE estimator presented previously. The primary difference is that the estimating equations are now constructed for the marginal CDF \(F_{a}(y)\) rather than the marginal quantiles \(q_{a}(p)\).

Since \(m_1(y,X;\boldsymbol{\xi})\) does not depend on \(\boldsymbol{\psi}\),
\[
\frac{\partial}{\partial \boldsymbol{\psi}^\top}
\phi_{1,y}
=
-
\frac{\mathbb I(A=1)}
{\pi(1 \mid \mathbf X;\boldsymbol{\psi})^2}
\left\{
\mathbb I(Y\le y)-m_1(y,\mathbf X;\boldsymbol{\xi})
\right\}
\frac{\partial}{\partial \boldsymbol{\psi}^\top}
\pi(1 \mid \mathbf X;\boldsymbol{\psi}).
\]
Under the logistic propensity score model, $\pi(1 \mid \mathbf X;\boldsymbol{\psi})
=
\operatorname{expit}(\boldsymbol{\psi}^\top \widetilde X)$ with $\widetilde{ \mathbf X}=(1,\mathbf X^\top)^\top$,
which yields
\[
\boldsymbol A_{1\psi}
=
\mathbb E
\left[
\mathbb I(A=1)
\frac{1-\pi(1 \mid \mathbf X;\boldsymbol{\psi})}
{\pi(1 \mid \mathbf X;\boldsymbol{\psi})}
\left\{
\mathbb I(Y\le y)-m_1(y,\mathbf X;\boldsymbol{\xi})
\right\}
\widetilde{\mathbf X}^\top
\right].
\]

Similarly, because \(\pi(1 \mid \mathbf X;\boldsymbol{\psi})\) does not depend on \(\boldsymbol{\xi}\),
\[
\boldsymbol A_{1\xi}
=
\mathbb E
\left[
\left\{
\frac{\mathbb I(A=1)}
{\pi(1 \mid \mathbf X;\boldsymbol{\psi})}
-1
\right\}
\frac{\partial}{\partial \boldsymbol{\xi}^\top}
m_1(y,\mathbf X;\boldsymbol{\xi})
\right],
\]
where $\frac{\partial}{\partial \boldsymbol{\xi}^\top}
m_1(y, \mathbf X;\boldsymbol{\xi})$ was derived in Section~\ref{sec qte var}.

Similarly, the corresponding expressions for the control group are
\[
\boldsymbol A_{0\psi}
=
-\mathbb E
\left[
\mathbb I(A=0)
\frac{1-\pi(0 \mid \mathbf X;\boldsymbol{\psi})}
{\pi(0 \mid \mathbf X;\boldsymbol{\psi})}
\left\{
\mathbb I(Y\le y)-m_0(y,\mathbf X;\boldsymbol{\xi})
\right\}
\widetilde{\mathbf X}^\top
\right],
\]
and
\[
\boldsymbol A_{0\xi}
=
\mathbb E
\left[
\left\{
\frac{\mathbb I(A=0)}
{\pi(0 \mid \mathbf X;\boldsymbol{\psi})}
-1
\right\}
\frac{\partial}{\partial \boldsymbol{\xi}^\top}
m_0(y,\mathbf X;\boldsymbol{\xi})
\right].
\]

\subsubsection*{ $A_{11}$ and $A_{00}$}
For a fixed threshold \(y\), the estimating equation for
\(F_{1}(y)\) is the EIF itself, thus,
\[
\phi_{1,y}
\left(
Z_i;
F_{1}(y),
\boldsymbol{\psi},
\boldsymbol{\xi}
\right)
=
\frac{\mathbb{I}(A=1)}
{\pi(A=1\mid \mathbf X;\boldsymbol{\psi})}
\left\{
\mathbb{I}(Y \le y)
-
m_1(y,\mathbf X;\boldsymbol{\xi})
\right\}
+
m_1(y, \mathbf X;\boldsymbol{\xi})
-
F_{1}(y).
\]

Then, $A_{11}
=
-
\mathbb{E}
\left[
\frac{\partial}{\partial F_{1}(y)}
\phi_{1,y}
\left(
Z_i;
F_{1}(y),
\boldsymbol{\psi},
\boldsymbol{\xi}
\right)
\right] = 1$. Similarly, $A_{00} = 1$.

\section{Simulation Study}
\FloatBarrier
\subsection{Empirical distribution of simulated outcome}
\FloatBarrier
\begin{figure}[ht]
    \centering
   \includegraphics[height=0.22\textheight]{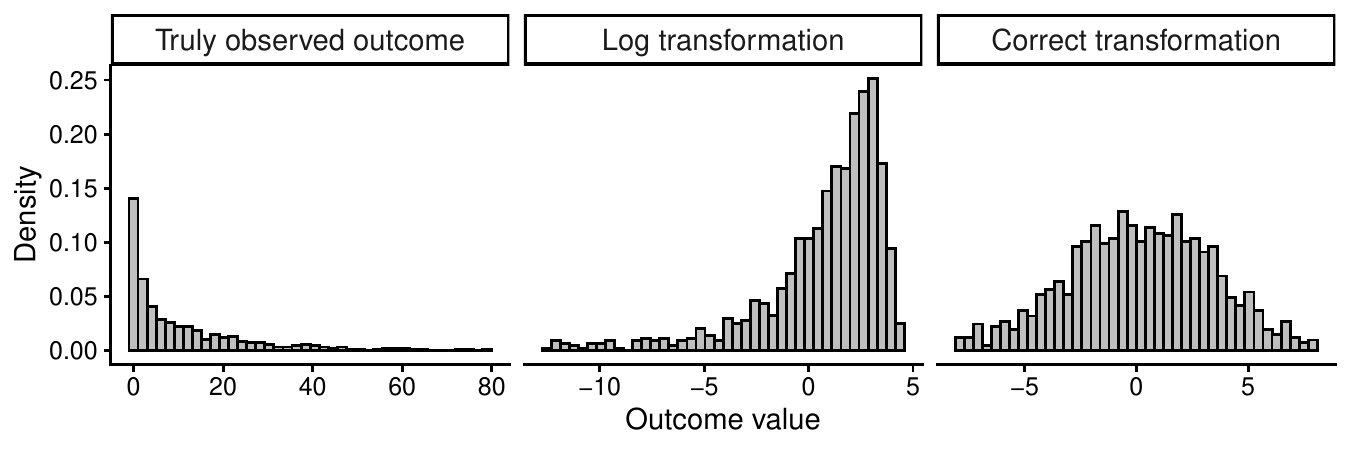}
\caption{
Empirical distributions of the simulated outcome under different transformations.
The left panel shows the distribution of the observed outcome, which is highly right-skewed.
The middle panel shows the distribution after a log transformation, which partially reduces skewness but still deviates from normality.
The right panel shows the distribution under the correct transformation implied by the data-generating mechanism, yielding an approximately symmetric distribution.
}
    \label{fig:transformation}
\end{figure}

\subsection{Different Quantile Levels Under Sample Sizes of 500 and 1,000}
For brevity, we use the same estimator notation and data-generating mechanisms as defined in the main manuscript. Specifically, \textbf{AIPW-CPM-icdf} and \textbf{AIPW-CPM} denote the proposed CPM-based doubly robust estimators based on the inverse-CDF and direct estimating-equation approaches, respectively. \textbf{OR-CPM-icdf} and \textbf{IPW-icdf} denote the corresponding outcome regression and inverse probability weighting estimators under the inverse-CDF framework. Competing estimators include \textbf{IPW-Firpo}, \textbf{AIPW}, \textbf{TMLE}, \textbf{TMLE-cqr}, and the proposed \textbf{TMLE-CPM}. Prefixes ``log-'' and ``ct-'' indicate estimators applied to the log-transformed outcome and correctly transformed latent outcome, respectively. Correct and misspecified nuisance-model settings follow the definitions in the main text. Tables S3.1--S3.9 present simulation results across the 10th, 25th, 50th, 75th, and 90th percentiles under sample sizes of 500 and 1,000. The number of simulation replicates was 1,000. In the main manuscript, we presented only the results for $\mathrm{QTE}(0.5)$ with sample size 1,000. The additional results reported here exhibit qualitatively similar patterns across different quantile levels and sample sizes.

\begin{table}[htbp]
\centering
\scriptsize
\setlength{\tabcolsep}{3.6pt}
\renewcommand{\arraystretch}{1.10}

\caption{Simulation performance metrics across nuisance-model specification scenarios. Target parameter is $QTE(0.1)$, true value is 0.23. Each scenario was evaluated using 1,000 simulation replicates. Sample size is 1,000.
\textit{OR}: outcome regression. 
\textit{PS}: propensity score. 
\textit{Var}: empirical variance over 1,000 simulation replicates. 
\textit{RMSE}: root mean squared error. 
\textit{MedAE}: median absolute error.}
\label{Stable1}

\begin{threeparttable}

\begin{adjustbox}{max width=\linewidth}
\begin{tabular}{@{}l
  *{4}{S[table-format=2.3,table-text-alignment=center]}
  @{\hspace{3.6em}}
  *{4}{S[table-format=2.3,table-text-alignment=center]}@{}}
\toprule
& \multicolumn{4}{c}{\textbf{Correct OR, Correct PS}}
& \multicolumn{4}{c}{\textbf{Correct OR, Misspecified PS}}\\
\cmidrule(r){2-5}\cmidrule(l){6-9}
\textbf{Estimators}
& {\textbf{Bias}}
& {\textbf{Var}}
& {\textbf{RMSE}}
& {\textbf{MedAE}}
& {\textbf{Bias}}
& {\textbf{Var}}
& {\textbf{RMSE}}
& {\textbf{MedAE}}\\
\midrule

AIPW                 & 0.011 & 0.004 & 0.067 & 0.042 & -0.066 & 0.003 & 0.086 & 0.076\\
TMLE                 & 0.012 & 0.005 & 0.071 & 0.043 & 0.066 & 0.006 & 0.102 & 0.064\\
TMLE-cqr             & 0.048 & 0.006 & 0.092 & 0.056 & 0.076 & 0.007 & 0.112 & 0.071\\
\midrule
log-AIPW             & 0.011 & 0.004 & 0.065 & 0.041 & 0.029 & 0.004 & 0.067 & 0.040\\
log-TMLE             & 0.012 & 0.004 & 0.065 & 0.040 & 0.021 & 0.004 & 0.068 & 0.041\\
log-TMLE-cqr         & 0.012 & 0.004 & 0.067 & 0.040 & 0.043 & 0.005 & 0.083 & 0.051\\
\midrule
ct-AIPW              & 0.010 & 0.004 & 0.064 & 0.040 & 0.012 & 0.004 & 0.060 & 0.038\\
ct-TMLE              & 0.010 & 0.004 & 0.064 & 0.040 & 0.010 & 0.004 & 0.064 & 0.040\\
ct-TMLE-cqr          & 0.011 & 0.004 & 0.064 & 0.040 & 0.010 & 0.004 & 0.064 & 0.040\\
\midrule
AIPW-CPM             & 0.010 & 0.004 & 0.064 & 0.040 & 0.010 & 0.004 & 0.061 & 0.039\\
AIPW-CPM-icdf        & 0.014 & 0.004 & 0.065 & 0.040 & 0.012 & 0.004 & 0.061 & 0.039\\
TMLE-CPM             & 0.011 & 0.004 & 0.065 & 0.040 & 0.011 & 0.004 & 0.064 & 0.040\\
\midrule
AIPW-CPM-mislink      & 0.011 & 0.004 & 0.065 & 0.040 & 0.010 & 0.004 & 0.061 & 0.039\\
AIPW-CPM-icdf-mislink & 0.014 & 0.004 & 0.065 & 0.040 & 0.012 & 0.004 & 0.061 & 0.039\\
TMLE-CPM-mislink      & 0.011 & 0.004 & 0.065 & 0.040 & 0.011 & 0.004 & 0.064 & 0.040\\
\midrule
OR-CPM-icdf          & 0.013 & 0.003 & 0.054 & 0.033 & 0.013 & 0.003 & 0.054 & 0.033\\
IPW-icdf             & 0.019 & 0.006 & 0.077 & 0.045 & 0.183 & 0.012 & 0.214 & 0.172\\
IPW-Firpo          & 0.013 & 0.005 & 0.075 & 0.045 & 0.176 & 0.012 & 0.208 & 0.165\\

\bottomrule
\end{tabular}
\end{adjustbox}

\vspace{7pt}

\begin{adjustbox}{max width=\linewidth}
\begin{tabular}{@{}l
  *{4}{S[table-format=2.3,table-text-alignment=center]}
  @{\hspace{3.6em}}
  *{4}{S[table-format=2.3,table-text-alignment=center]}@{}}
\toprule
& \multicolumn{4}{c}{\textbf{Misspecified OR, Correct PS}}
& \multicolumn{4}{c}{\textbf{Misspecified OR, Misspecified PS}}\\
\cmidrule(r){2-5}\cmidrule(l){6-9}
\textbf{Estimators}
& {\textbf{Bias}}
& {\textbf{Var}}
& {\textbf{RMSE}}
& {\textbf{MedAE}}
& {\textbf{Bias}}
& {\textbf{Var}}
& {\textbf{RMSE}}
& {\textbf{MedAE}}\\
\midrule

AIPW                 & 0.014 & 0.005 & 0.075 & 0.044 & 0.176 & 0.012 & 0.208 & 0.165\\
TMLE                 & 0.008 & 0.008 & 0.088 & 0.058 & 0.184 & 0.015 & 0.221 & 0.172\\
TMLE-cqr             & 0.014 & 0.005 & 0.074 & 0.044 & 0.177 & 0.012 & 0.208 & 0.165\\
\midrule
log-AIPW             & 0.014 & 0.005 & 0.075 & 0.044 & 0.176 & 0.012 & 0.208 & 0.165\\
log-TMLE             & 0.016 & 0.005 & 0.075 & 0.045 & 0.176 & 0.011 & 0.206 & 0.166\\
log-TMLE-cqr         & 0.014 & 0.005 & 0.074 & 0.044 & 0.177 & 0.012 & 0.208 & 0.165\\
\midrule
ct-AIPW              & 0.014 & 0.005 & 0.075 & 0.044 & 0.176 & 0.012 & 0.208 & 0.165\\
ct-TMLE              & 0.014 & 0.005 & 0.074 & 0.045 & 0.177 & 0.012 & 0.208 & 0.167\\
ct-TMLE-cqr          & 0.014 & 0.005 & 0.074 & 0.044 & 0.177 & 0.012 & 0.208 & 0.165\\
\midrule
AIPW-CPM             & 0.014 & 0.005 & 0.075 & 0.044 & 0.176 & 0.012 & 0.208 & 0.165\\
AIPW-CPM-icdf        & 0.017 & 0.005 & 0.075 & 0.044 & 0.182 & 0.012 & 0.213 & 0.171\\
TMLE-CPM             & 0.014 & 0.005 & 0.075 & 0.046 & 0.177 & 0.012 & 0.208 & 0.168\\
\midrule
AIPW-CPM-mislink      & 0.014 & 0.005 & 0.075 & 0.044 & 0.176 & 0.012 & 0.208 & 0.165\\
AIPW-CPM-icdf-mislink & 0.017 & 0.005 & 0.075 & 0.044 & 0.183 & 0.012 & 0.213 & 0.172\\
TMLE-CPM-mislink      & 0.014 & 0.005 & 0.075 & 0.045 & 0.177 & 0.012 & 0.208 & 0.168\\
\midrule
OR-CPM-icdf          & 0.182 & 0.008 & 0.203 & 0.168 & 0.182 & 0.008 & 0.203 & 0.168\\
IPW-icdf             & 0.019 & 0.006 & 0.077 & 0.045 & 0.183 & 0.012 & 0.214 & 0.172\\
IPW-Firpo          & 0.013 & 0.005 & 0.075 & 0.045 & 0.176 & 0.012 & 0.208 & 0.165\\

\bottomrule
\end{tabular}
\end{adjustbox}

\end{threeparttable}
\end{table}

\FloatBarrier
\begin{table}[htbp]
\centering
\scriptsize
\setlength{\tabcolsep}{3.6pt}
\renewcommand{\arraystretch}{1.10}

\caption{Simulation performance metrics across nuisance-model specification scenarios. Target parameter is $QTE(0.25)$, true value is 1.62. Each scenario was evaluated using 1,000 simulation replicates. Sample size is 1,000.
\textit{OR}: outcome regression. 
\textit{PS}: propensity score. 
\textit{Var}: empirical variance over 1,000 simulation replicates. 
\textit{RMSE}: root mean squared error. 
\textit{MedAE}: median absolute error.}
\label{Stable2}

\begin{threeparttable}

\begin{adjustbox}{max width=\linewidth}
\begin{tabular}{@{}l
  *{4}{S[table-format=2.3,table-text-alignment=center]}
  @{\hspace{3.6em}}
  *{4}{S[table-format=2.3,table-text-alignment=center]}@{}}
\toprule
& \multicolumn{4}{c}{\textbf{Correct OR, Correct PS}}
& \multicolumn{4}{c}{\textbf{Correct OR, Misspecified PS}}\\
\cmidrule(r){2-5}\cmidrule(l){6-9}
\textbf{Estimators}
& {\textbf{Bias}}
& {\textbf{Var}}
& {\textbf{RMSE}}
& {\textbf{MedAE}}
& {\textbf{Bias}}
& {\textbf{Var}}
& {\textbf{RMSE}}
& {\textbf{MedAE}}\\
\midrule

AIPW                 & 0.010 & 0.056 & 0.237 & 0.160 & 0.103 & 0.057 & 0.259 & 0.162\\
TMLE                 & 0.008 & 0.059 & 0.243 & 0.159 & 0.162 & 0.057 & 0.288 & 0.185\\
TMLE-cqr             & 0.012 & 0.054 & 0.233 & 0.161 & 0.017 & 0.053 & 0.232 & 0.161\\
\midrule
log-AIPW             & 0.010 & 0.056 & 0.238 & 0.159 & -0.056 & 0.051 & 0.233 & 0.162\\
log-TMLE             & 0.009 & 0.056 & 0.237 & 0.159 & 0.093 & 0.060 & 0.262 & 0.167\\
log-TMLE-cqr         & 0.008 & 0.057 & 0.238 & 0.162 & 0.111 & 0.057 & 0.264 & 0.166\\
\midrule
ct-AIPW              & 0.009 & 0.054 & 0.233 & 0.163 & 0.011 & 0.050 & 0.223 & 0.150\\
ct-TMLE              & 0.010 & 0.054 & 0.233 & 0.160 & 0.009 & 0.053 & 0.231 & 0.156\\
ct-TMLE-cqr          & 0.009 & 0.054 & 0.233 & 0.160 & 0.010 & 0.053 & 0.231 & 0.160\\
\midrule
AIPW-CPM             & 0.009 & 0.054 & 0.234 & 0.161 & 0.011 & 0.051 & 0.226 & 0.154\\
AIPW-CPM-icdf        & 0.016 & 0.054 & 0.233 & 0.161 & 0.013 & 0.051 & 0.226 & 0.154\\
TMLE-CPM             & 0.010 & 0.054 & 0.233 & 0.159 & 0.008 & 0.054 & 0.232 & 0.159\\
\midrule
AIPW-CPM-mislink      & 0.010 & 0.054 & 0.234 & 0.158 & 0.012 & 0.051 & 0.227 & 0.154\\
AIPW-CPM-icdf-mislink & 0.016 & 0.054 & 0.233 & 0.160 & 0.014 & 0.051 & 0.227 & 0.153\\
TMLE-CPM-mislink      & 0.009 & 0.054 & 0.233 & 0.160 & 0.010 & 0.054 & 0.232 & 0.159\\
\midrule
OR-CPM-icdf          & 0.016 & 0.040 & 0.200 & 0.135 & 0.016 & 0.040 & 0.200 & 0.135\\
IPW-icdf             & 0.019 & 0.068 & 0.261 & 0.171 & 0.865 & 0.131 & 0.938 & 0.854\\
IPW-Firpo          & 0.010 & 0.067 & 0.260 & 0.172 & 0.855 & 0.130 & 0.928 & 0.839\\

\bottomrule
\end{tabular}
\end{adjustbox}

\vspace{7pt}

\begin{adjustbox}{max width=\linewidth}
\begin{tabular}{@{}l
  *{4}{S[table-format=2.3,table-text-alignment=center]}
  @{\hspace{3.6em}}
  *{4}{S[table-format=2.3,table-text-alignment=center]}@{}}
\toprule
& \multicolumn{4}{c}{\textbf{Misspecified OR, Correct PS}}
& \multicolumn{4}{c}{\textbf{Misspecified OR, Misspecified PS}}\\
\cmidrule(r){2-5}\cmidrule(l){6-9}
\textbf{Estimators}
& {\textbf{Bias}}
& {\textbf{Var}}
& {\textbf{RMSE}}
& {\textbf{MedAE}}
& {\textbf{Bias}}
& {\textbf{Var}}
& {\textbf{RMSE}}
& {\textbf{MedAE}}\\
\midrule

AIPW                 & 0.010 & 0.068 & 0.260 & 0.174 & 0.855 & 0.130 & 0.928 & 0.838\\
TMLE                 & 0.021 & 0.068 & 0.261 & 0.176 & 0.847 & 0.136 & 0.924 & 0.829\\
TMLE-cqr             & 0.011 & 0.067 & 0.259 & 0.172 & 0.855 & 0.130 & 0.928 & 0.838\\
\midrule
log-AIPW             & 0.010 & 0.067 & 0.260 & 0.173 & 0.855 & 0.130 & 0.928 & 0.838\\
log-TMLE             & 0.008 & 0.067 & 0.259 & 0.173 & 0.848 & 0.130 & 0.922 & 0.833\\
log-TMLE-cqr         & 0.011 & 0.067 & 0.259 & 0.172 & 0.855 & 0.130 & 0.928 & 0.839\\
\midrule
ct-AIPW              & 0.010 & 0.068 & 0.260 & 0.175 & 0.855 & 0.130 & 0.928 & 0.838\\
ct-TMLE              & 0.014 & 0.067 & 0.259 & 0.174 & 0.855 & 0.131 & 0.928 & 0.839\\
ct-TMLE-cqr          & 0.011 & 0.067 & 0.259 & 0.172 & 0.855 & 0.131 & 0.928 & 0.839\\
\midrule
AIPW-CPM             & 0.010 & 0.068 & 0.260 & 0.175 & 0.855 & 0.130 & 0.928 & 0.839\\
AIPW-CPM-icdf        & 0.016 & 0.067 & 0.259 & 0.173 & 0.865 & 0.131 & 0.937 & 0.854\\
TMLE-CPM             & 0.012 & 0.067 & 0.260 & 0.177 & 0.854 & 0.133 & 0.928 & 0.836\\
\midrule
AIPW-CPM-mislink      & 0.010 & 0.068 & 0.260 & 0.175 & 0.855 & 0.130 & 0.928 & 0.839\\
AIPW-CPM-icdf-mislink & 0.015 & 0.067 & 0.259 & 0.173 & 0.864 & 0.131 & 0.937 & 0.851\\
TMLE-CPM-mislink      & 0.012 & 0.067 & 0.259 & 0.174 & 0.854 & 0.130 & 0.927 & 0.835\\
\midrule
OR-CPM-icdf          & 0.934 & 0.108 & 0.990 & 0.915 & 0.934 & 0.108 & 0.990 & 0.915\\
IPW-icdf             & 0.019 & 0.068 & 0.261 & 0.171 & 0.865 & 0.131 & 0.938 & 0.854\\
IPW-Firpo          & 0.010 & 0.067 & 0.260 & 0.172 & 0.855 & 0.130 & 0.928 & 0.839\\

\bottomrule
\end{tabular}
\end{adjustbox}

\end{threeparttable}
\end{table}

\FloatBarrier
\begin{table}[htbp]
\centering
\scriptsize
\setlength{\tabcolsep}{3.6pt}
\renewcommand{\arraystretch}{1.10}

\caption{Simulation performance metrics across nuisance-model specification scenarios. Target parameter is $QTE(0.75)$, true value is 12.34. Each scenario was evaluated using 1,000 simulation replicates. Sample size is 1,000.
\textit{OR}: outcome regression. 
\textit{PS}: propensity score. 
\textit{Var}: empirical variance over 1,000 simulation replicates. 
\textit{RMSE}: root mean squared error. 
\textit{MedAE}: median absolute error.}
\label{table3}

\begin{threeparttable}

\begin{adjustbox}{max width=\linewidth}
\begin{tabular}{@{}l
  *{4}{S[table-format=2.3,table-text-alignment=center]}
  @{\hspace{3.6em}}
  *{4}{S[table-format=2.3,table-text-alignment=center]}@{}}
\toprule
& \multicolumn{4}{c}{\textbf{Correct OR, Correct PS}}
& \multicolumn{4}{c}{\textbf{Correct OR, Misspecified PS}}\\
\cmidrule(r){2-5}\cmidrule(l){6-9}
\textbf{Estimators}
& {\textbf{Bias}}
& {\textbf{Var}}
& {\textbf{RMSE}}
& {\textbf{MedAE}}
& {\textbf{Bias}}
& {\textbf{Var}}
& {\textbf{RMSE}}
& {\textbf{MedAE}}\\
\midrule

AIPW                 & 0.044 & 1.274 & 1.130 & 0.758 & -0.205 & 1.300 & 1.159 & 0.736\\
TMLE                 & 0.021 & 1.203 & 1.097 & 0.733 & 1.071 & 1.185 & 1.527 & 1.105\\
TMLE-cqr             & 0.030 & 1.272 & 1.128 & 0.749 & 1.472 & 1.303 & 1.863 & 1.443\\
\midrule
log-AIPW             & 0.035 & 1.084 & 1.042 & 0.682 & 0.426 & 1.113 & 1.138 & 0.787\\
log-TMLE             & 0.026 & 1.080 & 1.039 & 0.662 & 0.389 & 1.100 & 1.119 & 0.776\\
log-TMLE-cqr         & 0.035 & 1.064 & 1.032 & 0.663 & 0.046 & 1.054 & 1.028 & 0.674\\
\midrule
ct-AIPW              & 0.033 & 1.065 & 1.033 & 0.670 & 0.068 & 1.085 & 1.044 & 0.682\\
ct-TMLE              & 0.036 & 1.066 & 1.033 & 0.675 & 0.031 & 1.056 & 1.028 & 0.677\\
ct-TMLE-cqr          & 0.033 & 1.061 & 1.031 & 0.660 & 0.032 & 1.048 & 1.024 & 0.653\\
\midrule
AIPW-CPM             & 0.030 & 1.065 & 1.032 & 0.672 & 0.037 & 1.039 & 1.020 & 0.668\\
AIPW-CPM-icdf        & 0.012 & 1.057 & 1.028 & 0.665 & 0.030 & 1.035 & 1.018 & 0.663\\
TMLE-CPM             & 0.031 & 1.051 & 1.026 & 0.654 & 0.030 & 1.053 & 1.027 & 0.666\\
\midrule
AIPW-CPM-mislink      & 0.031 & 1.073 & 1.036 & 0.676 & 0.046 & 1.044 & 1.023 & 0.681\\
AIPW-CPM-icdf-mislink & 0.014 & 1.060 & 1.030 & 0.672 & 0.037 & 1.035 & 1.018 & 0.665\\
TMLE-CPM-mislink      & 0.030 & 1.057 & 1.029 & 0.657 & 0.042 & 1.062 & 1.031 & 0.671\\
\midrule
OR-CPM-icdf          & 0.032 & 0.701 & 0.838 & 0.561 & 0.032 & 0.701 & 0.838 & 0.561\\
IPW-icdf             & -0.010 & 1.603 & 1.266 & 0.845 & 5.342 & 2.813 & 5.599 & 5.417\\
IPW-Firpo          & 0.020 & 1.615 & 1.271 & 0.866 & 5.365 & 2.836 & 5.623 & 5.427\\

\bottomrule
\end{tabular}
\end{adjustbox}

\vspace{7pt}

\begin{adjustbox}{max width=\linewidth}
\begin{tabular}{@{}l
  *{4}{S[table-format=2.3,table-text-alignment=center]}
  @{\hspace{3.6em}}
  *{4}{S[table-format=2.3,table-text-alignment=center]}@{}}
\toprule
& \multicolumn{4}{c}{\textbf{Misspecified OR, Correct PS}}
& \multicolumn{4}{c}{\textbf{Misspecified OR, Misspecified PS}}\\
\cmidrule(r){2-5}\cmidrule(l){6-9}
\textbf{Estimators}
& {\textbf{Bias}}
& {\textbf{Var}}
& {\textbf{RMSE}}
& {\textbf{MedAE}}
& {\textbf{Bias}}
& {\textbf{Var}}
& {\textbf{RMSE}}
& {\textbf{MedAE}}\\
\midrule

AIPW                 & 0.025 & 1.633 & 1.278 & 0.870 & 5.365 & 2.836 & 5.623 & 5.427\\
TMLE                 & 0.035 & 1.627 & 1.276 & 0.860 & 5.387 & 2.823 & 5.643 & 5.440\\
TMLE-cqr             & 0.027 & 1.607 & 1.268 & 0.866 & 5.363 & 2.835 & 5.621 & 5.423\\
\midrule
log-AIPW             & 0.020 & 1.628 & 1.276 & 0.870 & 5.365 & 2.837 & 5.624 & 5.428\\
log-TMLE             & 0.036 & 1.654 & 1.286 & 0.839 & 5.369 & 2.856 & 5.629 & 5.456\\
log-TMLE-cqr         & 0.027 & 1.607 & 1.268 & 0.867 & 5.363 & 2.835 & 5.621 & 5.424\\
\midrule
ct-AIPW              & 0.021 & 1.630 & 1.277 & 0.870 & 5.365 & 2.836 & 5.623 & 5.427\\
ct-TMLE              & 0.044 & 1.619 & 1.273 & 0.842 & 5.364 & 2.822 & 5.621 & 5.464\\
ct-TMLE-cqr          & 0.027 & 1.607 & 1.268 & 0.866 & 5.363 & 2.835 & 5.621 & 5.423\\
\midrule
AIPW-CPM             & 0.015 & 1.628 & 1.276 & 0.870 & 5.365 & 2.837 & 5.623 & 5.427\\
AIPW-CPM-icdf        & -0.009 & 1.636 & 1.279 & 0.864 & 5.346 & 2.812 & 5.603 & 5.417\\
TMLE-CPM             & 0.032 & 1.615 & 1.271 & 0.837 & 5.365 & 2.807 & 5.621 & 5.446\\
\midrule
AIPW-CPM-mislink      & 0.015 & 1.628 & 1.276 & 0.871 & 5.365 & 2.837 & 5.623 & 5.427\\
AIPW-CPM-icdf-mislink & -0.009 & 1.637 & 1.280 & 0.863 & 5.344 & 2.811 & 5.601 & 5.417\\
TMLE-CPM-mislink      & 0.031 & 1.613 & 1.271 & 0.849 & 5.365 & 2.821 & 5.621 & 5.423\\
\midrule
OR-CPM-icdf          & 5.529 & 2.243 & 5.728 & 5.569 & 5.529 & 2.243 & 5.728 & 5.569\\
IPW-icdf             & -0.010 & 1.603 & 1.266 & 0.845 & 5.342 & 2.813 & 5.599 & 5.417\\
IPW-Firpo          & 0.020 & 1.615 & 1.271 & 0.866 & 5.365 & 2.836 & 5.623 & 5.427\\

\bottomrule
\end{tabular}
\end{adjustbox}

\end{threeparttable}
\end{table}

\FloatBarrier

\begin{table}[htbp]
\centering
\scriptsize
\setlength{\tabcolsep}{3.6pt}
\renewcommand{\arraystretch}{1.10}

\caption{Simulation performance metrics across nuisance-model specification scenarios. Target parameter is $QTE(0.9)$, true value is 18.69. Each scenario was evaluated using 1,000 simulation replicates. Sample size is 1,000.
\textit{OR}: outcome regression. 
\textit{PS}: propensity score. 
\textit{Var}: empirical variance over 1,000 simulation replicates. 
\textit{RMSE}: root mean squared error. 
\textit{MedAE}: median absolute error.}
\label{table4}

\begin{threeparttable}

\begin{adjustbox}{max width=\linewidth}
\begin{tabular}{@{}l
  *{4}{S[table-format=2.3,table-text-alignment=center]}
  @{\hspace{3.6em}}
  *{4}{S[table-format=2.3,table-text-alignment=center]}@{}}
\toprule
& \multicolumn{4}{c}{\textbf{Correct OR, Correct PS}}
& \multicolumn{4}{c}{\textbf{Correct OR, Misspecified PS}}\\
\cmidrule(r){2-5}\cmidrule(l){6-9}
\textbf{Estimators}
& {\textbf{Bias}}
& {\textbf{Var}}
& {\textbf{RMSE}}
& {\textbf{MedAE}}
& {\textbf{Bias}}
& {\textbf{Var}}
& {\textbf{RMSE}}
& {\textbf{MedAE}}\\
\midrule

AIPW & 0.026 & 4.450 & 2.110 & 1.429 & 2.064 & 5.463 & 3.118 & 2.257\\
TMLE & 0.030 & 4.125 & 2.031 & 1.388 & 1.004 & 4.188 & 2.280 & 1.617\\
TMLE-cqr & 0.045 & 5.128 & 2.265 & 1.540 & 2.264 & 5.242 & 3.220 & 2.417\\
\midrule
log-AIPW & -0.021 & 4.340 & 2.083 & 1.397 & -3.107 & 4.706 & 3.789 & 3.139\\
log-TMLE & 0.002 & 4.577 & 2.139 & 1.487 & 1.325 & 4.547 & 2.510 & 1.792\\
log-TMLE-cqr & 0.067 & 3.912 & 1.979 & 1.389 & 0.348 & 3.728 & 1.962 & 1.360\\
\midrule
ct-AIPW & -0.018 & 3.774 & 1.943 & 1.389 & 0.163 & 3.833 & 1.965 & 1.356\\
ct-TMLE & -0.028 & 3.779 & 1.944 & 1.364 & -0.022 & 3.735 & 1.933 & 1.367\\
ct-TMLE-cqr & -0.030 & 3.734 & 1.933 & 1.328 & -0.022 & 3.686 & 1.920 & 1.311\\
\midrule
AIPW-CPM & -0.016 & 3.777 & 1.943 & 1.379 & -0.015 & 3.567 & 1.889 & 1.334\\
AIPW-CPM-icdf & -0.131 & 3.697 & 1.927 & 1.374 & -0.070 & 3.476 & 1.866 & 1.310\\
TMLE-CPM & -0.037 & 3.796 & 1.949 & 1.401 & -0.035 & 3.711 & 1.927 & 1.383\\
\midrule
AIPW-CPM-mislink & -0.007 & 3.775 & 1.943 & 1.377 & -0.022 & 3.558 & 1.886 & 1.330\\
AIPW-CPM-icdf-mislink & -0.129 & 3.692 & 1.926 & 1.377 & -0.084 & 3.472 & 1.865 & 1.309\\
TMLE-CPM-mislink & -0.023 & 3.784 & 1.945 & 1.381 & -0.009 & 3.744 & 1.935 & 1.361\\
\midrule
OR-CPM-icdf & -0.041 & 2.185 & 1.479 & 0.967 & -0.041 & 2.185 & 1.479 & 0.967\\
IPW-icdf & -0.151 & 6.598 & 2.573 & 1.776 & 7.136 & 8.969 & 7.739 & 7.219\\
IPW-Firpo & 0.019 & 6.800 & 2.608 & 1.731 & 7.303 & 9.050 & 7.898 & 7.357\\

\bottomrule
\end{tabular}
\end{adjustbox}

\vspace{7pt}

\begin{adjustbox}{max width=\linewidth}
\begin{tabular}{@{}l
  *{4}{S[table-format=2.3,table-text-alignment=center]}
  @{\hspace{3.6em}}
  *{4}{S[table-format=2.3,table-text-alignment=center]}@{}}
\toprule
& \multicolumn{4}{c}{\textbf{Misspecified OR, Correct PS}}
& \multicolumn{4}{c}{\textbf{Misspecified OR, Misspecified PS}}\\
\cmidrule(r){2-5}\cmidrule(l){6-9}
\textbf{Estimators}
& {\textbf{Bias}}
& {\textbf{Var}}
& {\textbf{RMSE}}
& {\textbf{MedAE}}
& {\textbf{Bias}}
& {\textbf{Var}}
& {\textbf{RMSE}}
& {\textbf{MedAE}}\\
\midrule

AIPW & 0.020 & 6.828 & 2.613 & 1.750 & 7.303 & 9.051 & 7.898 & 7.357\\
TMLE & 0.114 & 6.379 & 2.528 & 1.762 & 7.170 & 8.915 & 7.766 & 7.163\\
TMLE-cqr & 0.028 & 6.746 & 2.598 & 1.731 & 7.301 & 9.053 & 7.896 & 7.344\\
\midrule
log-AIPW & 0.024 & 6.790 & 2.606 & 1.742 & 7.303 & 9.050 & 7.898 & 7.356\\
log-TMLE & -0.006 & 6.929 & 2.632 & 1.840 & 7.289 & 9.092 & 7.888 & 7.233\\
log-TMLE-cqr & 0.028 & 6.746 & 2.597 & 1.731 & 7.301 & 9.052 & 7.896 & 7.344\\
\midrule
ct-AIPW & 0.020 & 6.814 & 2.610 & 1.750 & 7.303 & 9.051 & 7.898 & 7.357\\
ct-TMLE & 0.065 & 6.758 & 2.600 & 1.748 & 7.284 & 8.993 & 7.877 & 7.330\\
ct-TMLE-cqr & 0.028 & 6.746 & 2.597 & 1.731 & 7.301 & 9.053 & 7.896 & 7.344\\
\midrule
AIPW-CPM & 0.009 & 6.800 & 2.608 & 1.729 & 7.303 & 9.051 & 7.898 & 7.357\\
AIPW-CPM-icdf & -0.117 & 6.855 & 2.621 & 1.785 & 7.157 & 8.960 & 7.758 & 7.229\\
TMLE-CPM & 0.037 & 6.762 & 2.601 & 1.748 & 7.282 & 9.045 & 7.878 & 7.340\\
\midrule
AIPW-CPM-mislink & 0.012 & 6.792 & 2.606 & 1.726 & 7.303 & 9.051 & 7.898 & 7.357\\
AIPW-CPM-icdf-mislink & -0.116 & 6.848 & 2.619 & 1.785 & 7.156 & 8.956 & 7.757 & 7.259\\
TMLE-CPM-mislink & 0.028 & 6.769 & 2.602 & 1.769 & 7.273 & 9.023 & 7.869 & 7.360\\
\midrule
OR-CPM-icdf & 7.156 & 6.182 & 7.576 & 7.320 & 7.156 & 6.182 & 7.576 & 7.320\\
IPW-icdf & -0.151 & 6.598 & 2.573 & 1.776 & 7.136 & 8.969 & 7.739 & 7.219\\
IPW-Firpo & 0.019 & 6.800 & 2.608 & 1.731 & 7.303 & 9.050 & 7.898 & 7.357\\

\bottomrule
\end{tabular}
\end{adjustbox}

\end{threeparttable}
\end{table}


\begin{table}[htbp]
\centering
\scriptsize
\setlength{\tabcolsep}{3.6pt}
\renewcommand{\arraystretch}{1.10}

\caption{Simulation performance metrics across nuisance-model specification scenarios. Target parameter is $QTE(0.1)$, true value is 0.23. Sample size is 500.
\textit{OR}: outcome regression. 
\textit{PS}: propensity score. 
\textit{Var}: empirical variance over 1,000 simulation replicates. 
\textit{RMSE}: root mean squared error. 
\textit{MedAE}: median absolute error.}
\label{table5}

\begin{threeparttable}

\begin{adjustbox}{max width=\linewidth}
\begin{tabular}{@{}l
  *{4}{S[table-format=2.3,table-text-alignment=center]}
  @{\hspace{3.6em}}
  *{4}{S[table-format=2.3,table-text-alignment=center]}@{}}
\toprule
& \multicolumn{4}{c}{\textbf{Correct OR, Correct PS}}
& \multicolumn{4}{c}{\textbf{Correct OR, Misspecified PS}}\\
\cmidrule(r){2-5}\cmidrule(l){6-9}
\textbf{Estimators}
& {\textbf{Bias}}
& {\textbf{Var}}
& {\textbf{RMSE}}
& {\textbf{MedAE}}
& {\textbf{Bias}}
& {\textbf{Var}}
& {\textbf{RMSE}}
& {\textbf{MedAE}}\\
\midrule

AIPW                 & 0.016 & 0.010 & 0.100 & 0.062 & -0.059 & 0.007 & 0.103 & 0.086\\
TMLE                 & 0.016 & 0.011 & 0.104 & 0.065 & 0.069 & 0.013 & 0.132 & 0.070\\
TMLE-cqr             & 0.072 & 0.013 & 0.134 & 0.075 & 0.097 & 0.014 & 0.154 & 0.090\\
\midrule
log-AIPW             & 0.015 & 0.009 & 0.094 & 0.058 & 0.033 & 0.008 & 0.094 & 0.056\\
log-TMLE             & 0.017 & 0.009 & 0.094 & 0.058 & 0.026 & 0.009 & 0.099 & 0.059\\
log-TMLE-cqr         & 0.017 & 0.009 & 0.098 & 0.061 & 0.047 & 0.011 & 0.115 & 0.064\\
\midrule
ct-AIPW              & 0.016 & 0.009 & 0.094 & 0.058 & 0.017 & 0.008 & 0.088 & 0.055\\
ct-TMLE              & 0.016 & 0.008 & 0.093 & 0.058 & 0.016 & 0.008 & 0.093 & 0.057\\
ct-TMLE-cqr          & 0.016 & 0.008 & 0.093 & 0.059 & 0.017 & 0.008 & 0.093 & 0.057\\
\midrule
AIPW-CPM             & 0.016 & 0.008 & 0.093 & 0.058 & 0.017 & 0.008 & 0.089 & 0.055\\
AIPW-CPM-icdf        & 0.022 & 0.009 & 0.096 & 0.058 & 0.020 & 0.008 & 0.090 & 0.055\\
TMLE-CPM             & 0.016 & 0.008 & 0.093 & 0.058 & 0.016 & 0.008 & 0.092 & 0.058\\
\midrule
AIPW-CPM-mislink      & 0.016 & 0.008 & 0.093 & 0.058 & 0.017 & 0.008 & 0.089 & 0.055\\
AIPW-CPM-icdf-mislink & 0.022 & 0.009 & 0.096 & 0.057 & 0.020 & 0.008 & 0.091 & 0.055\\
TMLE-CPM-mislink      & 0.016 & 0.008 & 0.093 & 0.057 & 0.017 & 0.008 & 0.093 & 0.058\\
\midrule
OR-CPM-icdf          & 0.020 & 0.006 & 0.081 & 0.048 & 0.020 & 0.006 & 0.081 & 0.048\\
IPW-icdf             & 0.028 & 0.013 & 0.116 & 0.069 & 0.195 & 0.025 & 0.251 & 0.172\\
IPW-Firpo          & 0.018 & 0.012 & 0.111 & 0.069 & 0.181 & 0.024 & 0.239 & 0.159\\

\bottomrule
\end{tabular}
\end{adjustbox}

\vspace{7pt}

\begin{adjustbox}{max width=\linewidth}
\begin{tabular}{@{}l
  *{4}{S[table-format=2.3,table-text-alignment=center]}
  @{\hspace{3.6em}}
  *{4}{S[table-format=2.3,table-text-alignment=center]}@{}}
\toprule
& \multicolumn{4}{c}{\textbf{Misspecified OR, Correct PS}}
& \multicolumn{4}{c}{\textbf{Misspecified OR, Misspecified PS}}\\
\cmidrule(r){2-5}\cmidrule(l){6-9}
\textbf{Estimators}
& {\textbf{Bias}}
& {\textbf{Var}}
& {\textbf{RMSE}}
& {\textbf{MedAE}}
& {\textbf{Bias}}
& {\textbf{Var}}
& {\textbf{RMSE}}
& {\textbf{MedAE}}\\
\midrule

AIPW                 & 0.018 & 0.012 & 0.111 & 0.069 & 0.181 & 0.024 & 0.239 & 0.159\\
TMLE                 & 0.012 & 0.013 & 0.115 & 0.074 & 0.188 & 0.027 & 0.249 & 0.166\\
TMLE-cqr             & 0.018 & 0.012 & 0.111 & 0.069 & 0.182 & 0.024 & 0.239 & 0.159\\
\midrule
log-AIPW             & 0.018 & 0.012 & 0.111 & 0.069 & 0.181 & 0.024 & 0.239 & 0.159\\
log-TMLE             & 0.021 & 0.012 & 0.110 & 0.068 & 0.181 & 0.024 & 0.238 & 0.158\\
log-TMLE-cqr         & 0.018 & 0.012 & 0.111 & 0.069 & 0.182 & 0.024 & 0.239 & 0.159\\
\midrule
ct-AIPW              & 0.018 & 0.012 & 0.111 & 0.069 & 0.181 & 0.024 & 0.239 & 0.159\\
ct-TMLE              & 0.018 & 0.012 & 0.110 & 0.067 & 0.181 & 0.024 & 0.239 & 0.159\\
ct-TMLE-cqr          & 0.018 & 0.012 & 0.111 & 0.069 & 0.182 & 0.024 & 0.239 & 0.159\\
\midrule
AIPW-CPM             & 0.018 & 0.012 & 0.111 & 0.069 & 0.181 & 0.024 & 0.239 & 0.159\\
AIPW-CPM-icdf        & 0.024 & 0.012 & 0.112 & 0.067 & 0.193 & 0.025 & 0.250 & 0.170\\
TMLE-CPM             & 0.019 & 0.012 & 0.110 & 0.068 & 0.182 & 0.024 & 0.240 & 0.155\\
\midrule
AIPW-CPM-mislink      & 0.018 & 0.012 & 0.111 & 0.069 & 0.181 & 0.024 & 0.239 & 0.159\\
AIPW-CPM-icdf-mislink & 0.025 & 0.012 & 0.112 & 0.067 & 0.193 & 0.025 & 0.250 & 0.170\\
TMLE-CPM-mislink      & 0.019 & 0.012 & 0.111 & 0.069 & 0.182 & 0.024 & 0.239 & 0.158\\
\midrule
OR-CPM-icdf          & 0.195 & 0.017 & 0.235 & 0.179 & 0.195 & 0.017 & 0.235 & 0.179\\
IPW-icdf             & 0.028 & 0.013 & 0.116 & 0.069 & 0.195 & 0.025 & 0.251 & 0.172\\
IPW-Firpo          & 0.018 & 0.012 & 0.111 & 0.069 & 0.181 & 0.024 & 0.239 & 0.159\\

\bottomrule
\end{tabular}
\end{adjustbox}

\end{threeparttable}
\end{table}


\FloatBarrier

\begin{table}[htbp]
\centering
\scriptsize
\setlength{\tabcolsep}{3.6pt}
\renewcommand{\arraystretch}{1.10}

\caption{Simulation performance metrics across nuisance-model specification scenarios. Target parameter is $QTE(0.25)$, true value is 1.62. Each scenario was evaluated using 1,000 simulation replicates. Sample size is 500.
\textit{OR}: outcome regression. 
\textit{PS}: propensity score. 
\textit{Var}: empirical variance over 1,000 simulation replicates. 
\textit{RMSE}: root mean squared error. 
\textit{MedAE}: median absolute error.}
\label{table6}

\begin{threeparttable}

\begin{adjustbox}{max width=\linewidth}
\begin{tabular}{@{}l
  *{4}{S[table-format=2.3,table-text-alignment=center]}
  @{\hspace{3.6em}}
  *{4}{S[table-format=2.3,table-text-alignment=center]}@{}}
\toprule
& \multicolumn{4}{c}{\textbf{Correct OR, Correct PS}}
& \multicolumn{4}{c}{\textbf{Correct OR, Misspecified PS}}\\
\cmidrule(r){2-5}\cmidrule(l){6-9}
\textbf{Estimators}
& {\textbf{Bias}}
& {\textbf{Var}}
& {\textbf{RMSE}}
& {\textbf{MedAE}}
& {\textbf{Bias}}
& {\textbf{Var}}
& {\textbf{RMSE}}
& {\textbf{MedAE}}\\
\midrule

AIPW                 & 0.020 & 0.109 & 0.331 & 0.219 & 0.118 & 0.114 & 0.357 & 0.213\\
TMLE                 & 0.017 & 0.117 & 0.342 & 0.219 & 0.176 & 0.112 & 0.379 & 0.222\\
TMLE-cqr             & 0.022 & 0.105 & 0.324 & 0.207 & 0.030 & 0.106 & 0.327 & 0.212\\
\midrule
log-AIPW             & 0.016 & 0.114 & 0.338 & 0.220 & -0.046 & 0.103 & 0.325 & 0.220\\
log-TMLE             & 0.017 & 0.110 & 0.332 & 0.219 & 0.102 & 0.118 & 0.359 & 0.217\\
log-TMLE-cqr         & 0.016 & 0.110 & 0.332 & 0.222 & 0.122 & 0.115 & 0.361 & 0.220\\
\midrule
ct-AIPW              & 0.024 & 0.106 & 0.326 & 0.211 & 0.026 & 0.097 & 0.312 & 0.206\\
ct-TMLE              & 0.023 & 0.104 & 0.323 & 0.211 & 0.024 & 0.103 & 0.322 & 0.213\\
ct-TMLE-cqr          & 0.023 & 0.105 & 0.325 & 0.208 & 0.023 & 0.104 & 0.324 & 0.210\\
\midrule
AIPW-CPM             & 0.023 & 0.107 & 0.327 & 0.212 & 0.025 & 0.100 & 0.317 & 0.208\\
AIPW-CPM-icdf        & 0.035 & 0.106 & 0.328 & 0.213 & 0.031 & 0.100 & 0.318 & 0.205\\
TMLE-CPM             & 0.022 & 0.106 & 0.326 & 0.209 & 0.021 & 0.104 & 0.324 & 0.215\\
\midrule
AIPW-CPM-mislink      & 0.024 & 0.106 & 0.327 & 0.211 & 0.027 & 0.100 & 0.317 & 0.209\\
AIPW-CPM-icdf-mislink & 0.035 & 0.107 & 0.328 & 0.212 & 0.033 & 0.101 & 0.319 & 0.204\\
TMLE-CPM-mislink      & 0.022 & 0.106 & 0.326 & 0.214 & 0.023 & 0.104 & 0.324 & 0.214\\
\midrule
OR-CPM-icdf          & 0.046 & 0.081 & 0.287 & 0.190 & 0.046 & 0.081 & 0.287 & 0.190\\
IPW-icdf             & 0.037 & 0.136 & 0.370 & 0.235 & 0.888 & 0.259 & 1.023 & 0.843\\
IPW-Firpo          & 0.017 & 0.135 & 0.367 & 0.228 & 0.865 & 0.259 & 1.004 & 0.824\\

\bottomrule
\end{tabular}
\end{adjustbox}

\vspace{7pt}

\begin{adjustbox}{max width=\linewidth}
\begin{tabular}{@{}l
  *{4}{S[table-format=2.3,table-text-alignment=center]}
  @{\hspace{3.6em}}
  *{4}{S[table-format=2.3,table-text-alignment=center]}@{}}
\toprule
& \multicolumn{4}{c}{\textbf{Misspecified OR, Correct PS}}
& \multicolumn{4}{c}{\textbf{Misspecified OR, Misspecified PS}}\\
\cmidrule(r){2-5}\cmidrule(l){6-9}
\textbf{Estimators}
& {\textbf{Bias}}
& {\textbf{Var}}
& {\textbf{RMSE}}
& {\textbf{MedAE}}
& {\textbf{Bias}}
& {\textbf{Var}}
& {\textbf{RMSE}}
& {\textbf{MedAE}}\\
\midrule

AIPW                 & 0.017 & 0.134 & 0.366 & 0.228 & 0.865 & 0.259 & 1.004 & 0.824\\
TMLE                 & 0.030 & 0.130 & 0.362 & 0.231 & 0.861 & 0.257 & 0.999 & 0.813\\
TMLE-cqr             & 0.019 & 0.133 & 0.366 & 0.228 & 0.866 & 0.259 & 1.004 & 0.824\\
\midrule
log-AIPW             & 0.018 & 0.133 & 0.365 & 0.228 & 0.865 & 0.259 & 1.004 & 0.824\\
log-TMLE             & 0.015 & 0.133 & 0.366 & 0.233 & 0.858 & 0.256 & 0.996 & 0.813\\
log-TMLE-cqr         & 0.019 & 0.133 & 0.366 & 0.228 & 0.866 & 0.259 & 1.004 & 0.824\\
\midrule
ct-AIPW              & 0.017 & 0.134 & 0.366 & 0.228 & 0.865 & 0.259 & 1.004 & 0.824\\
ct-TMLE              & 0.022 & 0.134 & 0.367 & 0.236 & 0.866 & 0.256 & 1.003 & 0.826\\
ct-TMLE-cqr          & 0.019 & 0.133 & 0.366 & 0.228 & 0.866 & 0.259 & 1.004 & 0.824\\
\midrule
AIPW-CPM             & 0.016 & 0.134 & 0.367 & 0.229 & 0.865 & 0.259 & 1.004 & 0.824\\
AIPW-CPM-icdf        & 0.030 & 0.133 & 0.365 & 0.229 & 0.887 & 0.259 & 1.023 & 0.843\\
TMLE-CPM             & 0.022 & 0.132 & 0.364 & 0.232 & 0.867 & 0.257 & 1.004 & 0.830\\
\midrule
AIPW-CPM-mislink      & 0.016 & 0.134 & 0.367 & 0.229 & 0.865 & 0.259 & 1.004 & 0.824\\
AIPW-CPM-icdf-mislink & 0.030 & 0.132 & 0.365 & 0.228 & 0.887 & 0.259 & 1.023 & 0.841\\
TMLE-CPM-mislink      & 0.022 & 0.134 & 0.367 & 0.233 & 0.867 & 0.255 & 1.003 & 0.826\\
\midrule
OR-CPM-icdf          & 0.965 & 0.214 & 1.070 & 0.932 & 0.965 & 0.214 & 1.070 & 0.932\\
IPW-icdf             & 0.037 & 0.136 & 0.370 & 0.235 & 0.888 & 0.259 & 1.023 & 0.843\\
IPW-Firpo          & 0.017 & 0.135 & 0.367 & 0.228 & 0.865 & 0.259 & 1.004 & 0.824\\

\bottomrule
\end{tabular}
\end{adjustbox}

\end{threeparttable}
\end{table}
\FloatBarrier
\begin{table}[p]
\centering
\scriptsize
\setlength{\tabcolsep}{3.6pt}
\renewcommand{\arraystretch}{1.10}

\caption{Simulation performance metrics across nuisance-model specification scenarios. Target parameter is $QTE(0.5)$, true value is 5.9. Each scenario was evaluated using 1,000 simulation replicates. Sample size is 500.
\textit{OR}: outcome regression. 
\textit{PS}: propensity score. 
\textit{Var}: empirical variance over 1,000 simulation replicates. 
\textit{RMSE}: root mean squared error. 
\textit{MedAE}: median absolute error.}
\label{table7}

\begin{threeparttable}

\begin{adjustbox}{max width=\linewidth}
\begin{tabular}{@{}l
  *{4}{S[table-format=2.3,table-text-alignment=center]}
  @{\hspace{3.6em}}
  *{4}{S[table-format=2.3,table-text-alignment=center]}@{}}
\toprule
& \multicolumn{4}{c}{\textbf{Correct OR, Correct PS}}
& \multicolumn{4}{c}{\textbf{Correct OR, Misspecified PS}}\\
\cmidrule(r){2-5}\cmidrule(l){6-9}
\textbf{Estimators}
& {\textbf{Bias}}
& {\textbf{Var}}
& {\textbf{RMSE}}
& {\textbf{MedAE}}
& {\textbf{Bias}}
& {\textbf{Var}}
& {\textbf{RMSE}}
& {\textbf{MedAE}}\\
\midrule

AIPW                 & 0.039 & 0.641 & 0.802 & 0.519 & 0.548 & 0.648 & 0.974 & 0.667\\
TMLE                 & 0.029 & 0.597 & 0.773 & 0.517 & 0.494 & 0.621 & 0.930 & 0.636\\
TMLE-cqr             & 0.024 & 0.572 & 0.757 & 0.505 & 0.377 & 0.590 & 0.856 & 0.572\\
\midrule
log-AIPW             & 0.018 & 0.584 & 0.765 & 0.521 & 0.294 & 0.633 & 0.848 & 0.551\\
log-TMLE             & 0.022 & 0.571 & 0.756 & 0.498 & 0.287 & 0.597 & 0.824 & 0.547\\
log-TMLE-cqr         & 0.022 & 0.548 & 0.740 & 0.481 & 0.167 & 0.561 & 0.768 & 0.499\\
\midrule
ct-AIPW              & 0.018 & 0.537 & 0.733 & 0.475 & 0.022 & 0.542 & 0.736 & 0.474\\
ct-TMLE              & 0.015 & 0.540 & 0.735 & 0.479 & 0.022 & 0.547 & 0.740 & 0.466\\
ct-TMLE-cqr          & 0.016 & 0.541 & 0.736 & 0.480 & 0.022 & 0.542 & 0.736 & 0.467\\
\midrule
AIPW-CPM             & 0.018 & 0.541 & 0.736 & 0.475 & 0.023 & 0.539 & 0.735 & 0.476\\
AIPW-CPM-icdf        & 0.023 & 0.541 & 0.736 & 0.474 & 0.028 & 0.538 & 0.734 & 0.476\\
TMLE-CPM             & 0.018 & 0.540 & 0.735 & 0.475 & 0.014 & 0.541 & 0.735 & 0.474\\
\midrule
AIPW-CPM-mislink      & 0.020 & 0.542 & 0.736 & 0.477 & 0.028 & 0.534 & 0.732 & 0.474\\
AIPW-CPM-icdf-mislink & 0.023 & 0.540 & 0.735 & 0.471 & 0.034 & 0.536 & 0.733 & 0.473\\
TMLE-CPM-mislink      & 0.018 & 0.543 & 0.737 & 0.471 & 0.020 & 0.545 & 0.739 & 0.474\\
\midrule
OR-CPM-icdf          & 0.048 & 0.403 & 0.637 & 0.426 & 0.048 & 0.403 & 0.637 & 0.426\\
IPW-icdf             & 0.035 & 0.687 & 0.829 & 0.567 & 2.714 & 1.475 & 2.973 & 2.647\\
IPW-Firpo          & 0.027 & 0.670 & 0.819 & 0.550 & 2.696 & 1.479 & 2.958 & 2.634\\

\bottomrule
\end{tabular}
\end{adjustbox}

\vspace{7pt}

\begin{adjustbox}{max width=\linewidth}
\begin{tabular}{@{}l
  *{4}{S[table-format=2.3,table-text-alignment=center]}
  @{\hspace{3.6em}}
  *{4}{S[table-format=2.3,table-text-alignment=center]}@{}}
\toprule
& \multicolumn{4}{c}{\textbf{Misspecified OR, Correct PS}}
& \multicolumn{4}{c}{\textbf{Misspecified OR, Misspecified PS}}\\
\cmidrule(r){2-5}\cmidrule(l){6-9}
\textbf{Estimators}
& {\textbf{Bias}}
& {\textbf{Var}}
& {\textbf{RMSE}}
& {\textbf{MedAE}}
& {\textbf{Bias}}
& {\textbf{Var}}
& {\textbf{RMSE}}
& {\textbf{MedAE}}\\
\midrule

AIPW                 & 0.033 & 0.680 & 0.825 & 0.551 & 2.696 & 1.479 & 2.958 & 2.634\\
TMLE                 & 0.044 & 0.672 & 0.821 & 0.540 & 2.707 & 1.481 & 2.968 & 2.640\\
TMLE-cqr             & 0.038 & 0.673 & 0.821 & 0.550 & 2.696 & 1.477 & 2.958 & 2.634\\
\midrule
log-AIPW             & 0.034 & 0.677 & 0.824 & 0.551 & 2.696 & 1.479 & 2.958 & 2.634\\
log-TMLE             & 0.035 & 0.683 & 0.827 & 0.567 & 2.699 & 1.482 & 2.961 & 2.626\\
log-TMLE-cqr         & 0.038 & 0.673 & 0.821 & 0.550 & 2.696 & 1.477 & 2.957 & 2.634\\
\midrule
ct-AIPW              & 0.036 & 0.680 & 0.825 & 0.549 & 2.696 & 1.479 & 2.958 & 2.634\\
ct-TMLE              & 0.043 & 0.676 & 0.824 & 0.544 & 2.691 & 1.460 & 2.950 & 2.646\\
ct-TMLE-cqr          & 0.038 & 0.673 & 0.821 & 0.549 & 2.696 & 1.477 & 2.958 & 2.634\\
\midrule
AIPW-CPM             & 0.030 & 0.680 & 0.825 & 0.551 & 2.696 & 1.479 & 2.958 & 2.634\\
AIPW-CPM-icdf        & 0.032 & 0.677 & 0.823 & 0.564 & 2.716 & 1.474 & 2.975 & 2.647\\
TMLE-CPM             & 0.038 & 0.671 & 0.820 & 0.547 & 2.694 & 1.463 & 2.953 & 2.639\\
\midrule
AIPW-CPM-mislink      & 0.031 & 0.681 & 0.826 & 0.551 & 2.696 & 1.479 & 2.958 & 2.634\\
AIPW-CPM-icdf-mislink & 0.032 & 0.676 & 0.823 & 0.564 & 2.715 & 1.475 & 2.975 & 2.643\\
TMLE-CPM-mislink      & 0.040 & 0.677 & 0.824 & 0.554 & 2.693 & 1.470 & 2.953 & 2.628\\
\midrule
OR-CPM-icdf          & 2.902 & 1.270 & 3.114 & 2.828 & 2.902 & 1.270 & 3.114 & 2.828\\
IPW-icdf             & 0.035 & 0.687 & 0.829 & 0.567 & 2.714 & 1.475 & 2.973 & 2.647\\
IPW-Firpo          & 0.027 & 0.670 & 0.819 & 0.550 & 2.696 & 1.479 & 2.958 & 2.634\\

\bottomrule
\end{tabular}
\end{adjustbox}

\end{threeparttable}
\end{table}

\FloatBarrier
\begin{table}[p]
\centering
\scriptsize
\setlength{\tabcolsep}{3.6pt}
\renewcommand{\arraystretch}{1.10}

\caption{Simulation performance metrics across nuisance-model specification scenarios. Target parameter is $QTE(0.75)$, true value is 12.34. Each scenario was evaluated using 1,000 simulation replicates. Sample size is 500.
\textit{OR}: outcome regression. 
\textit{PS}: propensity score. 
\textit{Var}: empirical variance over 1,000 simulation replicates. 
\textit{RMSE}: root mean squared error. 
\textit{MedAE}: median absolute error.}
\label{table8}

\begin{threeparttable}

\begin{adjustbox}{max width=\linewidth}
\begin{tabular}{@{}l
  *{4}{S[table-format=2.3,table-text-alignment=center]}
  @{\hspace{3.6em}}
  *{4}{S[table-format=2.3,table-text-alignment=center]}@{}}
\toprule
& \multicolumn{4}{c}{\textbf{Correct OR, Correct PS}}
& \multicolumn{4}{c}{\textbf{Correct OR, Misspecified PS}}\\
\cmidrule(r){2-5}\cmidrule(l){6-9}
\textbf{Estimators}
& {\textbf{Bias}}
& {\textbf{Var}}
& {\textbf{RMSE}}
& {\textbf{MedAE}}
& {\textbf{Bias}}
& {\textbf{Var}}
& {\textbf{RMSE}}
& {\textbf{MedAE}}\\
\midrule

AIPW                 & -0.023 & 2.336 & 1.529 & 1.040 & -0.239 & 2.364 & 1.556 & 1.078\\
TMLE                 & -0.003 & 2.212 & 1.487 & 1.013 & 1.049 & 2.277 & 1.838 & 1.244\\
TMLE-cqr             & -0.007 & 2.299 & 1.516 & 1.035 & 1.458 & 2.490 & 2.148 & 1.521\\
\midrule
log-AIPW             & -0.002 & 1.976 & 1.406 & 0.953 & 0.408 & 2.059 & 1.492 & 0.989\\
log-TMLE             & -0.004 & 1.974 & 1.405 & 0.975 & 0.390 & 2.055 & 1.486 & 1.033\\
log-TMLE-cqr         & -0.020 & 1.918 & 1.385 & 0.933 & 0.014 & 1.922 & 1.387 & 0.943\\
\midrule
ct-AIPW              & -0.013 & 1.939 & 1.392 & 0.923 & 0.048 & 2.009 & 1.418 & 0.957\\
ct-TMLE              & -0.013 & 1.918 & 1.385 & 0.942 & -0.005 & 1.940 & 1.393 & 0.947\\
ct-TMLE-cqr          & -0.013 & 1.907 & 1.381 & 0.935 & -0.006 & 1.928 & 1.389 & 0.938\\
\midrule
AIPW-CPM             & -0.011 & 1.957 & 1.399 & 0.946 & -0.006 & 1.898 & 1.378 & 0.956\\
AIPW-CPM-icdf        & -0.056 & 1.912 & 1.384 & 0.899 & -0.021 & 1.880 & 1.371 & 0.895\\
TMLE-CPM             & -0.010 & 1.944 & 1.394 & 0.904 & -0.013 & 1.939 & 1.393 & 0.936\\
\midrule
AIPW-CPM-mislink      & -0.004 & 1.971 & 1.404 & 0.942 & -0.004 & 1.907 & 1.381 & 0.948\\
AIPW-CPM-icdf-mislink & -0.054 & 1.925 & 1.389 & 0.900 & -0.013 & 1.889 & 1.375 & 0.903\\
TMLE-CPM-mislink      & -0.023 & 1.956 & 1.399 & 0.910 & -0.003 & 1.933 & 1.390 & 0.946\\
\midrule
OR-CPM-icdf          & 0.032 & 1.362 & 1.167 & 0.795 & 0.032 & 1.362 & 1.167 & 0.795\\
IPW-icdf             & -0.080 & 2.889 & 1.702 & 1.174 & 5.271 & 5.487 & 5.768 & 5.203\\
IPW-Firpo          & -0.005 & 2.972 & 1.724 & 1.169 & 5.324 & 5.564 & 5.823 & 5.207\\

\bottomrule
\end{tabular}
\end{adjustbox}

\vspace{7pt}

\begin{adjustbox}{max width=\linewidth}
\begin{tabular}{@{}l
  *{4}{S[table-format=2.3,table-text-alignment=center]}
  @{\hspace{3.6em}}
  *{4}{S[table-format=2.3,table-text-alignment=center]}@{}}
\toprule
& \multicolumn{4}{c}{\textbf{Misspecified OR, Correct PS}}
& \multicolumn{4}{c}{\textbf{Misspecified OR, Misspecified PS}}\\
\cmidrule(r){2-5}\cmidrule(l){6-9}
\textbf{Estimators}
& {\textbf{Bias}}
& {\textbf{Var}}
& {\textbf{RMSE}}
& {\textbf{MedAE}}
& {\textbf{Bias}}
& {\textbf{Var}}
& {\textbf{RMSE}}
& {\textbf{MedAE}}\\
\midrule

AIPW                 & 0.003 & 2.974 & 1.724 & 1.166 & 5.324 & 5.563 & 5.823 & 5.207\\
TMLE                 & 0.019 & 2.981 & 1.727 & 1.175 & 5.348 & 5.547 & 5.844 & 5.288\\
TMLE-cqr             & 0.017 & 2.945 & 1.716 & 1.157 & 5.324 & 5.562 & 5.823 & 5.207\\
\midrule
log-AIPW             & 0.008 & 2.996 & 1.731 & 1.161 & 5.324 & 5.564 & 5.823 & 5.207\\
log-TMLE             & 0.021 & 2.969 & 1.723 & 1.174 & 5.304 & 5.542 & 5.803 & 5.259\\
log-TMLE-cqr         & 0.017 & 2.945 & 1.716 & 1.157 & 5.324 & 5.561 & 5.823 & 5.208\\
\midrule
ct-AIPW              & 0.007 & 3.003 & 1.733 & 1.157 & 5.324 & 5.563 & 5.823 & 5.207\\
ct-TMLE              & 0.038 & 2.958 & 1.720 & 1.167 & 5.325 & 5.554 & 5.823 & 5.237\\
ct-TMLE-cqr          & 0.017 & 2.945 & 1.716 & 1.157 & 5.324 & 5.561 & 5.823 & 5.207\\
\midrule
AIPW-CPM             & -0.004 & 3.002 & 1.733 & 1.157 & 5.324 & 5.563 & 5.823 & 5.207\\
AIPW-CPM-icdf        & -0.056 & 3.018 & 1.738 & 1.169 & 5.284 & 5.497 & 5.780 & 5.211\\
TMLE-CPM             & 0.022 & 2.912 & 1.707 & 1.145 & 5.320 & 5.479 & 5.812 & 5.237\\
\midrule
AIPW-CPM-mislink      & -0.003 & 2.992 & 1.730 & 1.167 & 5.324 & 5.563 & 5.823 & 5.207\\
AIPW-CPM-icdf-mislink & -0.056 & 3.016 & 1.738 & 1.165 & 5.285 & 5.498 & 5.782 & 5.219\\
TMLE-CPM-mislink      & 0.019 & 2.911 & 1.706 & 1.181 & 5.310 & 5.496 & 5.804 & 5.220\\
\midrule
OR-CPM-icdf          & 5.503 & 4.428 & 5.892 & 5.402 & 5.503 & 4.428 & 5.892 & 5.402\\
IPW-icdf             & -0.080 & 2.889 & 1.702 & 1.174 & 5.271 & 5.487 & 5.768 & 5.203\\
IPW-Firpo          & -0.005 & 2.972 & 1.724 & 1.169 & 5.324 & 5.564 & 5.823 & 5.207\\

\bottomrule
\end{tabular}
\end{adjustbox}

\end{threeparttable}
\end{table}

\FloatBarrier
\begin{table}[p]
\centering
\scriptsize
\setlength{\tabcolsep}{3.6pt}
\renewcommand{\arraystretch}{1.10}

\caption{Simulation performance metrics across nuisance-model specification scenarios. Target parameter is $QTE(0.9)$, true value is 18.69. Each scenario was evaluated using 1,000 simulation replicates. Sample size is 500.
\textit{OR}: outcome regression. 
\textit{PS}: propensity score. 
\textit{Var}: empirical variance over 1,000 simulation replicates. 
\textit{RMSE}: root mean squared error. 
\textit{MedAE}: median absolute error.}
\label{table9}

\begin{threeparttable}

\begin{tabular}{
@{}
p{3.8cm}
*{4}{S[
table-format=2.3,
table-number-alignment=center]}
@{\hspace{2.5em}}
*{4}{S[
table-format=2.3,
table-number-alignment=center]}
@{}
}
\toprule
& \multicolumn{4}{c}{\textbf{Correct OR, Correct PS}}
& \multicolumn{4}{c}{\textbf{Correct OR, Misspecified PS}}\\
\cmidrule(r){2-5}\cmidrule(l){6-9}
\textbf{Estimators}
& {\textbf{Bias}}
& {\textbf{Var}}
& {\textbf{RMSE}}
& {\textbf{MedAE}}
& {\textbf{Bias}}
& {\textbf{Var}}
& {\textbf{RMSE}}
& {\textbf{MedAE}}\\
\midrule

AIPW                 & -0.040 & 8.176 & 2.860 & 2.058 & 1.989 & 9.616 & 3.684 & 2.533\\
TMLE                 & -0.010 & 7.469 & 2.733 & 1.975 & 0.944 & 7.532 & 2.902 & 2.068\\
TMLE-cqr             & -0.065 & 9.235 & 3.040 & 2.220 & 2.112 & 9.313 & 3.711 & 2.628\\
\midrule
log-AIPW             & -0.077 & 8.103 & 2.848 & 1.996 & -3.169 & 9.874 & 4.463 & 3.314\\
log-TMLE             & -0.053 & 8.161 & 2.857 & 2.085 & 1.201 & 7.963 & 3.067 & 2.223\\
log-TMLE-cqr         & 0.148  & 7.486 & 2.740 & 2.011 & 0.404 & 7.343 & 2.740 & 2.008\\
\midrule
ct-AIPW              & 0.025  & 6.834 & 2.614 & 1.902 & 0.199 & 7.053 & 2.663 & 1.956\\
ct-TMLE              & 0.008  & 6.787 & 2.605 & 1.921 & 0.020 & 6.761 & 2.600 & 1.909\\
ct-TMLE-cqr          & 0.005  & 6.846 & 2.616 & 1.863 & 0.034 & 6.662 & 2.581 & 1.875\\
\midrule
AIPW-CPM             & 0.010  & 6.970 & 2.640 & 1.918 & -0.012 & 6.445 & 2.539 & 1.865\\
AIPW-CPM-icdf        & -0.214 & 6.615 & 2.581 & 1.901 & -0.141 & 6.195 & 2.493 & 1.802\\
TMLE-CPM             & -0.004 & 6.807 & 2.609 & 1.891 & 0.013 & 6.808 & 2.609 & 1.917\\
\midrule
AIPW-CPM-mislink      & 0.010  & 6.873 & 2.622 & 1.919 & -0.022 & 6.391 & 2.528 & 1.836\\
AIPW-CPM-icdf-mislink & -0.215 & 6.561 & 2.570 & 1.894 & -0.155 & 6.155 & 2.486 & 1.774\\
TMLE-CPM-mislink      & 0.017  & 6.706 & 2.590 & 1.865 & 0.030 & 6.682 & 2.585 & 1.883\\
\midrule
OR-CPM-icdf          & 0.062 & 4.100 & 2.026 & 1.483 & 0.062 & 4.100 & 2.026 & 1.483\\
IPW-icdf             & -0.442 & 12.116 & 3.509 & 2.489 & 6.840 & 15.066 & 7.865 & 6.918\\
IPW-Firpo          & -0.122 & 12.372 & 3.520 & 2.546 & 7.157 & 15.309 & 8.157 & 7.296\\

\bottomrule
\end{tabular}

\vspace{7pt}

\begin{tabular}{
@{}
p{3.8cm}
*{4}{S[
table-format=2.3,
table-number-alignment=center]}
@{\hspace{2.5em}}
*{4}{S[
table-format=2.3,
table-number-alignment=center]}
@{}
}
\toprule
& \multicolumn{4}{c}{\textbf{Misspecified OR, Correct PS}}
& \multicolumn{4}{c}{\textbf{Misspecified OR, Misspecified PS}}\\
\cmidrule(r){2-5}\cmidrule(l){6-9}
\textbf{Estimators}
& {\textbf{Bias}}
& {\textbf{Var}}
& {\textbf{RMSE}}
& {\textbf{MedAE}}
& {\textbf{Bias}}
& {\textbf{Var}}
& {\textbf{RMSE}}
& {\textbf{MedAE}}\\
\midrule

AIPW                 & -0.100 & 12.527 & 3.541 & 2.562 & 7.157 & 15.308 & 8.156 & 7.296\\
TMLE                 & 0.046  & 11.539 & 3.397 & 2.452 & 7.023 & 14.878 & 8.012 & 7.005\\
TMLE-cqr             & -0.097 & 12.287 & 3.507 & 2.578 & 7.157 & 15.309 & 8.157 & 7.296\\
\midrule
log-AIPW             & -0.103 & 12.302 & 3.509 & 2.548 & 7.157 & 15.308 & 8.156 & 7.296\\
log-TMLE             & -0.103 & 12.757 & 3.573 & 2.583 & 7.151 & 15.424 & 8.158 & 7.195\\
log-TMLE-cqr         & -0.096 & 12.287 & 3.507 & 2.579 & 7.157 & 15.309 & 8.157 & 7.295\\
\midrule
ct-AIPW              & -0.107 & 12.517 & 3.540 & 2.578 & 7.157 & 15.308 & 8.156 & 7.296\\
ct-TMLE              & -0.029 & 12.100 & 3.479 & 2.552 & 7.178 & 15.040 & 8.159 & 7.283\\
ct-TMLE-cqr          & -0.097 & 12.287 & 3.507 & 2.578 & 7.157 & 15.309 & 8.157 & 7.296\\
\midrule
AIPW-CPM             & -0.124 & 12.444 & 3.530 & 2.578 & 7.157 & 15.308 & 8.156 & 7.296\\
AIPW-CPM-icdf        & -0.350 & 12.678 & 3.578 & 2.605 & 6.897 & 14.965 & 7.907 & 6.984\\
TMLE-CPM             & -0.068 & 12.110 & 3.481 & 2.496 & 7.145 & 15.178 & 8.138 & 7.290\\
\midrule
AIPW-CPM-mislink      & -0.123 & 12.487 & 3.536 & 2.575 & 7.157 & 15.308 & 8.156 & 7.296\\
AIPW-CPM-icdf-mislink & -0.353 & 12.667 & 3.576 & 2.603 & 6.893 & 15.041 & 7.909 & 6.986\\
TMLE-CPM-mislink      & -0.080 & 12.072 & 3.475 & 2.535 & 7.105 & 15.282 & 8.110 & 7.167\\
\midrule
OR-CPM-icdf          & 7.148 & 11.121 & 7.888 & 7.040 & 7.148 & 11.121 & 7.888 & 7.040\\
IPW-icdf             & -0.442 & 12.116 & 3.509 & 2.489 & 6.840 & 15.066 & 7.865 & 6.918\\
IPW-Firpo          & -0.122 & 12.372 & 3.520 & 2.546 & 7.157 & 15.309 & 8.157 & 7.296\\

\bottomrule
\end{tabular}

\end{threeparttable}
\end{table}

\FloatBarrier
\subsection{Simulation Results of Sandwich Variance Estimation}

\subsubsection{QTE}
\FloatBarrier

As shown in Table~\ref{tab:aipw_cpm_sandwich_extra}, Sandwich-I generally produced variance estimates closer to the empirical Monte Carlo standard deviations across most simulation settings, whereas Sandwich-S exhibited greater variability, particularly in scenarios involving extreme quantiles. Although Sandwich-S occasionally achieved coverage probabilities closer to the nominal 95\% level, its performance was highly sensitive to the smoothing approximation used for the derivative of the indicator function. Thus, we suggest that Sandwich-I may provide a more reliable approach for practical inference.

\begin{table*}[p]
\centering
\caption{Sandwich Variance Estimation for the AIPW-CPM estimator under different nuisance-model specifications.}
\label{tab:aipw_cpm_sandwich_extra}

\begin{threeparttable}
\small
\setlength{\tabcolsep}{5pt}
\renewcommand{\arraystretch}{1.08}

\begin{tabular}{l
                S[table-format=2.2]
                S[table-format=1.3]
                S[table-format=1.3]
                S[table-format=1.3]
                S[table-format=2.1]
                S[table-format=1.3]
                S[table-format=3.1]}
\toprule
& & & &
\multicolumn{2}{c}{Sandwich-I} &
\multicolumn{2}{c}{Sandwich-S} \\
\cmidrule(lr){5-6} \cmidrule(lr){7-8}
Target & {Truth} & {Avg. Bias} & {Empirical SD}
& {SE} & {Cov., \%} & {SE} & {Cov., \%} \\
\midrule

\multicolumn{8}{c}{Correct OR, Correct PS} \\
QTE(0.10) & 0.23 & 0.010 & 0.063 & 0.042 & 81.9 & 1.021 & 100.0 \\
QTE(0.25) & 1.62 & 0.010 & 0.233 & 0.223 & 93.1 & 0.735 & 100.0 \\
QTE(0.50) & 5.90 & 0.020 & 0.519 & 0.502 & 91.9 & 0.497 & 92.9 \\
QTE(0.75) & 12.34 & 0.030 & 1.032 & 0.973 & 90.6 & 0.906 & 90.3 \\
QTE(0.90) & 18.69 & -0.016 & 1.944 & 1.700 & 86.3 & 1.694 & 90.1 \\
\midrule

\multicolumn{8}{c}{Correct OR, Misspecified PS} \\
QTE(0.10) & 0.23 & 0.010 & 0.060 & 0.040 & 78.1 & 0.351 & 100.0\\
QTE(0.25) & 1.62 & 0.011 & 0.226 & 0.210 & 91.5 & 1.908 & 100.0 \\
QTE(0.50) & 5.90 & 0.030 & 0.519 & 0.500 & 90.9 & 0.504 & 93.6 \\
QTE(0.75) & 12.34 & 0.037 & 1.019 & 1.000 & 90.8 & 1.000 & 94.1 \\
QTE(0.90) & 18.69 & -0.013 & 1.890 & 1.770 & 88.1 & 2.044 & 95.8 \\
\midrule

\multicolumn{8}{c}{Misspecified OR, Correct PS} \\
QTE(0.10) & 0.23 & 0.010 & 0.074 & 0.052 & 84.1 & 0.474 & 100.0 \\
QTE(0.25) & 1.62 & 0.010 & 0.260 & 0.248 & 92.1 & 0.538 & 100.0 \\
QTE(0.50) & 5.90 & 0.002 & 0.597 & 0.560 & 91.3 & 0.577 & 94.6 \\
QTE(0.75) & 12.34 & 0.015 & 1.276 & 1.210 & 91.0 & 1.087 & 90.0 \\
QTE(0.90) & 18.69 & 0.010 & 2.610 & 2.240 & 84.1 & 2.227 & 91.1 \\
\midrule

\multicolumn{8}{c}{Misspecified OR, Misspecified PS} \\
QTE(0.10) & 0.23 & 0.176 & 0.109 & 0.080 & 42.9 & 0.547 & 100.0 \\
QTE(0.25) & 1.62 & 0.855 & 0.361 & 0.350 & 32.2 & 0.731 & 94.1 \\
QTE(0.50) & 5.90 & 2.681 & 0.883 & 0.830 & 10.9 & 0.910 & 12.8 \\
QTE(0.75) & 12.34 & 5.365 & 1.684 & 1.590 & 11.8 & 1.504 & 6.1 \\
QTE(0.90) & 18.69 & 7.300 & 3.008 & 2.466 & 23.8 & 2.507 & 21.4 \\
\bottomrule
\end{tabular}

\begin{tablenotes}
\footnotesize
\item Note: OR denotes the outcome regression model and PS denotes the propensity score model. 
Empirical SD denotes the empirical Monte Carlo standard deviation. 
Cov. denotes the empirical coverage probability of the nominal 95\% confidence interval. Sandwich-I refers to the estimator obtained by interchanging differentiation and expectation, whereas Sandwich-S is based on direct differentiation
with a smoothing approximation. For each scenario, results were based on 1{,}000 Monte Carlo replicates with sample size $n=1000$.
\end{tablenotes}

\end{threeparttable}
\end{table*}

\FloatBarrier
\subsubsection{PTE}
Shown in Table~\ref{tab:pte_dr_variance} are the point estimation and variance estimation results for the doubly robust estimator of $\mathrm{PTE}(7.95)$, where the threshold value was set to $y = 7.95$. The simulation setup followed the same data generating mechanism, nuisance-model specification scenarios, and definitions of correct and misspecified models as those described for the QTE estimators in the main text. For each scenario, results were based on 1{,}000 Monte Carlo replicates with sample size $n=1000$. We compared the empirical Monte Carlo variance with variance estimates obtained from the nonparametric bootstrap, EIF--based, and sandwich approaches.

The results show patterns similar to those observed for the QTE estimators in the main text, as well as findings reported in \citealt{shook2025double}. In particular, the EIF-based variance estimator was sensitive to nuisance-model misspecification, especially when the outcome regression model was misspecified while the propensity score model was correctly specified. Under this scenario, the EIF-based estimator substantially overestimated the variance and produced overly conservative confidence interval coverage. In contrast, both the nonparametric bootstrap and sandwich variance estimators remained stable across all nuisance-model specification scenarios, yielding variance estimates close to the empirical Monte Carlo variance and coverage probabilities near the nominal level whenever at least one nuisance model was correctly specified. 
\begin{table*}[!t]
\centering
\caption{Simulation results for the doubly robust estimator of $\mathrm{PTE}(7.95)$ under different nuisance-model specifications.}
\label{tab:pte_dr_variance}

\begin{threeparttable}
\small
\setlength{\tabcolsep}{5pt}
\renewcommand{\arraystretch}{1.08}

\begin{tabular}{S[table-format=-1.2e-1]
                S[table-format=1.3]
                S[table-format=1.3]
                S[table-format=1.3]
                S[table-format=2.1]
                S[table-format=1.3]
                S[table-format=2.1]
                S[table-format=1.3]
                S[table-format=2.1]
                S[table-format=1.3]
                S[table-format=2.1]}
\toprule
& & & 
\multicolumn{2}{c}{Empirical} &
\multicolumn{2}{c}{Bootstrap} &
\multicolumn{2}{c}{EIF-based} &
\multicolumn{2}{c}{Sandwich} \\
\cmidrule(lr){4-5}
\cmidrule(lr){6-7}
\cmidrule(lr){8-9}
\cmidrule(lr){10-11}
{Bias} & {RMSE} & {MAE}
& {SD} & {Cov., \%}
& {SE} & {Cov., \%}
& {SE} & {Cov., \%}
& {SE} & {Cov., \%} \\
\midrule

\multicolumn{11}{c}{Correct OR, Correct PS} \\
-4.40e-04 & 0.019 & 0.015 & 0.019 & 95.1 & 0.019 & 93.8 & 0.019 & 94.4 & 0.019 & 94.4 \\
\midrule

\multicolumn{11}{c}{Correct OR, Misspecified PS} \\
-4.64e-04 & 0.018 & 0.014 & 0.018 & 94.9 & 0.018 & 93.6 & 0.017 & 93.9 & 0.017 & 93.9 \\
\midrule

\multicolumn{11}{c}{Misspecified OR, Correct PS} \\
-1.13e-04 & 0.021 & 0.017 & 0.021 & 95.1 & 0.021 & 94.6 & 0.031 & 99.7 & 0.021 & 94.8 \\
\midrule

\multicolumn{11}{c}{Misspecified OR, Misspecified PS} \\
-1.02e-01 & 0.106 & 0.102 & 0.028 & 3.5 & 0.028 & 4.5 & 0.028 & 3.7 & 0.028 & 3.7 \\
\bottomrule
\end{tabular}

\begin{tablenotes}
\footnotesize
\item Note: The target parameter is $\mathrm{PTE}(7.95)$, with true value $-0.23$. OR denotes the outcome regression model and PS denotes the propensity score model. Bias is the average estimate minus the truth. RMSE denotes root mean squared error, MAE denotes mean absolute error, and SD denotes the empirical Monte Carlo standard deviation. SE denotes the average estimated standard error. Cov. denotes the empirical coverage probability of the nominal 95\% confidence interval. Bootstrap confidence intervals are constructed using the percentile method, whereas all other intervals are based on normal approximations. The number of bootstrap is 500. For each scenario, results were based on 1{,}000 Monte Carlo replicates with sample size $n=1000$.
\end{tablenotes}

\end{threeparttable}
\end{table*}

\subsection{Computational time}
\FloatBarrier

\begin{figure}[ht]
    \centering
    \includegraphics[height=0.38\textheight]{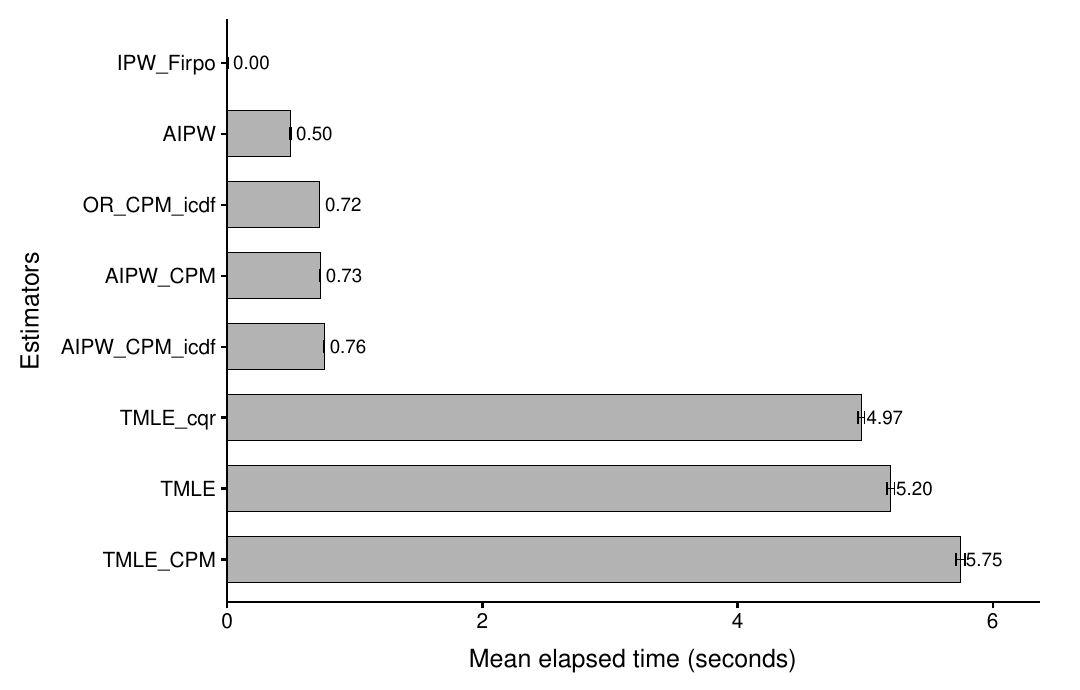}
   \caption{Average computational time of the considered estimators over 1,000 simulation replicates. 
Bars display the mean elapsed time (seconds), and error bars indicate the 95\% Monte Carlo confidence intervals. This simulation was conducted on a MacBook Air (M2, 2022) equipped with an Apple M2 CPU and 8 GB of RAM. The software environment was R version 4.4.2.}
\end{figure}

\FloatBarrier

\end{document}